\documentclass[12pt]{article}
\usepackage[dvips]{graphicx}

\newcommand{\be}{\begin{equation}}
\newcommand{\ee}{\end{equation}}
\newcommand{\beq}{\begin{equation}}
\newcommand{\eeq}{\end{equation}}
\newcommand{\bea}{\begin{eqnarray}}
\newcommand{\eea}{\end{eqnarray}}

\begin{document}
\baselineskip=15.5pt
\pagestyle{plain}
\setcounter{page}{1}


\def\del{{\partial}}
\def\vev#1{\left\langle #1 \right\rangle}
\def\cn{{\cal N}}
\def\co{{\cal O}}
\newfont{\Bbb}{msbm10 scaled 1200}     
\newcommand{\mathbb}[1]{\mbox{\Bbb #1}}
\def\IC{{\mathbb C}}
\def\IR{{\mathbb R}}
\def\IZ{{\mathbb Z}}
\def\RP{{\bf RP}}
\def\CP{{\bf CP}}
\def\Poincare{{Poincar\'e }}
\def\tr{{\rm tr}}
\def\tp{{\tilde \Phi}}

\def\TL{\hfil$\displaystyle{##}$}
\def\TR{$\displaystyle{{}##}$\hfil}
\def\TC{\hfil$\displaystyle{##}$\hfil}
\def\TT{\hbox{##}}
\def\HLINE{\noalign{\vskip1\jot}\hline\noalign{\vskip1\jot}}
\def\seqalign#1#2{\vcenter{\openup1\jot
  \halign{\strut #1\cr #2 \cr}}}
\def\lbldef#1#2{\expandafter\gdef\csname #1\endcsname {#2}}
\def\eqn#1#2{\lbldef{#1}{(\ref{#1})}%
\begin{equation} #2 \label{#1} \end{equation}}
\def\eqalign#1{\vcenter{\openup1\jot
    \halign{\strut\span\TL & \span\TR\cr #1 \cr
   }}}
\def\eno#1{(\ref{#1})}
\def\href#1#2{#2}
\def\half{{1 \over 2}}

\def\ads{{\it AdS}}
\def\adsp{{\it AdS}$_{p+2}$}
\def\cft{{\it CFT}}

\newcommand{\ber}{\begin{eqnarray}}
\newcommand{\eer}{\end{eqnarray}}

\newcommand{\beqar}{\begin{eqnarray}}
\newcommand{\cN}{{\cal N}}
\newcommand{\cO}{{\cal O}}
\newcommand{\cA}{{\cal A}}
\newcommand{\cT}{{\cal T}}
\newcommand{\cF}{{\cal F}}
\newcommand{\cC}{{\cal C}}
\newcommand{\cR}{{\cal R}}
\newcommand{\cW}{{\cal W}}
\newcommand{\eeqar}{\end{eqnarray}}
\newcommand{\tht}{\thteta}
\newcommand{\lm}{\lambda}\newcommand{\Lm}{\Lambda}
\newcommand{\eps}{\epsilon}


\newcommand{\nonu}{\nonumber}
\newcommand{\oh}{\displaystyle{\frac{1}{2}}}
\newcommand{\dsl}
  {\kern.06em\hbox{\raise.15ex\hbox{$/$}\kern-.56em\hbox{$\partial$}}}
\newcommand{\id}{i\!\!\not\!\partial}
\newcommand{\as}{\not\!\! A}
\newcommand{\ps}{\not\! p}
\newcommand{\ks}{\not\! k}
\newcommand{\D}{{\cal{D}}}
\newcommand{\dv}{d^2x}
\newcommand{\Z}{{\cal Z}}
\newcommand{\N}{{\cal N}}
\newcommand{\Dsl}{\not\!\! D}
\newcommand{\Bsl}{\not\!\! B}
\newcommand{\Psl}{\not\!\! P}
\newcommand{\eeqarr}{\end{eqnarray}}
\newcommand{\ZZ}{{\rm \kern 0.275em Z \kern -0.92em Z}\;}

                                                                                                    
\def\del{{\delta^{\hbox{\sevenrm B}}}} \def\ex{{\hbox{\rm e}}}
\def\azb{A_{\bar z}} \def\az{A_z} \def\bzb{B_{\bar z}} \def\bz{B_z}
\def\czb{C_{\bar z}} \def\cz{C_z} \def\dzb{D_{\bar z}} \def\dz{D_z}
\def\im{{\hbox{\rm Im}}} \def\mod{{\hbox{\rm mod}}} \def\tr{{\hbox{\rm Tr}}}
\def\ch{{\hbox{\rm ch}}} \def\imp{{\hbox{\sevenrm Im}}}
\def\trp{{\hbox{\sevenrm Tr}}} \def\vol{{\hbox{\rm Vol}}}
\def\rl{\Lambda_{\hbox{\sevenrm R}}} \def\wl{\Lambda_{\hbox{\sevenrm W}}}
\def\fc{{\cal F}_{k+\cox}} \def\vev{vacuum expectation value}
\def\nodiv{\mid{\hbox{\hskip-7.8pt/}}}
\def\ie{{\em i.e.}}
\def\ie{\hbox{\it i.e.}}

\def\CC{{\mathchoice
{\rm C\mkern-8mu\vrule height1.45ex depth-.05ex
width.05em\mkern9mu\kern-.05em}
{\rm C\mkern-8mu\vrule height1.45ex depth-.05ex
width.05em\mkern9mu\kern-.05em}
{\rm C\mkern-8mu\vrule height1ex depth-.07ex
width.035em\mkern9mu\kern-.035em}
{\rm C\mkern-8mu\vrule height.65ex depth-.1ex
width.025em\mkern8mu\kern-.025em}}}
                                                                                                    
\def\RR{{\rm I\kern-1.6pt {\rm R}}}
\def\NN{{\rm I\!N}}
\def\ZZ{{\rm Z}\kern-3.8pt {\rm Z} \kern2pt}
\def\IB{\relax{\rm I\kern-.18em B}}
\def\ID{\relax{\rm I\kern-.18em D}}
\def\II{\relax{\rm I\kern-.18em I}}
\def\IP{\relax{\rm I\kern-.18em P}}
\newcommand{\CS}{{\scriptstyle {\rm CS}}}
\newcommand{\CSs}{{\scriptscriptstyle {\rm CS}}}
\newcommand{\rc}{\nonumber\\}
\newcommand{\bear}{\begin{eqnarray}}
\newcommand{\eear}{\end{eqnarray}}
\newcommand{\W}{{\cal W}}
\newcommand{\F}{{\cal F}}
\newcommand{\x}{{\cal O}}
\newcommand{\LL}{{\cal L}}
                                                                                                    
\def\mani{{\cal M}}
\def\calo{{\cal O}}
\def\calb{{\cal B}}
\def\calw{{\cal W}}
\def\calz{{\cal Z}}
\def\cald{{\cal D}}
\def\calc{{\cal C}}
\def\to{\rightarrow}
\def\ele{{\hbox{\sevenrm L}}}
\def\ere{{\hbox{\sevenrm R}}}
\def\zb{{\bar z}}
\def\wb{{\bar w}}
\def\nodiv{\mid{\hbox{\hskip-7.8pt/}}}
\def\menos{\hbox{\hskip-2.9pt}}
\def\dr{\dot R_}
\def\drr{\dot r_}
\def\ds{\dot s_}
\def\da{\dot A_}
\def\dga{\dot \gamma_}
\def\ga{\gamma_}
\def\dal{\dot\alpha_}
\def\al{\alpha_}
\def\cl{{closed}}
\def\cls{{closing}}
\def\vev{vacuum expectation value}
\def\tr{{\rm Tr}}
\def\to{\rightarrow}
\def\too{\longrightarrow}


\def\a{\alpha}
\def\b{\beta}
\def\c{\gamma}
\def\d{\delta}
\def\e{\epsilon}           
\def\f{\phi}               
\def\vf{\varphi}  \def\tvf{\tilde{\varphi}}
\def\vp{\varphi}
\def\g{\gamma}
\def\h{\eta}
\def\i{\iota}
\def\j{\psi}
\def\k{\kappa}                    
\def\l{\lambda}
\def\m{\mu}
\def\n{\nu}
\def\o{\omega}  \def\w{\omega}
\def\q{\theta}  \def\th{\theta}                  
\def\r{\rho}                                     
\def\s{\sigma}                                   
\def\t{\tau}
\def\u{\upsilon}
\def\x{\xi}
\def\z{\zeta}
\def\pt{\tilde{\varphi}}
\def\tt{\tilde{\theta}}
\def\lab{\label}  
\def\6{\partial}
\def\wg{\wedge}
\def\atanh{{\rm arctanh}}
\def\bpsi{\bar{\psi}}
\def\bt{\bar{\theta}}
\def\bvf{\bar{\varphi}}

%
                                                                                                    
\newfont{\namefont}{cmr10}
\newfont{\addfont}{cmti7 scaled 1440}
\newfont{\boldmathfont}{cmbx10}
\newfont{\headfontb}{cmbx10 scaled 1728}
\renewcommand{\theequation}{{\rm\thesection.\arabic{equation}}}
\font\cmss=cmss10 \font\cmsss=cmss10 at 7pt
\par\hfill ITP-UU-08/79, SPIN-08/61

\begin{center}
{\LARGE{\bf The Klebanov-Strassler model with massive \\ \vskip 10pt dynamical flavors}}
\end{center}
\vskip 10pt
\begin{center}
{\large 
Francesco Bigazzi $^{a}$, Aldo L. Cotrone $^{b}$, Angel Paredes $^{c}$, \\
Alfonso V. Ramallo $^{d}$}
\end{center}
\vskip 10pt
\begin{center}
\textit{$^a$ Physique Th\'eorique et Math\'ematique and International Solvay
Institutes, Universit\'e Libre de Bruxelles; CP 231, B-1050
Bruxelles, Belgium.}\\
\textit{$^b$  Institute for theoretical physics, K.U. Leuven;
Celestijnenlaan 200D, B-3001 Leuven,
Belgium.}\\
\textit{$^c$ Institute for Theoretical Physics, Utrecht University; Leuvenlaan 4,
3584 CE Utrecht, The Netherlands.
}\\
\textit{$^d$ Departamento de  Fisica de Particulas, Universidade de Santiago de Compostela and Instituto Galego de 
Fisica de Altas Enerxias (IGFAE); E-15782, Santiago de Compostela, Spain.}\\
{\small fbigazzi@ulb.ac.be, Aldo.Cotrone@fys.kuleuven.be, A.ParedesGalan@uu.nl, alfonso@fpaxp1.usc.es}
\end{center}

\vspace{15pt}

\begin{center}
\textbf{Abstract}
\end{center}

\vspace{4pt}{\small \noindent We present a fully backreacted D3-D7 supergravity solution dual to the Klebanov-Strassler cascading gauge theory coupled to a large number of massive dynamical flavors in the Veneziano limit. The mass of the flavors can be larger or smaller than the dynamically generated scale.
The solution is always regular at the origin of the radial coordinate and as such it can be suitably employed to explore the rich IR physics of the dual gauge theory. In this paper we focus on the static quark-antiquark potential, the screening of chromoelectric charges induced by the dynamical flavors, the flux tube breaking and the mass spectrum of the first mesonic excitations. Moreover, we discuss the occurrence of quantum phase transitions in the connected part of the static quark-antiquark potential. Depending on the ratio of certain parameters, like the flavor mass, with respect to some critical values, we find a discontinuous (first order) or smooth transition from a Coulomb-like to a linear phase. We evaluate the related critical exponents finding that they take classical mean-field values and argue that this is a universal feature of analogous first order transitions occurring in the static potential for planar gauge theories having a dual supergravity description.}

\vfill

\newpage
\section{Introduction}
The quarks in QCD can be distinguished between light ($u,d,s$) and heavy ($c,b,t$), depending on their mass 
being smaller or (much) larger than the scale $\Lambda_{IR}$ dynamically generated via dimensional transmutation. 
The main vacuum polarization effects related to the flavors are due to the three light quarks, while the others, which can be mainly considered as ``probes'' of the theory, can be neglected in the path integral with a good approximation.

Nowadays, the only systematic, first-principles-based, non-perturbative approach to low energy QCD is provided by the study of extrapolations to the continuum of numerical simulations of the theory on a Euclidean space-time lattice of finite volume \cite{wilson}. On the lattice it is practically extremely hard to account for the vacuum polarization effects due to the light quarks and so many results are obtained using an approximation where they are treated as probes. This is the so-called ``quenched'' approximation. There are certainly indications about the possibility of overcoming this limit in the future and some partial results towards ``unquenching''  the lattice are known (see for example \cite{davies}). 

In the meantime we have to take in mind that most of what we know from the lattice is about quenched QCD. This theory is strictly confining, like pure Yang-Mills, while the same is not true for real QCD: due to the presence of the dynamical (light) quarks, an external quark-antiquark pair $\bar Q Q$ will not experience an indefinitely linear potential at large distances. Instead, since a $\bar q q$ pair of dynamical flavors can be popped out of the vacuum, the initial $\bar Q Q$ state will decay into a pair of heavy-light mesons $\bar Q q+ \bar q Q$ for distances larger than a certain ``screening length'' (at which the flux tube becomes sufficiently energetic for the decay to happen).
 
Remarkably, the string/gauge theory correspondence, which aims to be a complementary approach towards explaining the non-perturbative dynamics of QCD, offers a set of simple tools to analyze particular unquenched gauge theories in certain regimes. This is certainly not an ideal setting at the moment, since the present computational methods allow us to give a string dual description to theories which are at most (supersymmetric) extensions of planar ($N_c\rightarrow\infty$) QCD, where the phenomenologically interesting sector is coupled to spurious matter or has some higher dimensional UV completion. 
Nevertheless, the string approach can give valuable insights into many features of strongly coupled gauge theories, in regimes inaccessible by other methods.

In this paper we present an exact (and mostly analytic) fully backreacted D3-D7 supergravity solution dual to the 4d Klebanov-Strassler (KS) cascading gauge theory \cite{ks}, coupled to a large number of dynamical fundamental matter fields which are massive and non-chiral (relevant related studies are in Refs. \cite{localized}-\cite{alfonsoetal}). The unflavored theory is known to be confining and to share many notable properties with 4d ${\cal N}=1$ SYM. The flavored model flows in the IR to a SQCD-like theory. The dual supergravity solution allows one to consider, without limitations,  either light ($m<\Lambda_{IR}$) or heavy ($m>\Lambda_{IR}$) dynamical flavors. Remarkably, the solution is always regular at the origin of the radial coordinates and as such it is a very promising tool to explore the IR dynamics of the dual gauge theory.

With this aim, we begin to explore some relevant physical observables, concentrating in particular on their dependence on the flavor parameters.
We first focus on the static quark-antiquark potential and study its behavior as a function of the number $N_f$ and the masses of the flavors, keeping fixed different relevant field theory scales. We show how the solution accounts for the string breaking and screening effects due to the dynamical flavors. In particular it enables us to provide a first qualitative study of the behavior of the screening lengths as functions of the flavor parameters. We also start up an analysis of the mesonic spectrum, by considering fluctuations of the worldvolume gauge field of a $\kappa$-symmetric D7-brane probe corresponding to massless flavors.
We show how it is possible to study quantitatively how the spectrum varies w.r.t. to the number of sea flavors and their masses. Previous studies along these lines in models with flavor backreaction can be found in \cite{vaman,Paredes:2006wb,Bigazzi:2008gd,massivekw,varna,alfonsoetal}.

Finally, we analyze the interesting phenomenon of the occurrence  of first order quantum phase transitions in the ``connected'' part\footnote{In the dual picture this refers to the macroscopic open string describing the metastable $\bar Q Q$ state when the mixing with the final heavy-light mesons $\bar Q q + \bar q Q$, i.e. the ``disconnected'' string configurations, is artificially turned off.} of the static quark-antiquark potential. These transitions are discontinuous changes in the slope of the potential and actually occur in several models with and without flavors when at least two separate physical scales are present \cite{sfetsos,angelnc,avramista,avramis,Bigazzi:2008gd,massivekw,cargese,alfonsoetal}. For the case at hand the potential passes discontinuously from a Coulomb-like to a linear behavior. 
The transition disappears, i.e. the potential becomes smooth, for flavor masses above a certain critical value, or by varying some other mass parameters in the theory. In the present setup it is possible to evaluate the related critical exponents and to show that they take the classical mean-field values. 
We argue, by means of catastrophe theory, that this is a universal feature of every first order transition discovered, in different (also unflavored) models, in the static potential holographically evaluated by means of string theory. 
\subsection{Techniques and structure of the paper}
The addition of fundamental matter to the KS model is realized, following the general suggestion of \cite{kk}, by means of space-time filling D7-branes which are holomorphically embedded (so  as to preserve ${\cal N}=1$ supersymmetry) and wrapped on non-compact submanifolds of the transverse space \cite{kuper}. Just as in lattice gauge theory, the task of adding flavors in the stringy setup becomes computationally simpler if the vacuum polarization effects due to the fundamental matter fields are neglected. The quenched approximation is realized by neglecting the backreaction of the flavor D-branes on the background and treating them as external probes. 

In order to go beyond the quenched approximation we use a simple technique which was introduced in \cite{paris,cnp}: we homogeneously smear the flavor branes along their transverse directions. This operation is sensible only if the number $N_f$ of such branes is very large, as happens in the Veneziano regime (where $N_f, N_c\rightarrow\infty$ with $N_f/N_c$ fixed). In the smeared setups the flavor symmetry group is generically broken to a product of abelian ones, but this limitation does not spoil many features of physical interest in the related models.

When the flavors are massless the related smeared branes reach the origin of the transverse space. At this point the flavor symmetry is generically enhanced since the flavor branes overlap. In the known cases this is accompanied by the presence of a (good) singularity in the dual string solutions. This kind of singularity can be avoided in the massive case, where the smeared branes generically extend up to a certain finite distance from the origin along the radial direction and there is no special point where the flavor symmetry is fully enhanced. For this reason it is extremely interesting to focus on smeared-flavor-brane setups where the dual gauge theories are coupled to massive dynamical flavors.  

Using the smearing technique, the addition of massless dynamical flavors to the KS gauge theory, giving a solution with a (good) singularity in the IR, was considered in \cite{Benini:2007gx}. An \emph{approximate} solution corresponding to the inclusion of massive dynamical flavors to the Chamseddine-Volkov-Maldacena-Nunez confining theory \cite{mn}, was given in \cite{Bigazzi:2008gd} and was regular in the IR by construction.
In order to build up an \emph{exact} solution accounting for massive flavors in the KS case, we have to evaluate the density distribution of the smeared flavor D7-branes. This depends on a function of the radial variable, $N_f(\tau)$, which accounts for the effective number of flavor degrees of freedom at a given energy scale. 
Remarkably, despite the complicated setting, we are able to find the explicit expression of $N_f(\tau)$.
The precise knowledge of this function enables us to write up the fully backreacted D3-D7 solution.
Previous calculations of the analogous function and the derivation of the corresponding backgrounds in the singular conifold case were performed in \cite{massivekw,varna}.

This paper is organized as follows.
In section \ref{gendefcon} we derive the function $N_f(\tau)$, the main formulas being (\ref{nfdefcon1}), (\ref{nfdefcon2}), (\ref{nfdefcon2b}). In section \ref{gencons} we calculate the full supergravity solution dual to the KS model with dynamical massive flavors (the solution can be found in section \ref{thesol}).
In section \ref{wilson} we study the static quark-antiquark potential and the screening lengths. In section \ref{phases} we analyze the first order quantum phase transitions in the static quark-antiquark potential. In section \ref{mesons} we focus on some mesonic mass spectra. We end up in section \ref{conclu} with some concluding remarks and a sketch of possible future research lines. The paper includes various appendices where many details of the calculations and validity checks are provided.
\section{Massive non-chiral flavors and smeared D7-branes on the deformed conifold}
\label{gendefcon}
\setcounter{equation}{0}
The deformed conifold is a regular six dimensional non compact manifold defined by the equation $z_1\,z_2-z_3\,z_4 =\epsilon^2$ in $\IC^4$. When the complex deformation parameter $\epsilon$ is turned off, it reduces to the singular conifold, which is invariant under complex rescaling of the $z_i$, has $SU(2)\times SU(2)\times U(1)$ isometry and $S^2\times S^3$ topology. The deformation parameter breaks the scale invariance, produces a blown-up $S^3$ at the apex of the conifold and breaks the $U(1)$ isometry to  ${\mathbb Z}_2$. 

The low energy dynamics of $N$ regular and $M$ fractional D3-branes on the deformed conifold is described by a cascading ${\cal N}=1$ 4d gauge theory with gauge group $SU(N+M)\times SU(N)$ and bifundamental matter fields $A, B$ transforming as $SU(2)\times SU(2)$ doublets and interacting with a quartic superpotential $W_{KW}=\epsilon^{ij}\epsilon^{kl}A_iB_kA_jB_l$. The KS solution \cite{ks} is relevant for the $N=n\,M$ case, where $n$ is an integer. The related theory develops a Seiberg duality cascade which stops after $n-1$ steps when the gauge group is reduced to $SU(2M)\times SU(M)$. The regular KS solution precisely accounts for the physics of an $A\leftrightarrow B$-symmetric point in the baryonic branch of the latter theory, which exhibits confinement and $U(1)_R\rightarrow {\mathbb Z}_{2N}\rightarrow {\mathbb Z}_2$ breaking due to the formation of a gluino condensate $\langle\lambda\lambda\rangle\sim\Lambda^3_{IR}$. The complex parameter $\epsilon$ is the geometric counterpart of this condensate.

Let us consider the addition of fundamental degrees of freedom to the theory. This can be realized by means of suitably chosen D7-branes. A relevant example is given by D7-branes wrapping the holomorphic 4-cycle defined by an equation of the form 
\be
z_1-z_2=2\hat\mu\,.
\ee
It was shown in \cite{kuper} that this embedding is $\kappa$-symmetric and hence preserves the four supercharges of the deformed conifold theory.

A D7-brane wrapping the 4-cycle defined above is conjectured to add a massless (if $\hat\mu=0$) or massive (anti) fundamental flavor to a node of the KS model. The resulting gauge theory is ``non-chiral'' because the flavor mass terms do not break the classical flavor symmetry of the massless theory. The related perturbative superpotential is, just as in the singular conifold case \cite{ouyang},
\begin{equation}
W = W_{KW} +\hat h_1\, \tilde q_1 (A_1B_1-A_2B_2 )q_1 + \hat h_2\,\tilde q_2 (B_1A_1-B_2A_2)q_2 + k_i\,(\tilde q_i q_i)^2 + m\,(\tilde q_i q_i)\,,
\label{superpot}
\end{equation}
where we have considered a  ${\mathbb Z}_2$-invariant setup with two stacks of D7-branes adding the same number of fundamental degrees of freedom (with same masses) to both nodes.  Thus the complex mass parameter $m$ in $W$ is mapped to the geometrical parameter $\hat\mu$. 

In the singular conifold case the massive embedding $z_1-z_2=2\hat\mu$ explicitly breaks the scale invariance and the $U(1)$ isometry of the background geometry. This is related to the explicit breaking of conformal invariance and $U(1)_R$ symmetry due to the mass terms in the dual gauge theory. The embedding equation also breaks part of the non abelian symmetry group of the conifold to a diagonal $SU(2)$ subgroup. 
\subsection{The D7-brane profile}
The metric of the deformed conifold is usually written as
\be
ds_6^2 = \frac{1}{2} \epsilon^{4/3} K(\tau) \left[ \frac{1}{3K^3(\tau)}(d\tau^2+(g^5)^2)+\cosh^2(\frac{\tau}{2})((g^3)^2+(g^4)^2)+\sinh^2(\frac{\tau}{2})((g^1)^2+(g^2)^2)\right]\,,
\label{ds6unflav}
\ee
where 
\be
K(\tau) = \frac{(\sinh(2\tau)-2\tau)^{1/3}}{2^{1/3}\sinh(\tau)}\,,
\ee
and
\bear
g^1 &=& \frac{-\sin\theta_1\,d\varphi_1 -\cos\psi\sin\theta_2\,d\varphi_2 +\sin\psi\,d\theta_2}{\sqrt{2}}\,,\rc
g^2 &=& \frac{d\theta_1-\sin\psi\sin\theta_2\,d\varphi_2 -\cos\psi\,d\theta_2}{\sqrt{2}}\,,\rc
g^3 &=& \frac{-\sin\theta_1\,d\varphi_1 +\cos\psi\sin\theta_2\,d\varphi_2 -\sin\psi\,d\theta_2}{\sqrt{2}}\,,\rc
g^4 &=& \frac{d\theta_1+\sin\psi\sin\theta_2\,d\varphi_2 +\cos\psi\,d\theta_2}{\sqrt{2}}\,,\rc
g^5&=& d\psi +\cos\theta_1\,d\varphi_1 + \cos\theta_2\,d\varphi_2\,.
\label{gis}
\eear
The range of the angles is $\psi \in [0,4\pi)$, $\varphi_i \in [0,2\pi)$, $\theta_i \in [0,\pi]$, while $\tau\in [0,\infty)$. For $\tau\rightarrow\infty$ the metric asymptotes the singular conifold one. In terms of these coordinates the non-chiral embedding $z_1-z_2=2\hat\mu$ can be written as
\be
\Theta_1\,\sinh\frac{\tau}{2} -i\,\Theta_2\,\cosh\frac{\tau}{2} = \frac{\hat\mu}{\epsilon}\,,
\label{lockup}
\ee
where
\bear
\Theta_1&=&\sin\frac{\theta_1}{2}\sin\frac{\theta_2}{2}\cos\frac{\varphi_1+\varphi_2-\psi}{2}-\cos\frac{\theta_1}{2}\cos\frac{\theta_2}{2}\cos\frac{\varphi_1+\varphi_2+\psi}{2}\,,\nonumber \\
\Theta_2&=&\sin\frac{\theta_1}{2}\sin\frac{\theta_2}{2}\sin\frac{\varphi_1+\varphi_2-\psi}{2}+\cos\frac{\theta_1}{2}\cos\frac{\theta_2}{2}\sin\frac{\varphi_1+\varphi_2+\psi}{2}\,,
\eear
and $|\Theta_i|\le 1$. From these equations it follows that the profile of the D7-branes has a non trivial radial dependence: the branes extend all along the radial direction up to a minimum  distance $\tau_{min}$ which depends on the relative phase of the $\hat\mu$ and $\epsilon$ parameters. 

If $\hat\mu/\epsilon$ is purely imaginary (resp. real), the embedding equations imply $\Theta_1=0$ (resp. $\Theta_2=0$) and $\tau_{min}=\tau_a=2{\rm arc}\cosh(|\hat\mu|/|\epsilon|)$ (resp. $\tau_{min}=\tau_b=2{\rm arc}\sinh(|\hat\mu|/|\epsilon|)$). For generic phases $\tau_a<\tau_{min}<\tau_b$. Notice that in the ``completely misaligned'' case where $\hat\mu/\epsilon$ is purely imaginary, $\tau_{min}=0$ - i.e. the D7-brane reaches the tip of the deformed conifold - if  $|\hat\mu|<|\epsilon|$. 

Let us now define $\tau_q\equiv\tau_a$ as the absolute minimal value of $\tau_{min}$. Related to this, one can introduce a parameter $m_q$, the absolute minimum flavor ``constituent mass'',\footnote{This is how the $m_q$ parameter is usually called in the literature. In the present context we adopt the same name with an abuse of language.} defined as the energy of an hypothetical straight string stretched along the radial direction from $\tau=0$ to $\tau_q$. If  $|\hat\mu|<|\epsilon|$, then $m_q=0$. As discussed above, the dimensional parameters $\hat\mu$ and $\epsilon$ can be related to the bare flavor mass $m$ (see eq. (\ref{superpot})) and the fundamental scale $\Lambda_{IR}$ of the dual gauge theory. Thus the relation  $|\hat\mu|<|\epsilon|$ can be interpreted as $m<\Lambda_{IR}$. Though at the level of the ``constituent mass'' we do not see differences between the  $|\hat\mu|<|\epsilon|$ and the $\hat\mu=0$ cases, we will see that the non zero bare mass parameter influences the density distribution of the flavor branes also when $m<\Lambda_{IR}$. This will thus mark a difference with the $m=0$ setup.

\subsection{The density distribution for smeared D7-branes}
\label{smedefcon}
Acting with an $SO(4)\sim SU(2)\times SU(2)$ rotation on the embedding equation $z_1-z_2=2\hat\mu$, we obtain the generalized embedding equation 
\be
\bar p z_1 - p z_2 + \bar q z_3 + q z_4 = 2\hat\mu\equiv 2|\hat\mu|e^{i\beta}\,,
\label{genembeddingks}
\ee
where $p, q$ span a unit 3-sphere
\begin{equation}
p=\cos\frac{\theta}{2}e^{i(\frac{\chi+\phi}{2})}\,,\quad q=\sin\frac{\theta}{2}e^{i(\frac{\chi-\phi}{2})}\,,
\end{equation}
and $\chi \in [0,4\pi)$, $\phi \in [0,2\pi)$, $\theta \in [0,\pi]$, $\beta \in [0,2\pi]$.

Let us now consider a maximal symmetric smeared distribution of $N_f\gg1$ D7-branes generally embedded as above. By ``maximal'' we mean that the distribution will not only be invariant under the $SU(2) \times SU(2)$ isometry of the deformed conifold, but also under the symmetry $U(1)_{\psi}$, under shifts of the $\psi$ angle, which is broken (to a ${\mathbb Z}_2$ subgroup) by the deformed conifold geometry. We will thus homogeneously distribute the D7-branes along $\chi,\phi,\theta$ as well as along the phase of the mass term $\beta$. We will instead take the modulus $|\hat\mu|$ (hence the modulus of the flavor mass parameter in the dual field theory) to be fixed. Smearing along $\beta$ will cause different D7-branes to reach different minimal distances $\tau_{min}$ from the origin. The whole distribution will end up at the absolute minimal distance $\tau_q$.  

The density distribution $\Omega$ of the smeared D7-branes is given by
\be
\Omega = \int \sigma_{\theta,\phi,\chi,\beta} \left(\delta(f_1) \delta(f_2) df_1 \wedge df_2
\right) d\theta\, d\phi\, d\chi\, d\beta\,,
\label{generalOmega}
\ee
where we have introduced the properly normalized density function $\sigma_{\theta,\phi,\chi,\beta} = N_f \sin\theta/32 \pi^3$ and $f_1=0$, $f_2=0$ are the two real constraints implied by the complex equation (\ref{genembeddingks}) (see appendix \ref{appNf}).

The symmetries strongly constrain the form of $\Omega$. As was shown in  \cite{Benini:2006hh}, the only possibility for an exact two-form preserving $SU(2) \times SU(2) \times U(1)_\psi \times{\mathbb Z}_2$ is, in the present setup,
\begin{equation}
\Omega = \frac{N_f(\tau)}{4\pi}(\sin\theta_1 d\theta_1 \wedge d\varphi_1
+ \sin\theta_2 d\theta_2 \wedge d\varphi_2)
-\frac{\dot N_f(\tau)}{4\pi} d\tau\wedge(d\psi + \cos \theta_1 d\varphi_1 + 
\cos \theta_2 d\varphi_2)\,,
\label{massive_2form}
\end{equation}
where the the dot means derivative w.r.t. $\tau$. The function $N_f(\tau)$ counts the effective number of dynamical flavors at a given energy scale, holographically related to $\tau$. It crucially depends on the particular kind of smeared embedding. Referring to appendix \ref{appNf}  for the details of the non trivial and very instructive calculation and defining 
\be
x\equiv\cosh\tau,\qquad \qquad \mu \equiv \frac{|\hat\mu|}{|\epsilon|}\,,
\ee
we find that in the present setup the function $N_f(x)$ is the solution of the first order equation
\be
\frac{dN_f(x)}{dx}= \frac{N_f \mu^2}{2\pi}\frac{\Theta[x-2\mu^2+1]}{(x^2-1)^2}\left[I_1(x)-\frac{4}{\mu^2}\Theta[2\mu^2+1-x]\,I_2(x)\right]\,,
\label{nfprimeeq}
\ee 
where $\Theta[y]$ is the Heaviside step function and
\bear
I_1&=&4\pi(1+x^2)\,,\rc
I_2&=& x\sqrt{(1+2\mu^2-x)(1-2\mu^2+x)(x^2-1)}+2\mu^2(1+x^2)\arctan\left[\sqrt{\frac{(1+x)(1+2\mu^2-x)}{(x-1)(1-2\mu^2+x)}}\right]\,. \nonumber
\eear
In the massless case $\mu=0$ we have $N_f(\tau)={\rm const}=N_f$ \cite{Benini:2007gx}. Moreover, for $x\le 2\mu^2-1$ we have $N_f(x)={\rm const}=0$.

Let us now split (\ref{nfprimeeq}) into two regions. {\it In region I}, $x>2\mu^2+1$ (i.e. $\tau>\tau_b$), we find the following simple solution
\be
N^{(I)}_f(x) = N_f\left[1 - 2\mu^2 \frac{x}{x^2-1}\right]\,, 
\label{nfdefcon1}
\ee
where the integration constant is fixed by consistency so that $N_f(\infty)=N_f$. Notice that in the large $\tau$ limit (with $|\epsilon|^2 e^{\tau}\sim r^3=e^{3\rho}$), this function asymptotes to the expression found in \cite{varna}, for the flavored version of the singular conifold Klebanov-Witten (KW) model \cite{kw}.

{\it In region II}, $2\mu^2-1<x<2\mu^2+1$ (i.e. $\tau_a<\tau<\tau_b$), we have a complicated expression in terms of Elliptic integrals of the first and third kind (${\cal F}[a\,|\,b]$ and $\Pi[a;b\,|\,c]$ respectively)
\be
N^{(II)}_f(x) = N_f\left[1 - 2\mu^2 \frac{x}{x^2-1}-\frac{4\mu^2}{\pi}\left(A_1(x,\mu^2) + A_2(x,\mu^2) - A_2(2\mu^2+1,\mu^2)\right)      \right]\,,
\label{nfdefcon2}
\ee
where
\bear
A_1(x,\mu^2)&=& -\frac{1}{4\mu^2}\sqrt{\frac{(x+1-2\mu^2)(2\mu^2+1-x)}{(x+1)(x-1)}} -\frac{x}{x^2-1}\arctan\left[\sqrt{\frac{(1+x)(1+2\mu^2-x)}{(x-1)(x+1-2\mu^2)}}\right]\,,\nonumber\\
A_2(x,\mu^2)&=& -\frac{i}{2\mu^4}(\mu^2-1){\cal F}\left[\arcsin\left(\sqrt{\frac{\mu^2(1+x)}{(1+\mu^2)(x-1)}}\right)\,|\,\frac{\mu^4-1}{\mu^4}\right]+\nonumber \\
&&-\frac{i}{\mu^4}{\Pi}\left[\frac{\mu^2+1}{\mu^2};\,\arcsin\left(\sqrt{\frac{\mu^2(1+x)}{(1+\mu^2)(x-1)}}\right)\,|\,\frac{\mu^4-1}{\mu^4}\right]\,.
\label{nfdefcon2b}
\eear
In (\ref{nfdefcon2}) we have fixed the integration constant by imposing continuity at $x=2\mu^2+1$, i.e. $N_f^{(I)}(x=2\mu^2+1)=N_f^{(II)}(x=2\mu^2+1)$. This condition is satisfied since $A_1(2\mu^2+1,\mu^2)=0$.

If $\mu>1$ and so $x_{min}=2\mu^2-1>1$, we also have that $N_f(x)$ is vanishing for $x<x_{min}$. Let us check continuity at $x=2\mu^2-1$. First, notice that
\be
A_1(2\mu^2-1,\mu^2)= -\frac{\pi}{8}\frac{(2\mu^2-1)}{\mu^2(\mu^2-1)}\,,
\ee
so that the above mentioned continuity condition amounts to having
\be
1-\frac{4\mu^2}{\pi}\left[A_2(2\mu^2-1,\mu^2) - A_2(2\mu^2+1,\mu^2)\right] = 0\,.
\ee
We have checked numerically that this condition is indeed satisfied and, more precisely, that
\be
1-\frac{4\mu^2}{\pi}\left[A_2(2\mu^2-1,\mu^2) - A_2(2\mu^2+1,\mu^2)\right] = -2\Theta[1-\mu]\,,
\ee
which thus vanishes for $\mu>1$.

\begin{figure}
\centering
\includegraphics[width=0.4\textwidth]{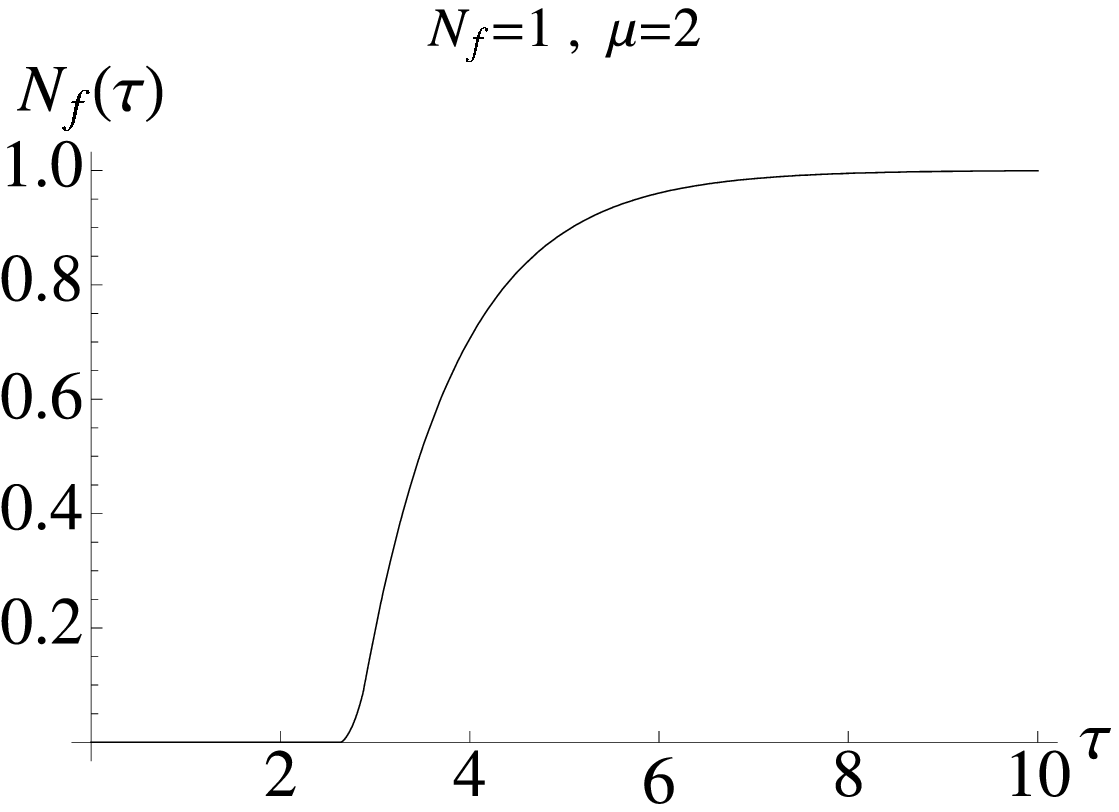} \hfill
\includegraphics[width=0.4\textwidth]{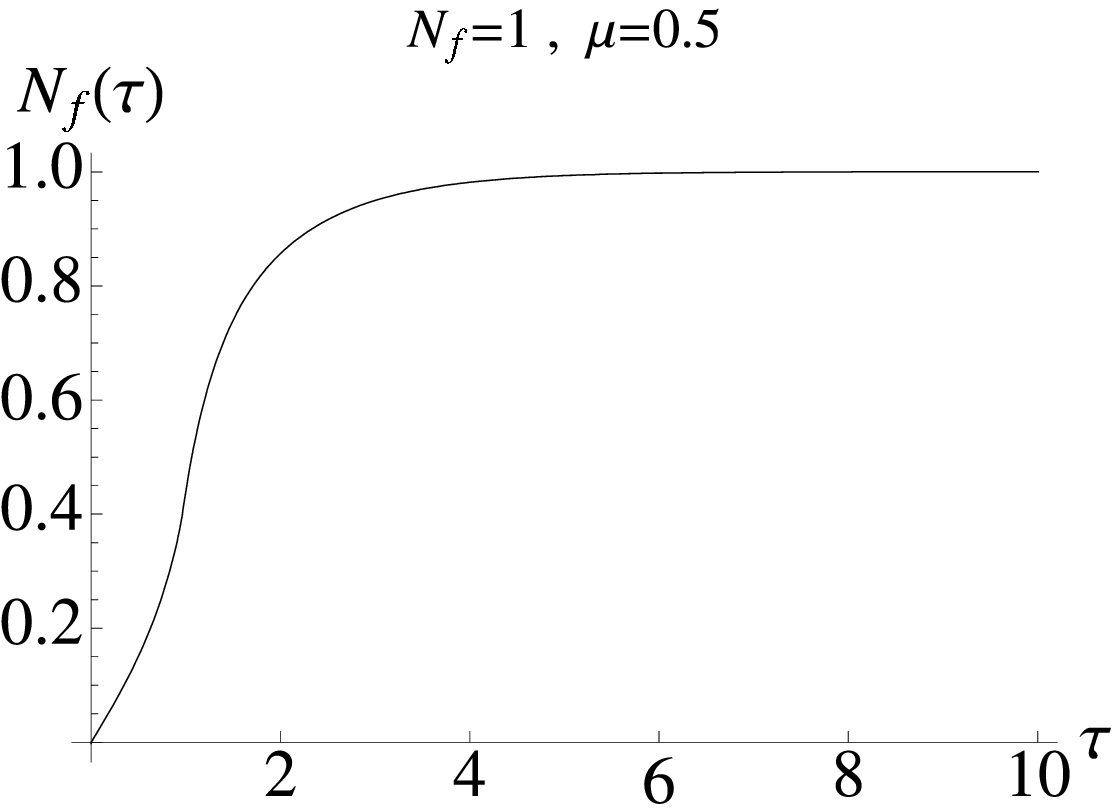}
\caption{Plots of $N_f(\tau)$ for two examples with $\mu>1$ and $\mu<1$.}
\label{nfplots}
\end{figure}

Relevant plots of $N_f(\tau)$ can be found in figure \ref{nfplots}.
For $\mu>1$ the shape of $N_f(\tau)$ resembles that of a ``smoothed-out'' Heaviside step function $N_f(\tau)\sim N_f \Theta[\tau-\tau_q]$. In fact, as we have anticipated, the function $N_f(\tau)$ counts the effective number of flavor degrees of freedom at a given energy scale. At energies smaller than the flavor mass, the fundamental fields can be integrated out and the theory resembles the unflavored one. At higher energies the masses can be neglected and the theory looks like the massless-flavored one (for which $N_f(\tau)=N_f$). In other contexts where the density distribution of the flavor branes is difficult  to evaluate, the Heaviside step function can be fruitfully   used to construct approximate solutions. This is what was done in \cite{Bigazzi:2008gd} for the massive-flavored CVMN solution.

Let us stress that also when $\mu<1$ and so $m<\Lambda_{IR}$, $m_q=0$, there is a non trivial density distribution of branes. This marks a difference with the $m=0$ case studied in \cite{Benini:2007gx}. Of course approximating $N_f(\tau)$ with an Heaviside step function $N_f\Theta[\tau-\tau_q]$ in this massless case would simply replace our solution with the massless one for every $\tau$.

\section{The backreacted KS solution with massive flavors}
\setcounter{equation}{0}
\label{gencons}
In this section we present the new supergravity solutions accounting for the full backreaction of fractional and regular color D3-branes, as well as of smeared flavor D7-branes on the deformed conifold. The solutions follow from an action which is the sum of the bulk type IIB supergravity and the flavor brane actions. Following a suggestion in \cite{km} the action for the D7-branes is just taken as the sum of the Dirac-Born-Infeld (DBI) and Wess-Zumino (WZ) terms. This is actually an approximation which is sensible only if the effective coupling $g_sN_f$ is small. This is the case in localized setups if $N_f\ll N_c$ or in the smeared setups (where $N_f$ can be of the same order of $N_c$) due to the effective suppression of the coupling by the large transverse volume \cite{massivekw,carlosetal}. 
\subsection{The ansatz}
\label{ansatz}
In order to present the ansatz for the full background, let us first introduce the one-forms $\sigma_i$ and $\omega_i$ ($i=1,2,3$) 
as follows
\bear
&& \sigma_1\,=\,d\theta_1 \, , \qquad \qquad
\sigma_2\,=\,\sin{\theta_1}\,
d\varphi_1 \, , \qquad \qquad
\sigma_3\,=\,\cos{\theta_1}\,d\varphi_1 \, , \rc
&&\omega_1\,=\,\sin{\psi} \sin{\theta_2}\,
d\varphi_2\,+\,\cos{\psi}\,d\theta_2\,\, ,
\qquad\qquad
\omega_2\,=\,-\cos{\psi}
\sin{\theta_2}\, d\varphi_2\,+\,\sin{\psi}\,d\theta_2\,\, , \rc
&&\omega_3\,=\,d\psi\,+\,\cos{\theta_2}\,d\varphi_2 \,\, .
\label{sigma-omega}
\eear
Here the angles are the same as in the deformed conifold. 

The Einstein frame metric ansatz has the same warped form as in the massless case \cite{Benini:2007gx}
\bear
ds^2 &=& h^{-1/2}(\tau)\,dx_\mu\,dx^{\mu} + h^{1/2}(\tau)\,ds_6^2\,, \nonumber \\
ds_6^2 &=& \frac{1}{9}e^{2G_3(\tau)}(d\tau^2 + g_5^2) + e^{2G_2(\tau)}(1-g(\tau))(g_1^2+g_2^2) + e^{2G_2(\tau)}(1+g(\tau))(g_3^2+g_4^2)\,,\nonumber
\label{metric}
\eear
where $dx^2_{1,3}$ denotes the four-dimensional Minkowski metric  and
$G_i=G_i(\tau)$ ($i=1,2,3$), $g=g(\tau)$ and $h(\tau)$ are five unknown
radial functions. Quite nicely, the embedding equation expressed in terms the ``deformed conifold $\tau$ variable'' looks the same in terms of the ``backreacted ansatz $\tau$ variable''. See appendix \ref{radial} for details. 

As for the dilaton and the forms we will adopt the same ansatz as in  \cite{Benini:2007gx}, modulo the substitution of $N_f$ with the function $N_f(\tau)$ evaluated in the previous section. In units $g_s=1$ we have
\bear
F_5&=&d h^{-1}(\tau)\wedge dx^0\wedge\cdots\wedge dx^3\,+\,{\rm Hodge\,\,dual} \,,\qquad \phi=\phi(\tau)\,,\rc
B_2 &=& \alpha'\frac{M}{2} \Bigl[ f\, g^1 \wedge g^2\,+\,k\, g^3 \wedge 
g^4 \Bigr]\,,\rc
H_3&=&  \alpha'\frac{M}{2} \, \Bigl[ d\tau \wedge (\dot f \,g^1 \wedge g^2\,+\,
\dot k\,g^3 \wedge g^4)\,+\,{1 \over 2}(k-f)\, g^5 \wedge (g^1
\wedge g^3\,+\,g^2 \wedge g^4) \Bigr]\,,\rc
F_1&=&{N_f(\tau)\over 4\pi}\,\,g^5\,,\rc
F_3&=& \alpha'\frac{M}{2} \Big\{ g^5\wedge \Big[ \big( F+\frac{N_f(\tau)}{4\pi}f\big)g^1\wedge g^2 + \big(1- F+\frac{N_f(\tau)}{4\pi}k\big)g^3\wedge g^4 \Big] +\rc
&&+\dot F d\tau \wedge \big(g^1\wedge g^3 + g^2\wedge g^4   \big)\Big\} \,,
 \label{theansatz}
\eear
where $M$ is the fractional D3-brane Page charge and  $f=f(\tau)$, $k=k(\tau)$, $F=F(\tau)$ are functions of the radial coordinate (and
where the dot denotes derivative with respect to $\tau$).

Notice that, consistently, $dF_1=-\Omega$, where $\Omega$ is the D7-brane density distribution form given in eq. (\ref{massive_2form}). This and the other modified Bianchi identities 
\bear
&&dF_3\,=\,H_3\wedge F_1\,-\,\Omega\wedge B_2\,\,,\rc\rc
&&dF_5\,=\,H_3\wedge F_3\,-\,{1\over 2}\,\Omega\wedge
 B_2\wedge  B_2\,,
 \label{newBianchi}
\eear
follow from the WZ term of the smeared D7-brane action (see appendix \ref{smedac}).

\subsection{The BPS equations}\label{kssolution}
The modified Bianchi identity for $F_3$ in  (\ref{newBianchi}) is automatically satisfied by the ansatz, while that for $F_5$ reduces to a first order differential equation for the warp factor
\begin{equation}
 \dot h\, e^{2G_1+2G_2}\,=\,-
{{\alpha'}^2 \over 4}M^2\Big[f-(f-k)F+{N_f(\tau)\over 4\pi}fk\Big] + N_0\,,
\label{bpswarp}
\end{equation}
where $N_0$ is an integration constant that we will set to zero as in \cite{ks} and \cite{Benini:2007gx}. The previous equation is the same as the one obtained in \cite{Benini:2007gx} with the substitution $N_f\to N_f(\tau)$. In general, the BPS equations following from the bulk fermionic supersymmetric variations and from the Bianchi identities are of exactly the same form as those in \cite{Benini:2007gx}, with the only substitution of $N_f$ with the function $N_f(\tau)$. This is also due to the fact that, despite the modified Bianchi identities of the forms $F_{i}$ in the massive setup differing from those in the massless case, the fermionic supersymmetric variations only contain the $F_{i}$ and not the $d F_{i}$. 

In this way one arrives at the same algebraic constraint for $g$ as in \cite{Benini:2007gx}: $\,g [ g^2-1+e^{2(G_1-G_2)}]=0$. Its two solutions $g=0$ and $g^2=1-e^{2(G_1-G_2)}$ correspond to the singular (and resolved) conifold and to the deformed conifold respectively. Here we focus only on this latter case since we want to have a regular solution at $\tau\rightarrow 0$.

The BPS equations for the 6d metric functions are
\bear
\dot G_1&=&\frac{1}{18}e^{2G_3-G_1-G_2}+\frac{1}{2}e^{G_2-G_1}-\frac{1}{2}e^{G_1-G_2}\,,\nonumber \\
\dot G_2&=&\frac{1}{18}e^{2G_3-G_1-G_2}-\frac{1}{2}e^{G_2-G_1}+\frac{1}{2}e^{G_1-G_2}\,,\nonumber \\
\dot G_3 &=&-\frac{1}{9}e^{2G_3-G_1-G_2}+e^{G_2-G_1}-\frac{N_f(\tau)}{8\pi}e^{\phi}\,,
\label{bpsmetric}
\eear
while for the dilaton we have
\be
\dot\phi=\frac{N_f(\tau)}{4\pi}e^{\phi}\,.
\label{bpsdil}
\ee
Notice that, just as in the massless case, by defining $\lambda_1=G_1-G_2$ we get the simple equation $\dot\lambda_1+2\sinh\lambda_1=0$, from which, up to an integration constant (that we fix to zero as in the massless case) it follows that
\be
e^{G_1-G_2}=\tanh\tau,\,\,\rightarrow\,\,g^{-1}=\cosh\tau\,.
\ee
Taking this result into account, for the flux functions we have
\bear
\dot k &=& e^{\phi}\left(F+\frac{N_f(\tau)}{4\pi}\,f\right)\coth^2\frac{\tau}{2}\,,\nonumber \\
\dot f &=& e^{\phi}\left(1-F+\frac{N_f(\tau)}{4\pi}\,k\right)\tanh^2\frac{\tau}{2}\,,\nonumber \\
\dot F &=& \frac{1}{2}e^{-\phi}(k-f)\,,
\label{bpsflux}
\eear
supported by the algebraic constraint
\be
e^{-\phi}(k-f)=\tanh\frac{\tau}{2}- 2F\,\coth\tau + \frac{N_f(\tau)}{4\pi}\left[k\,\tanh\frac{\tau}{2} -f\,\coth\frac{\tau}{2}\right]\,.
\label{fluxconstr}
\ee
In order to solve the above set of equations we will have to distinguish between the two possible cases: 1) $\mu>1$, i.e. $m>\Lambda_{IR}$ and $m_q\neq 0$, with the running function $N_f(\tau)$ being equal to zero for $\tau\le \tau_q$; 2) $\mu<1$, i.e. $m<\Lambda_{IR}$ and $m_q=0$ with the running function $N_f(\tau)$ being non trivial up to $\tau=0$. 
\subsection{The solution}
\label{thesol}
Let us start by considering the $|\hat\mu|>|\epsilon|$ (i.e. $\mu>1$) case. Correspondingly, there is a region $\tau\in[0,\tau_q]$ where the effective D7-brane charge is zero. In that region, requiring regularity, the solution is just (a slight generalization of) the unflavored KS one. Since $N_f(\tau)=0$, the dilaton does not run (see eq. (\ref{bpsdil})) and the flux functions are just the KS ones, modulo an overall constant
\bear
e^{\phi}&=&e^{\phi_{IR}}={\rm constant}\,,\qquad F=\frac{\sinh \tau - \tau}{2 \sinh \tau}\,,\rc
f&=& e^{\phi_{IR}}\,\frac{\tau \coth \tau -1}{2 \sinh \tau} (\cosh \tau -1)\equiv  e^{\phi_{IR}}\,f_{KS} \,,\rc
k&=& e^{\phi_{IR}}\,\frac{\tau \coth \tau -1}{2 \sinh \tau} (\cosh \tau +1)\equiv  e^{\phi_{IR}}\,k_{KS} \,.
\label{fkFexplicitIR}
\eear
The metric is a warped product of 4d Minkowski and the deformed conifold with deformation parameter $\epsilon=\epsilon_{IR}$ (\ref{ds6unflav}), since
\be
e^{2G_1}=e^{2G_2}\tanh^2\tau\,,\quad e^{2G_2(\tau)}=\frac{\cosh\tau}{4}\epsilon_{IR}^{4/3}K(\tau)\,,\quad e^{2G_3(\tau)}=\frac{3}{2}\frac{\epsilon_{IR}^{4/3}}{K^2(\tau)}\,.
\ee
The warp factor is given by
\be
h(\tau) = \frac{2^\frac23 {\alpha'}^2M^2}
{\epsilon_{IR}^\frac83}\left[ h_0\,-\, e^{\phi_{IR}} \int_{0}^{\tau}\frac{(\xi \coth \xi -1)(\sinh 2\xi -2 \xi)^\frac13}{\sinh^2 \xi}d\xi\right]\,,
\label{dhunflav}
\ee
where $h(0)=h_0$ is an integration constant. In \cite{ks}, $h_0$ was fixed by imposing $h(\tau = \infty)=0$. Since the solution we are considering is only valid up to $\tau=\tau_q$ we cannot fix the integration constant in the same way.

The above solution has to be continuously  glued to the one obtained in the region $\tau>\tau_q$ where the effective D7-brane charge is non zero. The function $N_f(\tau)$ has a very non trivial expression in general, so we will have to perform some numerical integration. 
For the dilaton, for example, from eq. (\ref{bpsdil}), it follows that
\be
e^{-\phi(\tau)}=\frac{1}{4\pi}\int_{\tau}^{\tau_0} N_f(\xi)d\xi\,,
\ee
where the $\tau_0$ is a point where $e^\phi$ blows up. A simple analytic expression can be obtained in the 
$\tau > \rm{arc}\cosh(2\mu^2+1)$ region where $N_f(\tau)$ is given by eq. (\ref{nfdefcon1})
\be
e^{\phi(\tau)}=\frac{4\pi}{N_f}\frac{1}{(\tau_0-\tau)+2\mu^2(1/\sinh(\tau_0)-1/\sinh(\tau))}\,,\qquad (\tau > \rm{arc}\cosh(2\mu^2+1))\,.
\ee
Just as in the massless-flavored KS \cite{Benini:2007gx} or in the flavored KW cases \cite{Benini:2006hh,massivekw,varna}, where $\tau_0$ was related to a Landau pole in the dual gauge theories, our solution cannot be continued up to infinity: $\tau\le\tau_0$. 

Requiring continuity at $\tau_q$ we find 
\be
e^{-\phi_{IR}}=\frac{1}{4\pi}\int_{\tau_q}^{\tau_0} N_f(\xi)d\xi\,,
\ee
which explicitly depends on $N_f$ and $\mu$.

A remarkable feature of the present setup is that also in the effectively flavored region most of the solution can be given in an analytic way. In fact after some algebra we find the following results for the metric functions
\be
e^{2G_1}={\epsilon_{UV}^{{4\over 3}}\over 4}\,\,e^{-{\phi\over 3}}\,
{\sinh^2\tau\over \cosh\tau}\,\,{\cal K}(\tau)\,,\quad e^{2G_2}={\epsilon_{UV}^{{4\over 3}}\over 4}\,e^{-{\phi\over 3}}\,\cosh\tau\,{\cal K}(\tau)\,,\quad e^{2G_3}={3\over 2}\,\,\epsilon_{UV}^{{4\over 3}}\,\,{e^{-{\phi\over 3}}\over
{\cal K}(\tau)^2}\,,
\ee
where
\beq
{\cal K}(\tau)\,\equiv\,{\big[\sinh 2\tau-2\tau\,+\,\eta(\tau)\big]^{{1\over 3}}\over
2^{{1\over 3}}\,\sinh\tau}\,\,,
\eeq
and $d(4\pi e^{-\phi}\eta)/d\tau = (\sinh2\tau-2\tau)N_f(\tau)$. The function $\eta(\tau)$ is thus a constant in the unflavored region. By requiring continuity at $\tau_q$ we fix this constant to zero so that
\beq
\eta(\tau)\,=\,{e^{\phi}\over 4\pi}\,\int_{\tau_q}^{\tau}\,
\big(\sinh 2 \xi-2 \xi\,\big)\,N_f(\xi)\,d\xi\,,
\label{eta-def}
\eeq
and
\be\label{gluingcond}
\epsilon_{UV}=\epsilon_{IR}\,e^{\phi_{IR}/4}, \qquad \qquad {\cal K}(\tau_q)=K(\tau_q)\,.
\ee
The metric in the $\tau>\tau_q$ region is thus a warped product of Minkowski 4d and a slight deformation (driven by $\eta(\tau)$) of the deformed conifold metric 
\bear
&&ds^2_{6\,UV}\,=\,{1\over 2}\,\,\epsilon_{UV}^{{4\over 3}}\,\,e^{-{\phi(\tau)\over 3}}\,{\cal K}(\tau)\,\,
\Bigg[\,{1\over 
3{\cal K}^3(\tau)}\,\,\big(\,d\tau^2\,+\,(g^5)^2\,\big)\,+\,
  \cosh^2\Big({\tau\over 2}\big)\,\Big(\,(g^3)^2\,+\, 
(g^4)^2\,\Big)\,+\,\,\rc
  &&\qquad\qquad\qquad\qquad\qquad\qquad\qquad
  +\,\sinh^2\Big({\tau\over 2}\Big)\,
\Big(\,(g^1)^2\,+\, (g^2)^2\,\Big)\,
\,\Bigg]\,\,.
\label{flav-def-metric}
\eear
In appendix \ref{matchmas} we will verify that this metric reduces to the one found in \cite{Benini:2007gx} when the  flavors are massless. The warp factor is given by
\be
h(\tau)=-\frac{4M^2{\alpha'}^2}{\epsilon_{UV}^{8/3}}\int_{\tau_q}^{\tau}\frac{e^{\frac23\phi(\xi)}}{\sinh^2\xi\,{\cal K(\xi)}}\left[f(\xi)-(f(\xi)-k(\xi))F(\xi)+\frac{N_f(\xi)}{4\pi}f(\xi)k(\xi)\right] d\xi\,+ h_1\,,
\ee
where the integration constant $h_1$ is constrained by the continuity condition at $\tau_q$
\be
h_1 = \frac{2^\frac23 {\alpha'}^2M^2}
{\epsilon_{IR}^\frac83}\left[ h_0 - e^{\phi_{IR}}\int_{0}^{\tau_q}\frac{(\xi \coth \xi -1)(\sinh 2\xi -2 \xi)^\frac13}{\sinh^2 \xi}d\xi\right]\,.
\ee
There is something important to notice here: for large values of $\tau \to \tau_0$, $\dot h$ diverges as $ - (\tau_0 - \tau)^{-2}$.
Thus, $h(\tau_0) = - \infty$ (this happens also in the massless-flavored KS case). Since the metric is only well defined
for $h > 0$, we conclude that there exist some {\it maximal value of the radial coordinate $\tau_{max} < \tau_0$}
where $h$ vanishes and a singularity appears. 
This behavior at $\tau_{max}$ could be connected to the presence of a duality wall \cite{Benini:2007gx}.

For the flux functions things are simpler. By using the constraint equation (\ref{fluxconstr}) and imposing continuity at $\tau_q$, we promptly get
\bear
f&=& e^{\phi}\,\frac{\tau \coth \tau -1}{2 \sinh \tau} (\cosh \tau -1)\,\,,\rc
k&=& e^{\phi}\,\frac{\tau \coth \tau -1}{2 \sinh \tau} (\cosh \tau +1)\,\,,\rc
F&=& \frac{\sinh \tau - \tau}{2 \sinh \tau}\,.
\label{fkFexplicit}
\eear

If the D7-branes reach the tip of the cone, i.e. if $|\hat\mu|\le|\epsilon|$ (i.e $\mu\le 1$), there is no effectively unflavored region, and thus the solution has the non trivial $N_f(\tau)$ dependence in the whole $\tau\in[0,\tau_{max}<\tau_0]$ region. The 6d part of the warped metric is again a generalized ($\eta(\tau)\neq0$) deformed conifold with parameter $\hat\epsilon$.
In principle in this case we would not need a condition analogous to the first equation in (\ref{gluingcond}).
In practice, by continuity with the $|\hat\mu|>|\epsilon|$ solutions we will rescale $\hat\epsilon$ as in (\ref{gluingcond}).

\begin{figure}
 \centering
\includegraphics[width=.32\textwidth]{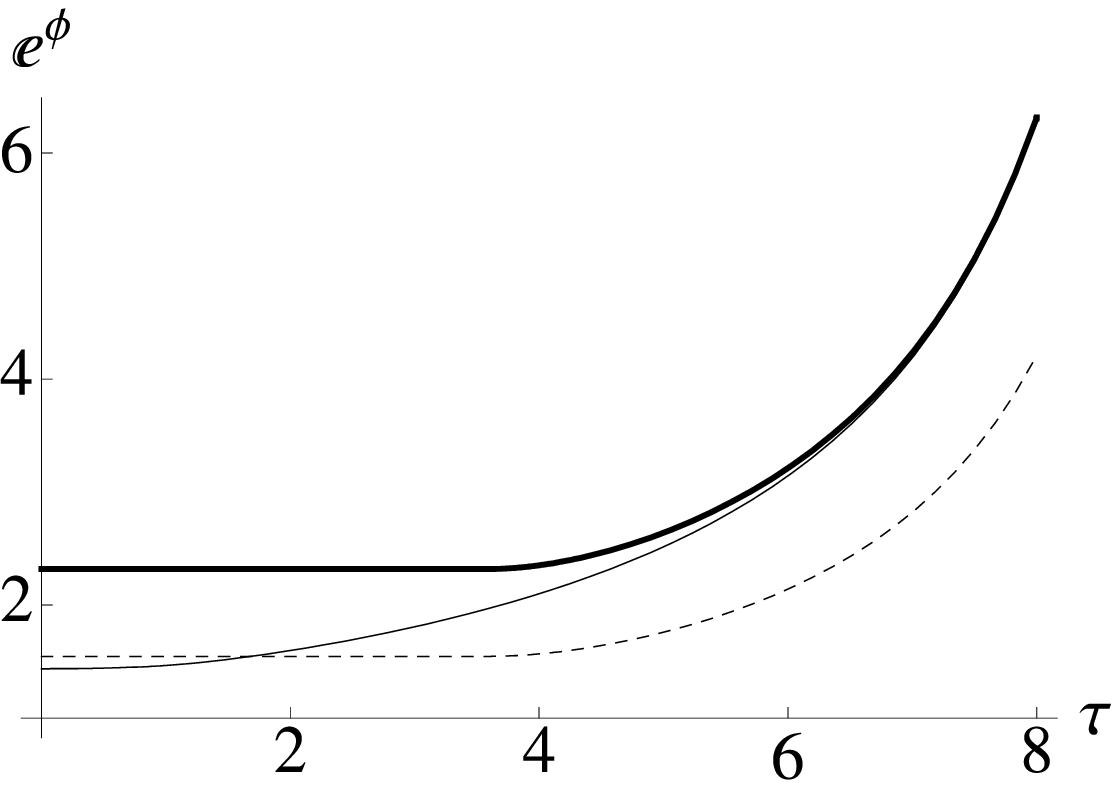}
\includegraphics[width=.32\textwidth]{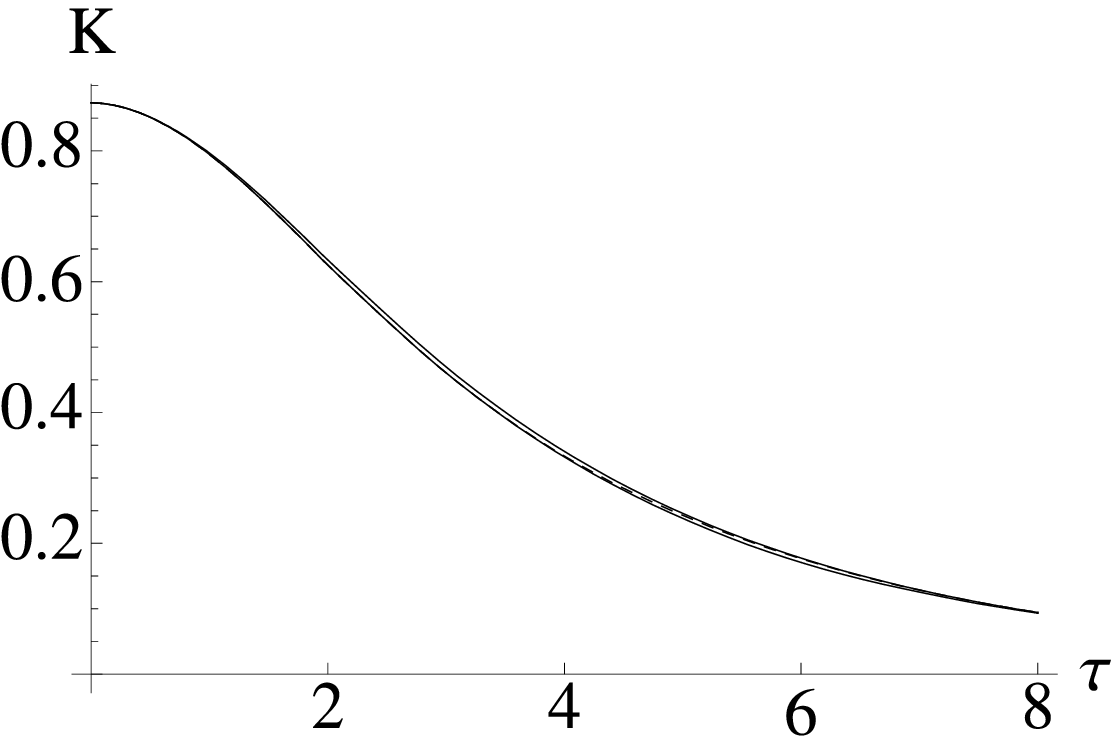}
\includegraphics[width=.32\textwidth]{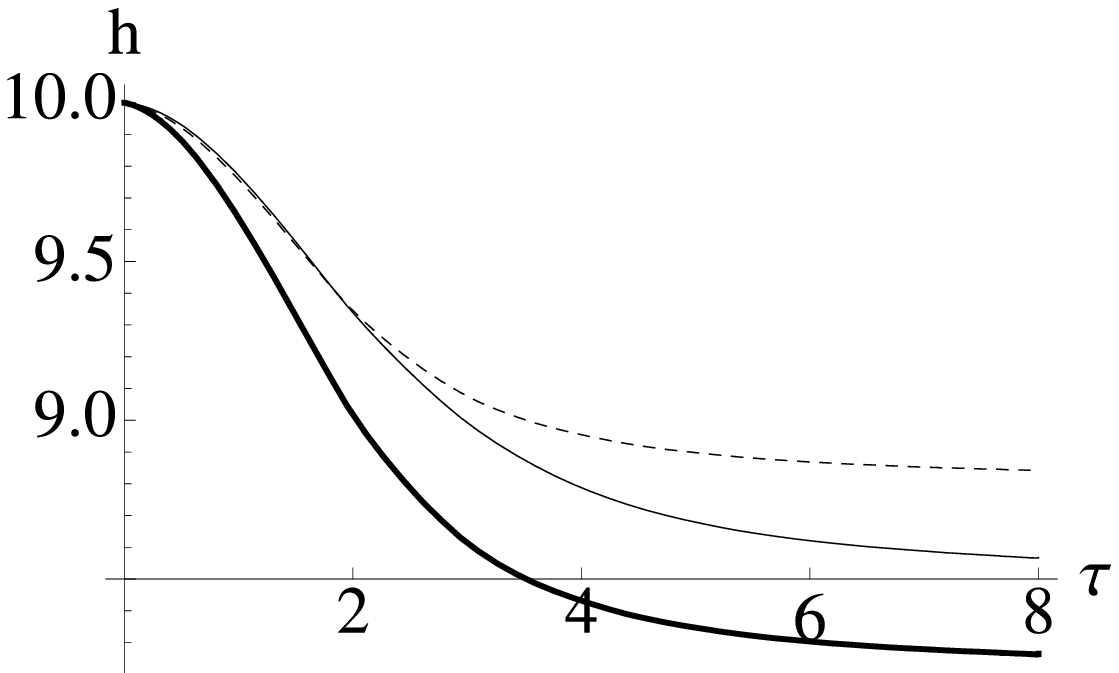}
\caption{Representative plots of the solution for fixed $h_0$ and: $\mu=3$, $N_f=1$ (thick line): $\mu=3$, $N_f=1.5$ (dashed); $\mu=0.5$, $N_f=1$ (normal). In the plots it is not visible the region near the Landau pole, where the dilaton blows up and $h\rightarrow -\infty$. Notice that ${\cal K}$ is almost not varying with $\mu$ and $N_f$.}
\label{plotsol}
\end{figure}

Sample plots of the relevant functions in the solution are given in figure \ref{plotsol}.
In order to completely verify that these results are correct one has to check that the BPS equations solve the equations of motion of gravity, dilaton and forms. To this aim a general result in \cite{kotsi} helps: it was shown in that paper that, for general ${\cal N}=1$ backreacted solutions with metric having a warped $(3,1)\times 6$ form, the solutions of the first order equations following from the bulk fermionic supersymmetry variations and from the (source modified) Bianchi identities, are also solutions of the Einstein equations and of the second order equations of motion for the dilaton and the form fields, provided the sources (smeared or localized) are supersymmetric (i.e. satisfy ${\cal N}=1$ preserving $\kappa$-symmetry conditions). Our setup indeed satisfies these conditions. In appendix \ref{eom} we give the relevant ingredients to perform the consistency check explicitly. See also \cite{Benini:2007kg} for relevant related comments.
\subsection{Validity and regularity of the solution}
We can now study the Ricci scalar $R$ and the square of the Ricci tensor $R_{MN}R^{MN}$ in the string frame, to check the validity and regularity of the solutions we have found above.
As a first thing, one can check that since $e^{\phi}\sim 1/N_f,\ h \sim M^2/N_f$ (where for $h$ we used the rescaling in (\ref{gluingcond})), we have that $R \sim 1/\sqrt{h e^{\phi}} \sim N_f/M$.
Thus, as usual in D3-D7 systems, the supergravity solution is reliable in the regime $1\ll N_f\ll M$. The same observation applies also to the massless-flavored KS solution.

Both curvature invariants diverge at $\tau_{max}$, though they stay ``small'' up to values of $\tau$ very close to $\tau_{max}$ (for example up to $\tau=9.5$ if we fix $\tau_0=10$).\footnote{Note added: there is actually another particular point $\tau_a<\tau_{max}$ (but very close to $\tau_{max}$) where the curvature invariants are finite but the holographic $a$-function has a singularity. One can argue the 
need to place a UV cutoff, smaller than $\tau_a$, when computing observables in this background \cite{noimesons}.} In order to study the behavior near $\tau=0$ we need to know the behaviors of the functions $h,{\cal K},\ e^{\phi}$ in the case $\mu<1$ (for $\mu>1$ the IR solution is the KS one, so the the solution is certainly regular at $\tau=0$).
Clearly everything depends on the behavior of $N_f(\tau)$, which goes as $n_f \tau $ for some constant $n_f$ in the limit. With this, one can verify that the relevant functions have expansions of the form
\be
e^{\phi}\sim e^{\phi_0} + e^{\phi_2}\,\tau^2\,,\qquad  h\sim h_0 - h_2\,\tau^2\,,\qquad \eta \sim \eta_5\, \tau^5\,,\qquad
{\cal K}\sim {\cal K}_0 + {\cal K}_2\,\tau^2\,,
\label{asy}
\ee
where the explicit expressions for the coefficients $\phi_0, \phi_2, h_2, \eta_5, {\cal K}_0 ,{\cal K}_2$ are not relevant here. Using these results it is easy to show that the components of the string frame metric have expansions in even powers of $\tau$ around the origin, exactly as in the unflavored KS case. More precisely, our massive-flavored solutions and the KS ones share the same IR topology. It is interesting to outline the differences with the singular massless case \cite{Benini:2007gx}: there $N_f=$const, $e^{\phi}\sim e^{\phi_0} + e^{\phi_1}\tau$, $\eta\sim\eta_4 \,\tau^4$, ${\cal K}\sim {\cal K}_0 +{\cal K}_1\, \tau$ and the string frame metric expansion also contains odd powers of $\tau$.  

The asymptotics (\ref{asy}), when inserted in the expressions for $R$ and $R_{MN}R^{MN}$, imply that the curvature invariants go to a constant at $\tau=0$ and so \emph{the solutions are regular in the IR for every $\mu>0$}. 

\section{Quark-antiquark potential and screening lengths}
\label{wilson}
\setcounter{equation}{0}
In the previous section we have constructed a string dual of a flavored version of the Klebanov-Strassler theory with a large number $N_f$ of massive dynamical flavors.
In this section we begin to study how the dynamical flavors affect the non-perturbative dynamics of the gauge theory.
We are going to probe the latter with an external quark-antiquark $\bar Q, Q$ pair with ``constituent mass'' $M_Q$. The idea is to study how the static quark-antiquark potential, as well as the screening lengths, depends on the sea quark parameters $\mu, N_f$.

As we have observed in section \ref{smedefcon}, the dynamical flavors have all the same bare mass $m$ (related to $\mu$), in modulus, but the corresponding D7-branes have different minimal distances $\tau_{min}$ from the origin, since they correspond to different phases of the mass parameter. The whole smeared distribution ends up at the absolute minimal distance $\tau_q$ which we have associated to a minimal ``constituent mass'' parameter $m_q$. 
The mass of the probe quarks $\bar Q, Q$ is required to be much larger than $m_q$: $M_Q\gg m_q$.
The $\bar QQ$ bound state is dual to an open string with the extrema lying on a probe D7-brane embedded in such a way that it reaches a minimal distance $\tau_Q\gg \tau_q$, related in the usual way to the ``constituent mass'' $M_Q$, from the bottom of the space.\footnote{More precisely, in order for the corresponding quarks to be non dynamical, we have to take $\tau_Q$ much larger that the maximal possible value of $\tau_{min}$ in the smeared distribution. As we have discussed in section \ref{gendefcon} this maximal value is $\tau_b=2{\rm arc}\sinh(|\hat\mu|/|\epsilon|)$.} Due to the existence of a maximal value of $\tau$, $\tau_{max}<\tau_0$ in the supergravity solution, $M_Q$ cannot be taken to be infinite and $\tau_Q$ must lie below $\tau_{max}$.

The open string attached to the probe D7-brane bends in the bulk and reaches a minimal radial position $\tau_{tip}$. The Minkowski separation $L$ between the test quarks, as well as the potential $V(L)$ (i.e. the total energy renormalized by subtraction of the static quark masses) depend on $\tau_{tip}$. For an open string embedding given by $t=\hat\tau,
y=\sigma, \tau=\tau(y)$ where $y\in [-L/2,L/2]$ is one of the
spatial Minkowski directions, one finds  \cite{maldawilson}
 \bear L(\tau_{tip})&=&2
\int_{\tau_{tip}}^{\tau_{Q}} \frac{G P_{tip}}{P\sqrt{P^2
-P_{tip}^2}}d\tau\,,\rc 
V (\tau_{tip})&=&\frac{2}{2\pi\alpha'} \Bigl[\int_{\tau_{tip}}^{\tau_{Q}}
\frac{G P}{\sqrt{P^2 -P_{tip}^2}}d\tau -
\int_{0}^{\tau_{Q}} G \ d\tau \Bigr]\,, 
\label{maldafor}
\eear
where $P, G$ are expressed in terms of the string frame metric
\be
P = \sqrt{g_{tt}g_{yy}}=e^{\phi/2}h^{-1/2}\,,\quad\qquad G = \sqrt{g_{tt}g_{\tau\tau}}=\frac{1}{3}e^{G_3}e^{\phi/2}\,,
\label{sonnendefs}
\ee 
and the ``$tip$'' subindex means that the quantity is evaluated at $\tau=\tau_{tip}$. It is not difficult
to check the relation $\frac{dV}{d\tau_{tip}} = \frac{P_{tip}}{2\pi\alpha'}
\frac{dL}{d\tau_{tip}}$. This for instance means that
$\frac{dV}{d\tau_{tip}}$ and $\frac{dL}{d\tau_{tip}}$ have the same sign.
In some cases they can change sign simultaneously at some
value of $\tau_{tip}$ so the $V(L)$ plot turns around \cite{sfetsos}. 

We can now use the background solutions found in the previous sections to study how the external quark interaction depends on the dynamical massive flavors. 
In doing so, we vary one of the physical parameters $N_f, \mu\, (m_q), M_Q$ while keeping the others fixed.
In the numerical analysis below we have set $2^{2/3}M^2{\alpha'}^2\epsilon_{IR}^{-8/3}=1$ in the expression for $h$. The masses are measured in units of $6^{-1/2}\epsilon_{IR}^{2/3}/(2\pi\alpha')$ and we have put $2\pi\alpha'=1$. Finally, remember that we are working in units $g_s=1$ and so what will be denoted by $N_f$ in the following numerical studies has to be read as $g_s N_f$ in standard units.

As in \cite{massivekw}, we have studied both the exact solutions and those obtained in the Heaviside approximation where $N_f(\tau)$ is replaced by $N_f\Theta(\tau-\tau_q)$. 
Apart from a mismatch in few results for the screening lengths, the overall qualitative agreement of the two solutions is general.

\subsection{How to compare physical observables}\label{possibilities}

The solution of section \ref{gencons} allows to vary the flavor number $N_f$ and mass $\mu$.
In general, there is no obvious energy scale or coupling which is expected to stay fixed as these parameters are varied.
Thus, the comparison of various physical observables will crucially depend on the \emph{choice} of the energy scale or coupling which is kept fixed.

There are actually several possibilities we can consider, since there are many scales in the theory at hand. One can decide to fix, for example, the mass of some state in the spectrum of the theory.  From a lattice perspective a sensible possibility would be fixing the mass of the lightest ($0^{++}$) glueball.\footnote{We are grateful to Biagio Lucini and Massimo D'Elia for their comments on these issues.} In our case, this is computationally very hard: the holographic calculation of that mass is extremely complicated also in the unflavored KS background \cite{Berg:2006xy}. Besides, the lightest glueball excitation in the KS model is actually a massless pseudo-scalar \cite{Gubser:2004qj}. For these reasons, we will not consider this scenario.
Another possibility is to keep fixed the mass of the lightest meson. We plan to provide in the near future an accurate analysis of the mesonic spectrum (which is partly missing also in the unflavored case), so we defer this possibility for a separate publication. We refer to section \ref{mesons} for some preliminary results on the spectrum. Finally, one could also consider the possibility of taking fixed, say, the ``effective 't Hooft coupling'' $Me^{\phi}$ at some scale, for example fixing either its IR value at $\tau=0$ or its UV one at the duality wall
$\tau=\tau_{max}$.

In this paper we will consider two alternative possibilities, which also have the advantage of being computationally easier to realize. They imply different ways of choosing the integration constant $h_0$. This constant sets ``the scale'' of the glueball and KK masses $h_0 \sim 1/m_{glue}^2$, as can be deduced from the metric.\footnote{Clearly, fixing $h_0$ is not enough to fix precisely the numerical value of the mass of the $0^{++}$ glueball, which depends also on other features of the background.} Moreover, it enters in the expression for the IR string tension $T \sim (e^{\phi_{IR}}/h_0)^{1/2}$ (where $e^{\phi_{IR}}$ explicitly depends on $\tau_0,\ N_f,\ \mu$).
We will thus consider the following possibilities:
\begin{itemize}
\item {\bf Possibility 1: constant glueball scale $h_0$.} Keeping constant $h_0$ means fixing the glueball and KK scale. The string tension $T$ will change as we change $N_f,\ \mu$. 
\item {\bf Possibility 2: constant string tension $T$.} Keeping constant the string tension $T$ means performing the rescaling $h_0 \rightarrow h_0 e^{\phi_{IR}}$. The glueball scale will thus change as we change $N_f,\ \mu$.
\end{itemize}
Let us notice that the string tension is defined from the large distance behavior of the heavy (static) quark-antiquark potential in the theory. Since there are dynamical flavors, the heavy quark-antiquark bound state is metastable towards decay into a pair of heavy-light mesons. Thus one has to keep in mind that for Possibility 2, the string configuration which defines $T$ is metastable.

In both ``Possibilities'', the radial value $\tau_0$ at which the dilaton diverges (a kind of UV Landau pole in the dual gauge theory picture) is kept fixed.
Clearly, other choices (other ``Possibilities'') could include the variation of this scale too.
We hope to analyze these scenarios in the near future.

\begin{figure}
 \centering
\includegraphics[width=.3\textwidth]{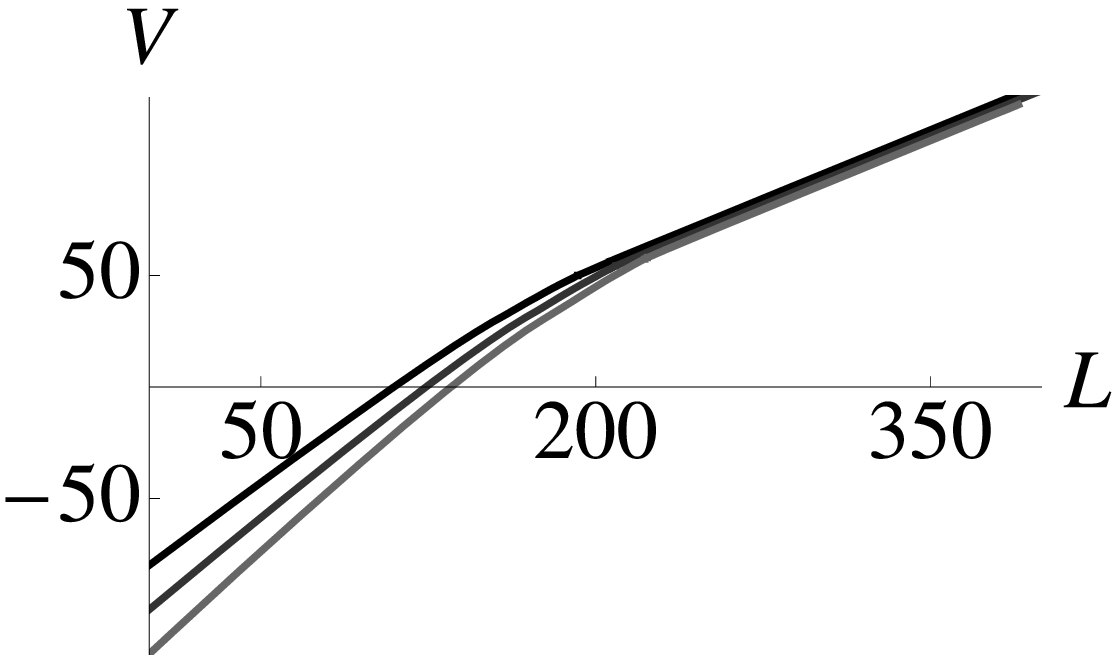}
\includegraphics[width=.3\textwidth]{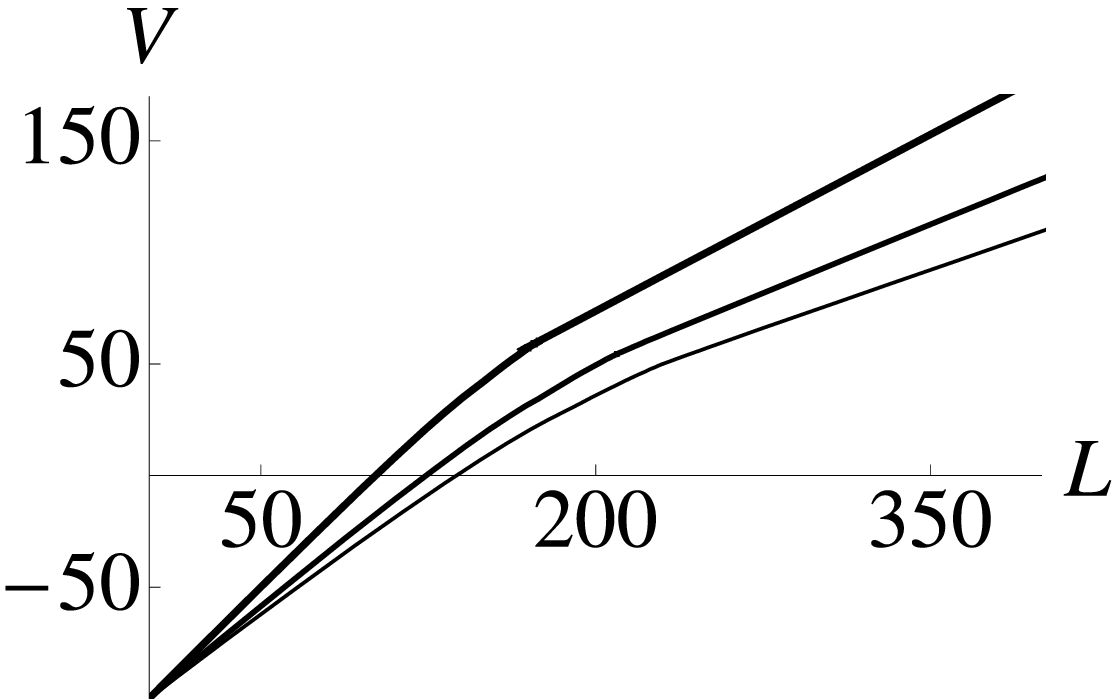}
\includegraphics[width=.3\textwidth]{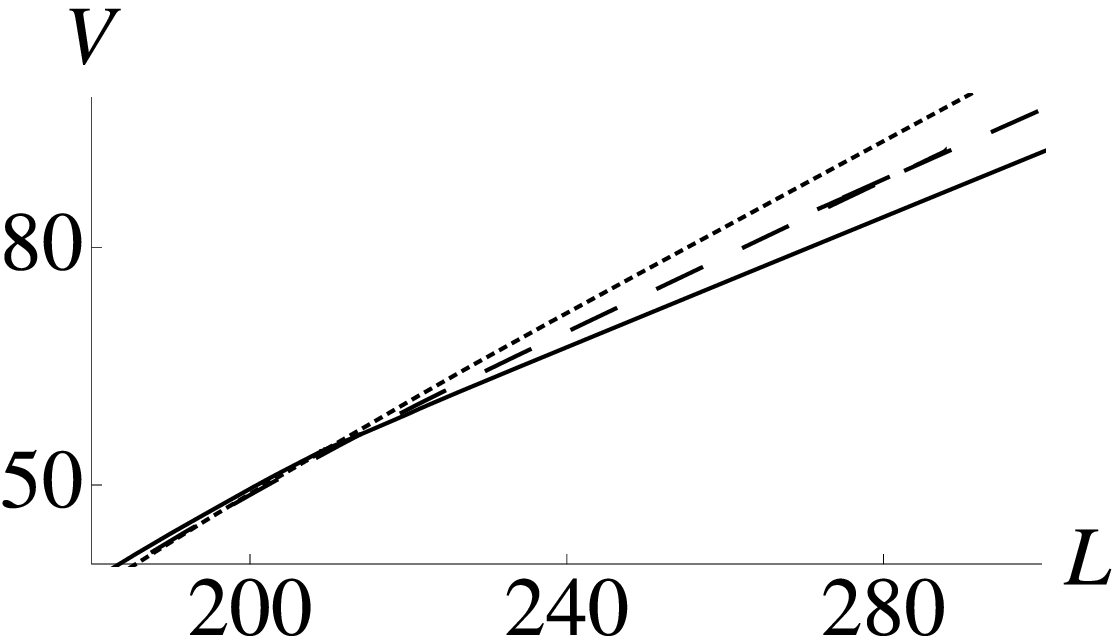}
\caption{The potential at fixed glueball scale $h_0=10$. From left to right: $N_f=1$, $m_q=1$, $M_Q=40,50,60$ (dark to pale); $M_Q=50$, $m_q=1$, $N_f=0.6, 1, 1.4$ (thick to thin); $M_Q=50$, $N_f=1$, $m_q=1,5,10$ (continuous, dashed, dotted).}
\label{vp1}
\end{figure}

\begin{figure}
 \centering
\includegraphics[width=.3\textwidth]{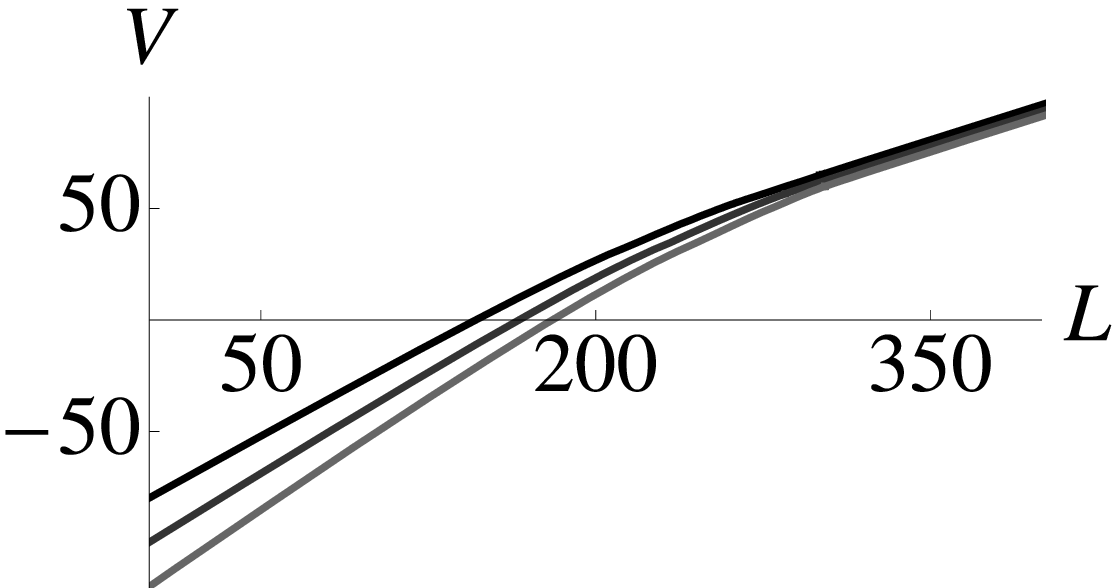}
\includegraphics[width=.3\textwidth]{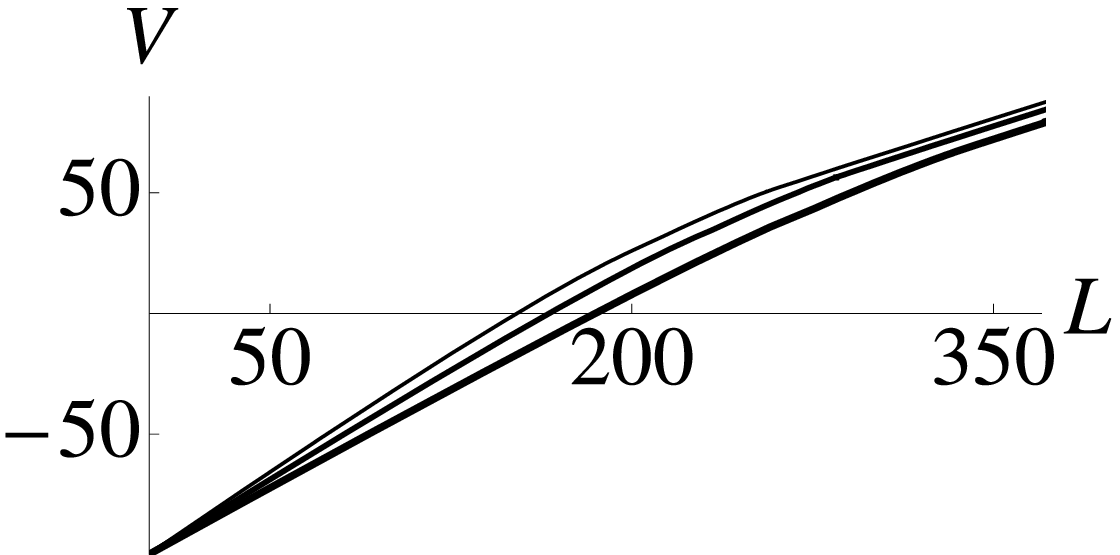}
\includegraphics[width=.3\textwidth]{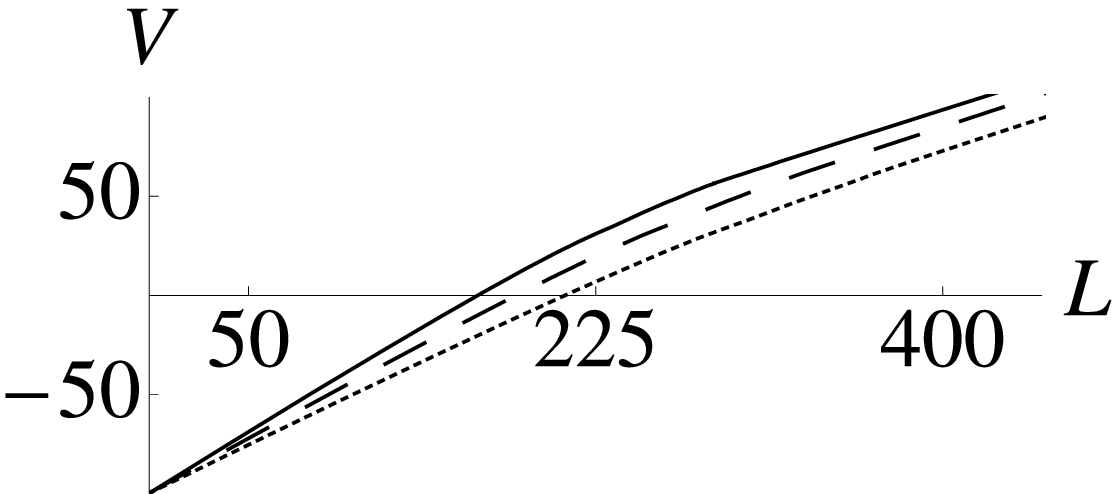}
\caption{The potential at fixed IR string tension. From left to right: $N_f=1$, $m_q=1$, $M_Q=40,50,60$ (dark to pale); $M_Q=50$, $m_q=1$, $N_f=0.6, 1, 1.4$ (thick to thin); $M_Q=50$, $N_f=1$, $m_q=1,5,10$ (continuous, dashed, dotted).}
\label{vp2}
\end{figure}

In figures \ref{vp1} and \ref{vp2} we present various plots of the static potential $V(L)$ as the flavor parameters are varied using the two prescriptions described above.
The potential, which has generically a Cornell-like shape, decreases as $M_Q$ is increased. The behavior w.r.t. to $N_f$ and the mass parameter, instead, depends on the chosen prescription. For Possibility 1 $V(L)$ decreases with $N_f$ and increases with $m_q$ (with a crossing of potentials for intermediate values of $L$). For Possibility 2 it behaves in the opposite way.
\subsection{The screening length}
As we have already observed, due to the presence of the dynamical flavors, the $\bar Q Q$ state is metastable, since a quark-antiquark  dynamical pair $\bar q, q$ can be popped out from the vacuum causing the decay $\bar Q Q\rightarrow \bar Q q+\bar q Q$. In our setup, the lightest possible heavy-light mesons which can arise from the decay of $\bar Q Q$ are nearly massless. This is due to the fact that, since the dynamical flavor branes are smeared, some of them will intersect the probe one. The corresponding heavy-light meson is holographically a string living at the intersection of the two branes and its mass is roughly given, at leading order, by $M_Q/\lambda$ \cite{Herzog:2008bp}, where $\lambda\gg1$ is the bare 't Hooft coupling of the theory. 

\begin{figure}
 \centering
\includegraphics[width=.22\textwidth]{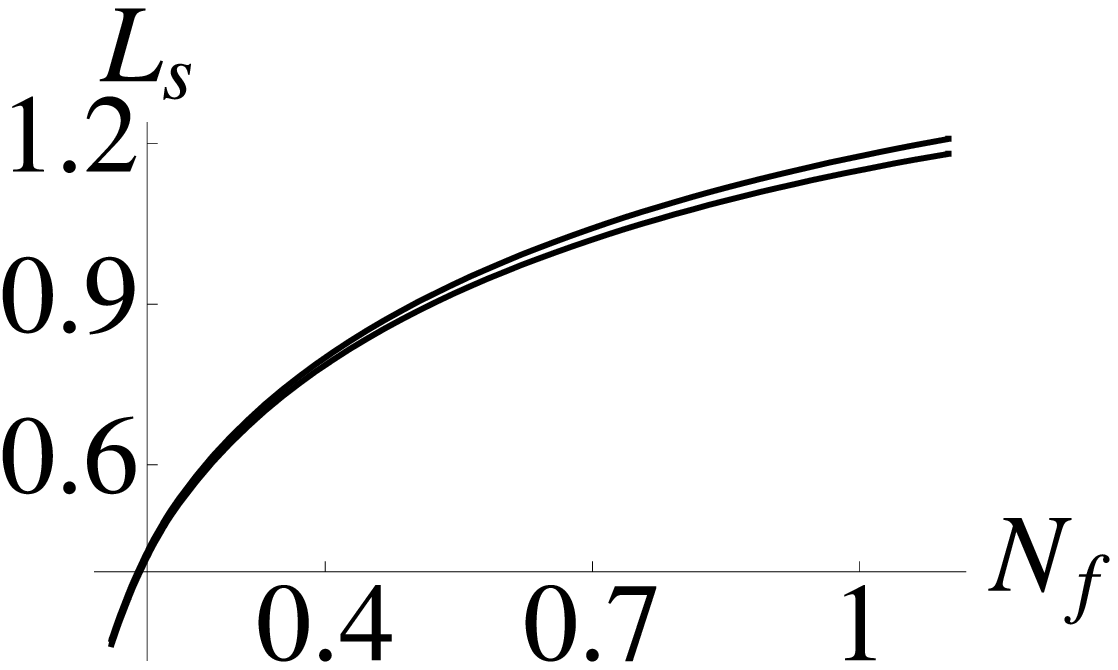}
\includegraphics[width=.22\textwidth]{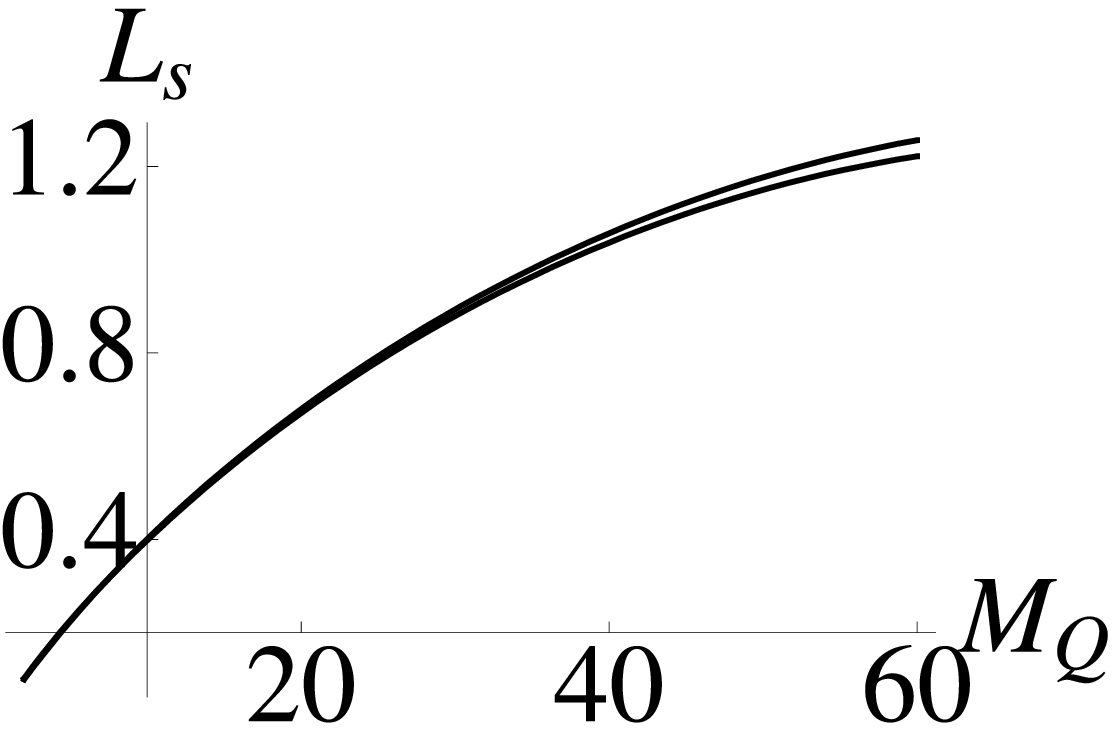}
\includegraphics[width=.22\textwidth]{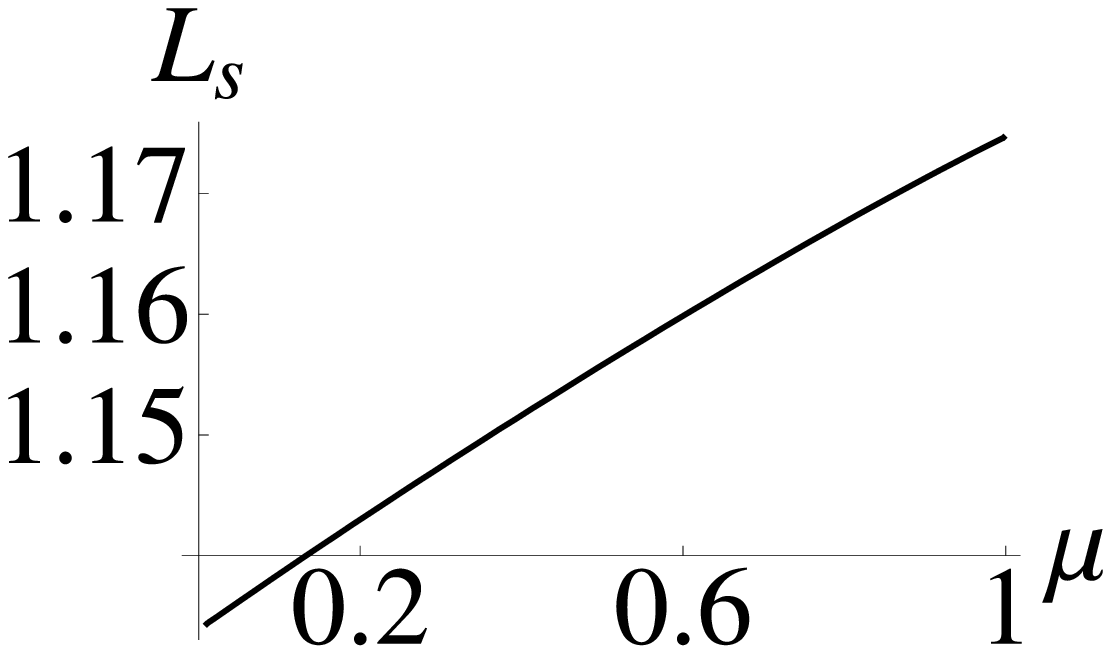}
\includegraphics[width=.22\textwidth]{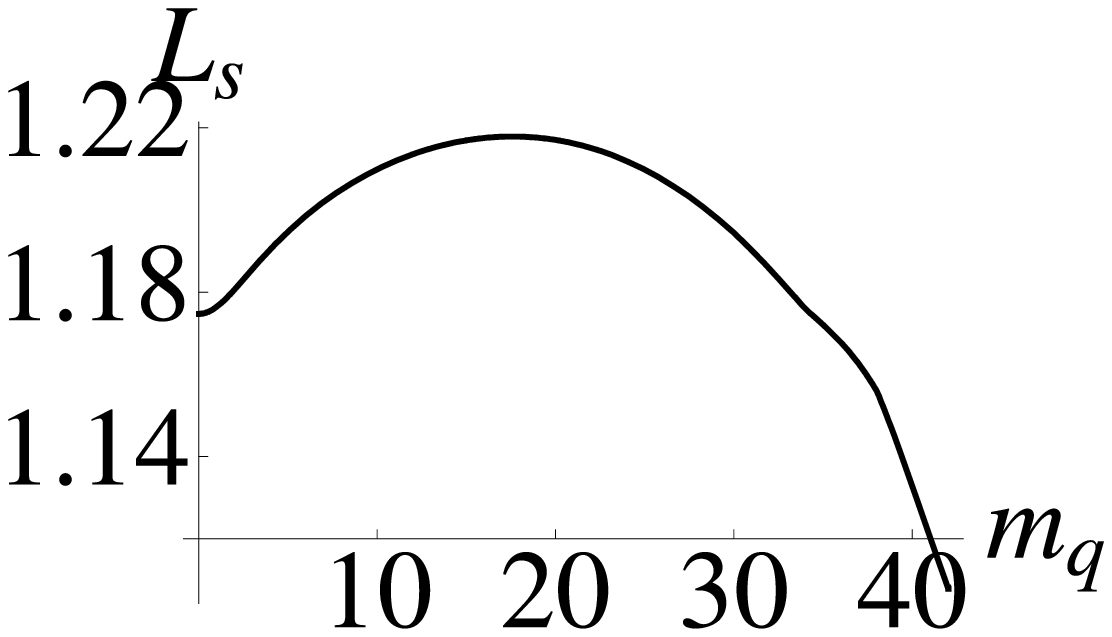}
\caption{The screening length for constant glueball scale $h_0$. In the first plot $M_Q=50$ and the higher (lower) line corresponds to $m_q=1$ ($\mu^2=0.125$). In the second plot $N_f=1$ and the higher (lower) line corresponds to $m_q=1$ ($\mu^2=0.125$). In the third and fourth plots $M_Q=50$ and $N_f=1$.}
\label{lsposs1}
\end{figure}

\begin{figure}
 \centering
\includegraphics[width=.22\textwidth]{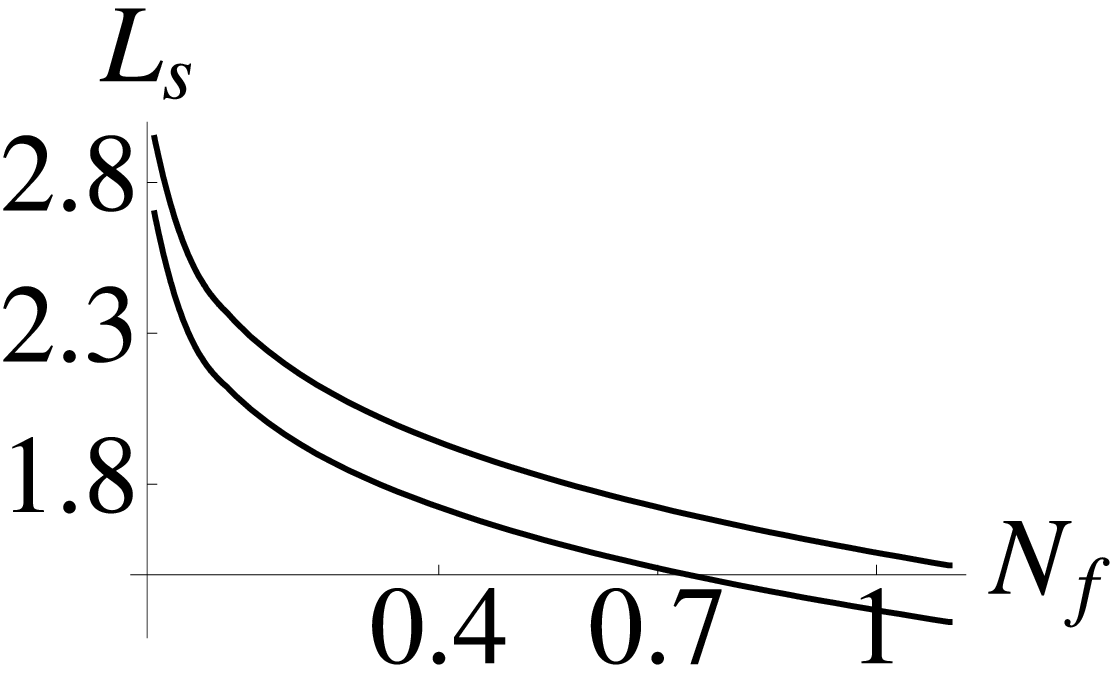}
\includegraphics[width=.22\textwidth]{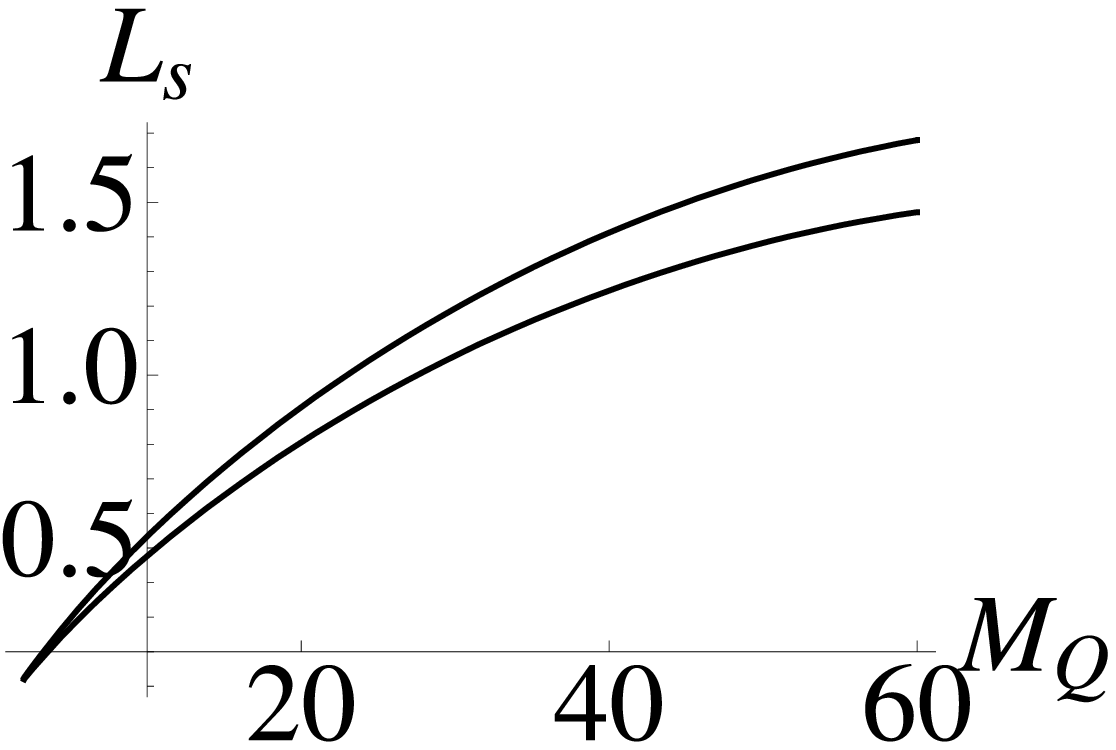}
\includegraphics[width=.22\textwidth]{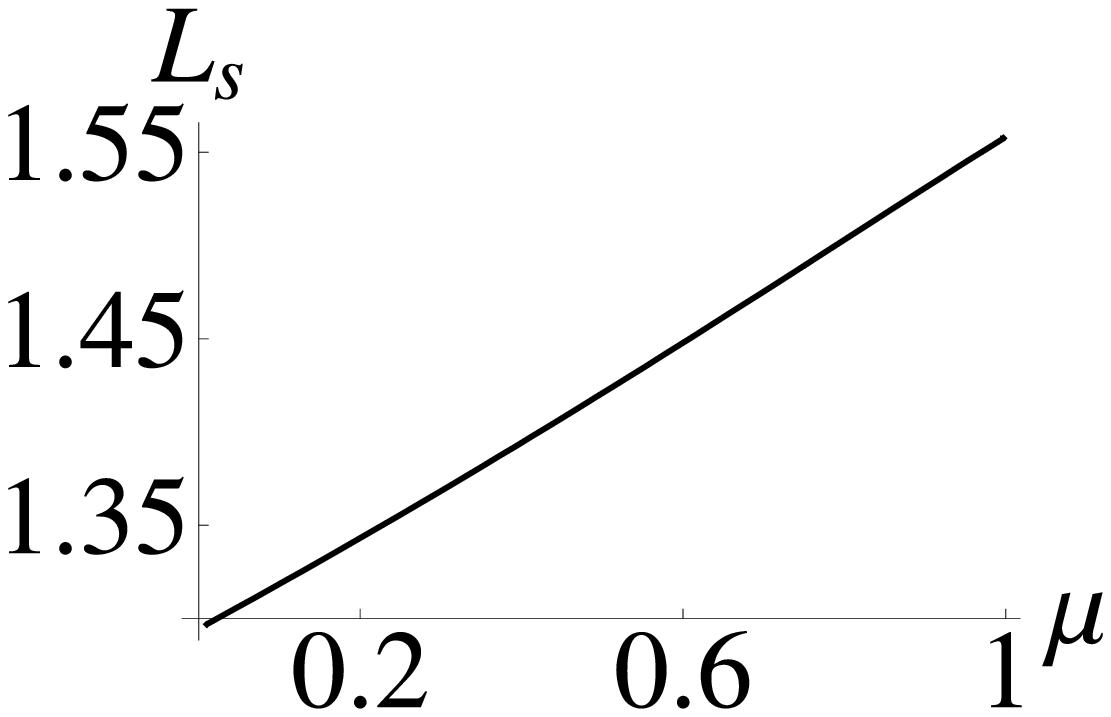}
\includegraphics[width=.22\textwidth]{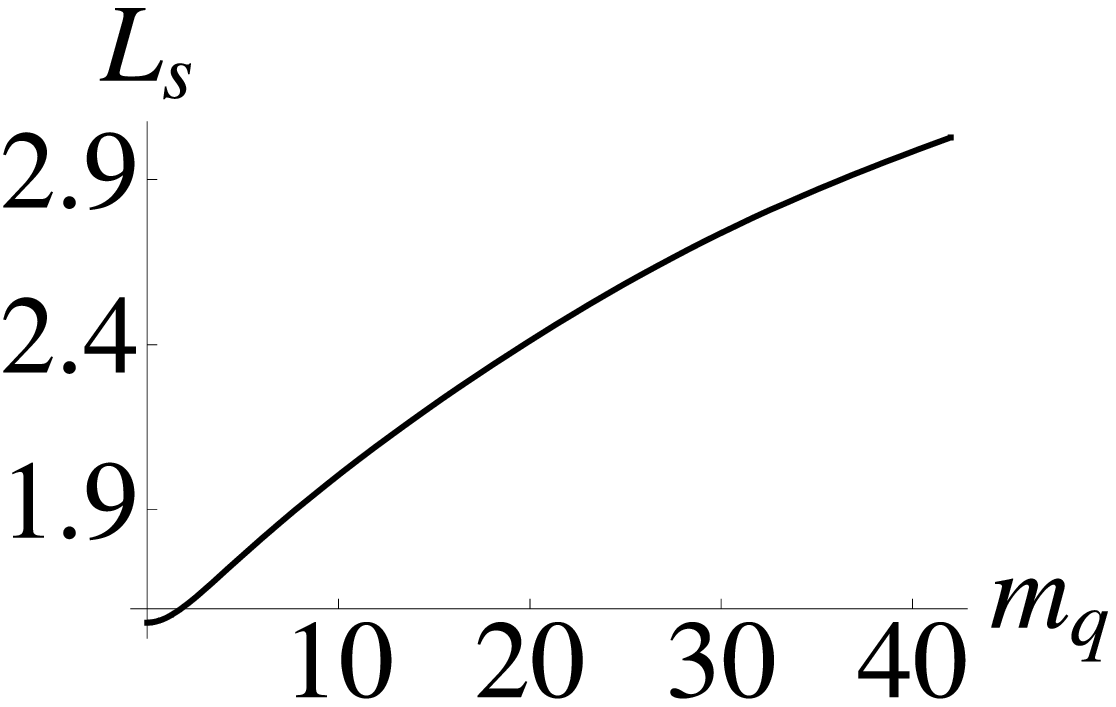}
\caption{The screening length for constant string tension $T$. In the first plot $M_Q=50$ and the  higher  (lower) line corresponds to $m_q=1$ ($\mu^2=0.125$). In the second plot $N_f=1$ and the  higher (lower) line corresponds to $m_q=1$ ($\mu^2=0.125$). In the third and fourth plots $M_Q=50$ and $N_f=1$.}
\label{lsposs2}
\end{figure}

The minimal static quark distance at which a pair of these nearly massless mesons can be produced is called ``screening length'' $L_s$. This is thus defined as
\be\label{screening}
V(L_{s})=-2M_Q + 2\frac{M_Q}{\lambda}\,.
\ee
Notice that from the relations (\ref{maldafor}) it follows that $V(0)=-2M_Q$. A non trivial value for $L_s$ can thus be obtained only without neglecting the $1/\lambda$ suppressed mass of the heavy-light mesons.  We have used a conventional value $\lambda=100$ in our numerical analysis, which is intended to give a \emph{qualitative} picture of how the screening length depends on the flavor parameters.\footnote{We have neglected for simplicity the produced meson interactions as well as the dependence of $\lambda$ on the flavor parameters. It would be interesting to include these contributions in the calculation. Notice that, instead, the first $\alpha'$ corrections to the background will induce subleading corrections to the quark-antiquark potential.}
The results are in figures \ref{lsposs1} (constant glueball scale) and \ref{lsposs2} (constant string tension)  for the two ``Possibilities'' discussed in section \ref{possibilities}.
For consistency with notations in previous studies \cite{Bigazzi:2008gd,massivekw,varna}, we have used $m_q$ (determined by $\tau_q$) as mass parameter in the $\mu>1$ region, where $2\mu^2=\cosh{(\tau_q)}+1$.
For $\mu<1$, on the other hand, $m_q$ is zero, and we use directly $\mu$ as flavor mass parameter.

As anticipated, it is evident from the plots in figure \ref{lsposs1} and \ref{lsposs2} that the behavior of the physical observables in the theory crucially depends on the choice of the fixed scale.
The screening length $L_s$ is in fact a monotonically increasing function of $N_f$ if the glueball scale is kept fixed (fig. \ref{lsposs1}, first plot), while it is monotonically decreasing if it is the string tension to be kept fixed (fig. \ref{lsposs2}, first plot).
For both Possibilities, the screening length $L_s$ is monotonically increasing with the static quark mass $M_Q$ (figs. \ref{lsposs1}, \ref{lsposs2}, second plots).
The same is true for the behavior with the dynamical flavor mass $\mu$ in the regime $\mu<1$ (figs. \ref{lsposs1}, \ref{lsposs2}, third plots).
But the behavior for $\mu>1$ is again very different: while in the constant glueball scale case $L_s$ has a local maximum in the intermediate regime of dynamical flavor masses (fig. \ref{lsposs1}, fourth plot),\footnote{This feature is not present in the Heaviside approximation, where the behavior is always increasing. A possible explanation is that for ``sufficiently large'' $m_q$ the Heaviside is not a very good approximation ($N_f(\tau)$ is smoothly increasing).} in the constant string tension case it continues to be monotonically increasing (fig. \ref{lsposs2}, fourth plot).

Naively, one would expect $L_{s}$ to be a decreasing function of $N_f$ (the more flavors there are, the more the screening is effective), and an increasing function of $\mu$ and $m_q$ (the more the flavors are massive, the less the screening is effective). We see that the constant string tension ``Possibility 2'' realizes these expectations.

Due to the smearing, the decay defining the screening length studied in this section is suppressed as $1/N_c$.
In appendix \ref{sbl} we study a different critical length, the ``string breaking length'', defined in such a way as to reduce this huge suppression to order $N_f/N_c$. 
The results for the string breaking length are qualitatively similar to the ones for the screening length, apart from a different dependence on $m_q$.

\section{The quantum phase transitions}
\label{phases}
\setcounter{equation}{0}
The presence of a quantum phase transition in the heavy quark potential $V(L)$ calculated in string theory, appearing as a discontinuity of the first derivative of $V(L)$, is a very common phenomenon \cite{sfetsos,angelnc,avramista,avramis,Bigazzi:2008gd,massivekw,cargese,alfonsoetal}.
It is tempting to conjecture that its occurrence is generic in string duals of theories with at least two mass scales, for example confining theories with some additional mass scale.\footnote{Confining theories with supergravity duals have actually already at least two distinct mass scales, which can be identified with the confining string tension and the glueball scale. Nevertheless, the ratio of these two scales cannot be varied at will while remaining in the regime where the gravity description is reliable (the string tension must remain much larger than the glueball scale). Due to this reason, for our present discussion these two scales cannot be considered independent.}
The flavored KS theory is not an exception to this behavior.
In this section, we discuss the presence of the quantum phase transition in the (deformation of the) unflavored case \cite{ks}, in the massless case of Ref. \cite{Benini:2007gx} and in the massive case presented in this paper.
We also discuss the universality properties of the phase transition in all the theories with a supergravity dual, arguing that the critical exponents are always classical in the regime of parameter where the gravity description is reliable. 

\subsection{Phase transitions in the (flavored) KS theory}

The KS theory is a confining theory with a dynamical scale $\Lambda_{IR}$.
The integration constant $h_0$ in the gravity dual was fixed in \cite{ks} to some specific value $h_{0KS}$ by requiring that $h(\infty$)=0.
There are indications that a departure of $h_0$ from the value $h_{0KS}$ in this type of theories is dual to a non-trivial source of a higher dimensional operator \cite{Klebanov:2000nc}.
Thus, values of $h_0$ different from $h_{0KS}$ introduce a second mass scale in the theory.
Variations of this scale with respect to $\Lambda_{IR}$ can cause the heavy quark potential to develop a quantum phase transition.
In fact, it can be explicitly verified that this is what happens when $h_0$ is tuned above some critical value $h_c$ which is larger than $h_{0KS}$.
Precisely the same phenomenon happens in the CVMN solution \cite{mn}, as described in \cite{cargese}.
The features of the potential are always the same in these cases and we are going to describe them in the following.
This example just indicates that also in the flavored case the phase transition is to be expected.
  
\begin{figure}
 \centering
\includegraphics[width=.4\textwidth]{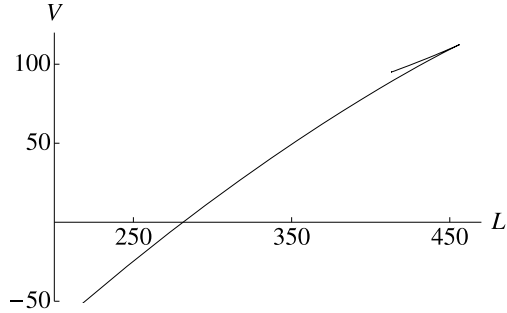}
\caption{The static potential in the massless case for large $h_0$. The IR singularity causes $L$ to have a maximal value. At $L=L_{max}$, $V(L)$ turns around.}
\label{m0ks}
\end{figure}

In the massless-flavored KS case \cite{Benini:2007gx}, the backreacted D3-D7 solution is singular at $\tau=0$. Just as in the massless-flavored CVMN case \cite{cnp}, the singularity causes the static quark-antiquark distance $L$ to have a maximum and the $V(L)$ plot to have a turnaround at $L=L_{max}$. In figure \ref{m0ks} we give a sketch of this behavior for a ``large'' value of the integration constant $h_0$. Crucially, for sufficiently small values of $h_0$, provided the latter is still larger than a certain critical value, there is also a second turnaround and a phase transition appears.

\begin{figure}
 \centering
\includegraphics[width=.32\textwidth]{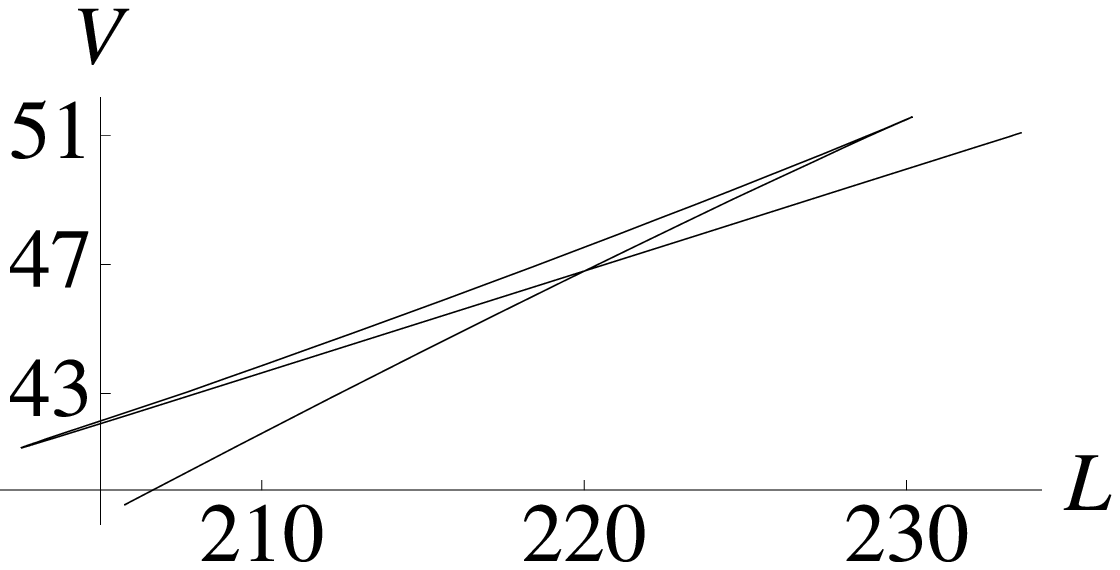}
\includegraphics[width=.32\textwidth]{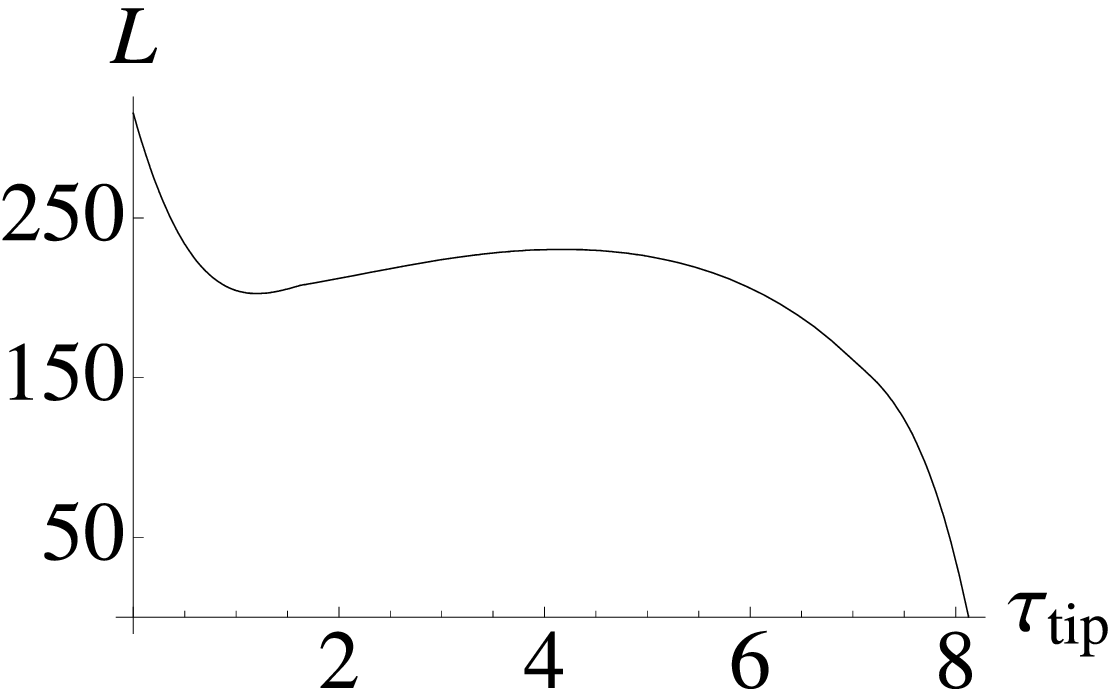}
\includegraphics[width=.32\textwidth]{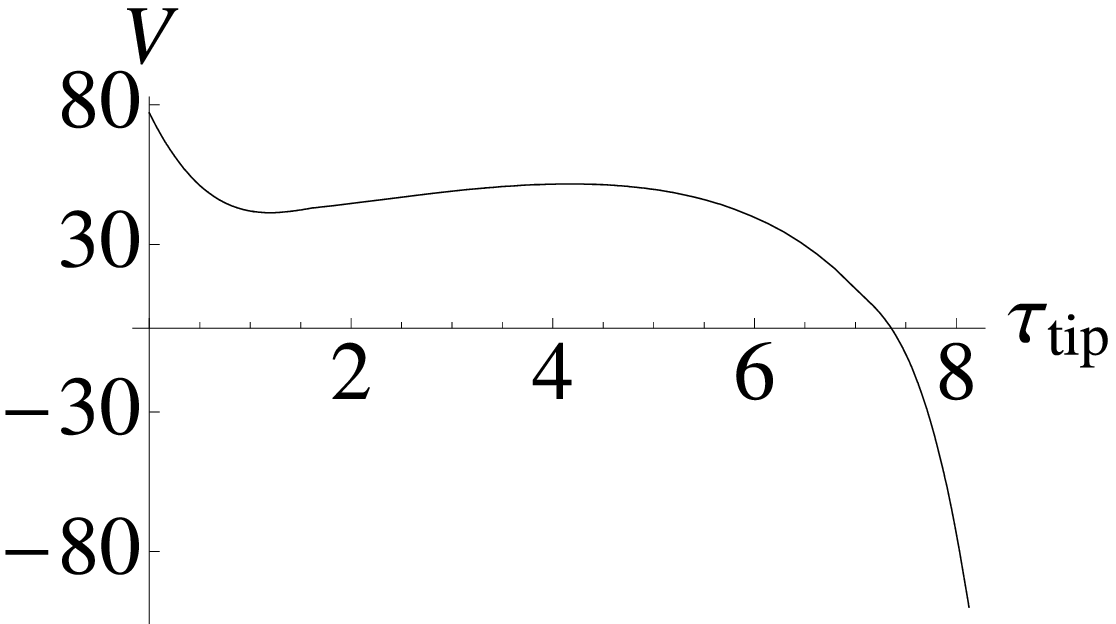}
\caption{The phase transition in the potential $V(L)$ (left) is implied by the presence of a local minimum in the $L(\tau_{tip})$ (center), or equivalently $V(\tau_{tip})$ (right), plot.}
\label{dtaP2}
\end{figure}

Though the massive case is not continuously connected with the massless one (since our gravity solution is always regular), we can figure out which form we should expect for $V(L)$ due to the previous behavior. For $\mu>1$, the IR solution is just the unflavored one: for this reason we see an unbounded linear behavior for the ``connected part'' of $V(L)$ at large $L$. This behavior can be glued with the massless one allowing the potential to have a double turnaround and an asymptotic linear increase. This is actually what it was found in \cite{Bigazzi:2008gd} in the massive-flavored CVMN theory (the plots are analogous to the ones in figure \ref{dtaP2}). The double turn around is such that there is a quantum phase transition between a Coulomb-like behavior for the potential at small $L$ and the linear behavior at large $L$. In \cite{Bigazzi:2008gd} this quantum phase transition was recognized to be a first order Van der Waals-like   transition, occurring for flavor mass parameters smaller than a certain critical value. For larger values the Coulomb and the linear phase of $V(L)$ are smoothly connected.
At the critical point for the disappearance of the phase transition, the latter is of second order.

\begin{figure}
\centering
\includegraphics*[width=.4\textwidth]{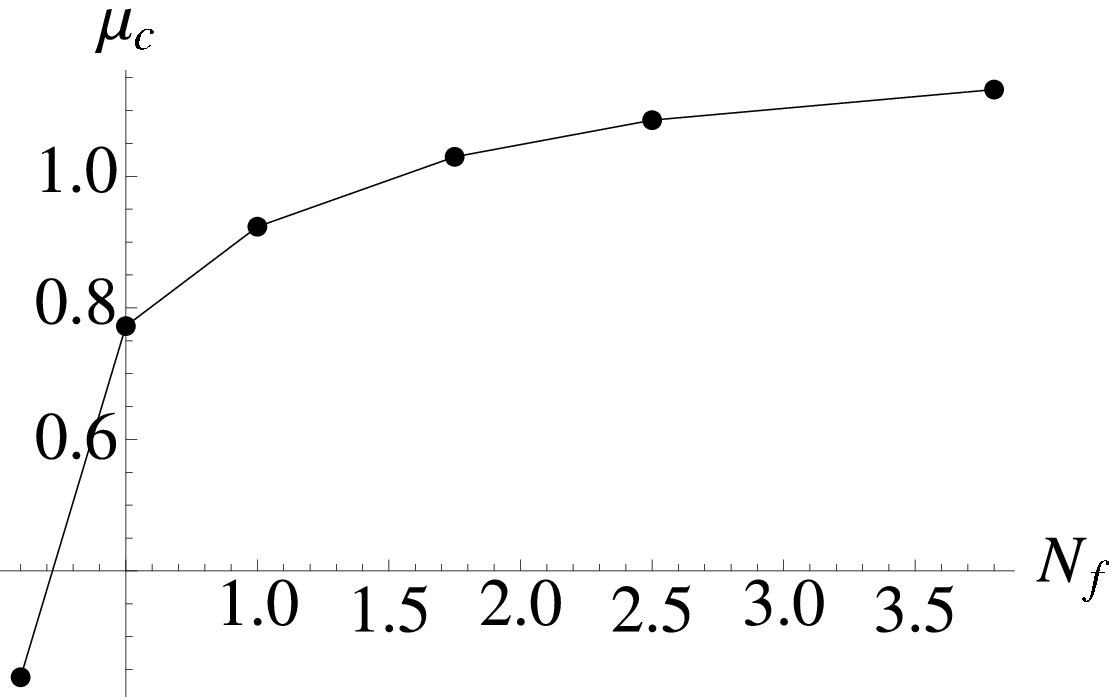}\quad
\includegraphics*[width=.4\textwidth]{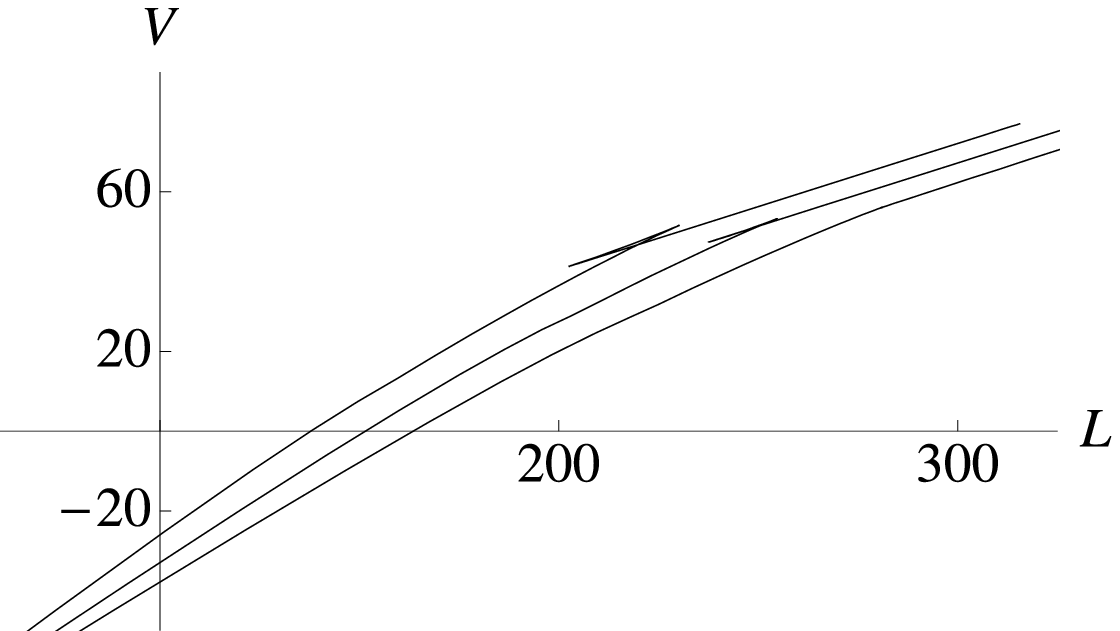}
\caption{Left: the critical value of the flavor mass $\mu_c$ for the appearance of the phase transition, as a function of the number of flavors $N_f$ in the constant IR string tension case. Right: in the same case, the $V(L)$ plot for $N_f=1$ and increasing values of $\mu$ (from left to right), showing the disappearance of the phase transition above $\mu_c$.}
\label{kcritP2}
\end{figure}

The solution presented in this paper is dual to a theory with many scales. 
Besides $\Lambda_{IR}$, there is a UV Landau pole-like scale related to $\tau_0$, the scale $\mu$ of the dynamical quarks and finally the higher dimensional operator scale set by $h_0$.
Thus, there is room for a quite rich pattern of appearance of phase transitions in the heavy quark potential.
In fact, these phase transitions appear if the parameters are varied in suitable ways.
In some cases, there are actually {\it two distinct phase transitions}, for example if we keep constant the IR string tension and we vary the mass of the dynamical flavors.
The first transition happens for mass smaller than a critical value $\mu_c$ which is an increasing function of $N_f$, as can be seen in figure \ref{kcritP2} (see also figure \ref{dtaP2}). 
The values of $\mu_c$ can be smaller or larger than $\mu=1$, so the latter point is not at all peculiar from this point of view.\footnote{Let us note that in the Heaviside approximation one has the very same result, even if of course the phase transition happens for $\mu_c$ above $\mu=1$.}
The second phase transition happens for mass above a second, larger critical value, i.e. when $m_q$ approaches $M_Q$.
As can be seen from the second plot in figure \ref{dtaP2}, which is equivalent to the pressure-volume graph of the Van der Waals system, the presence of a local minimum in the $L(\tau_{tip})$ plot implies the existence of the phase transition. 

It would be very interesting to give an exhaustive description of the patterns of the phase transitions in this complicated theory.
Our present aim is just to argue that, being the phase transition such a generic phenomenon, it is worthwhile to study its properties.
In the next section we are going to discuss its universality class.

\subsection{Critical exponents, universality classes and cusp catastrophes}
\label{critexp}
As we have argued above, the presence of the phase transition in our models is signaled by a local minimum in the behavior of the quark separation $L$ as a function of the position $\tau_{tip}$ of the tip of the string describing the Wilson loop. Let us focus on the family of functions $L(\tau_{tip},\mu)$, where we consider for simplicity just the behavior of the flavored KS system as we vary the mass $\mu$ of the dynamical flavors (equivalently, we could use $h_0$, for example). Near the critical point at $\mu=\mu_c$
we have been able to verify numerically, for a selected choice of parameters\footnote{We have used $g_s=1, N_f=1, \tau_0=10, h_0=10$ in the constant IR string tension case.} that the function $L(\tau_{tip},\mu)$ is well approximated by 
\be
\frac{L}{L_c}-1\approx -\left(\frac{\tau_{tip}}{\tau_c}-1\right)^3 - \left(\frac{\mu}{\mu_c}-1\right)\left(\frac{\tau_{tip}}{\tau_c}-1\right)\,.
\label{laround}
\ee
The behavior of the curve near the critical point allows us to extract the related critical exponents (for their definition see, for example, \cite{lebellac}) and to determine the universality class of our phase transitions. Along the critical line $\mu=\mu_c$ we find that
$(L/L_c)-1\approx [(\tau_{tip}/\tau_c)-1]^{\delta}$ with the critical exponent $\delta=3$.  At $L=L_c$, $\mu<\mu_c$ we find $[(\tau_{tip}/\tau_c)-1]\approx [1-(\mu/\mu_c)]^{\beta}$ with critical exponent $\beta=1/2$. The values we have found for $\delta$ and $\beta$ are thus the {\it classical ones}. The same results hold for the transition in the static quark-antiquark potential appearing in the unflavored KS model when varying $h_0$.

Due to the scaling relations, two critical exponents are sufficient to determine completely the universality class of the system, once its dimension is known. The scaling relations read
\be
\alpha+2\beta+\gamma=2\ , \qquad \gamma=\beta(\delta-1)\ , \qquad 2-\alpha=\nu d\ , \qquad \gamma=\nu(2-\eta)\ .
\ee 
The dimension $d$ enters the scaling relations of the critical exponents related to the correlation functions, which we cannot calculate; so, the ``effective dimension'' of our system is undetermined.
On the other hand, the scaling relations let us conclude that, given $\delta=3,\beta=1/2$, the other critical exponents are $\alpha=0$ and $\gamma=1$, as in the Landau (mean field) theory, and do not depend on the dimension.

Considering that both in \cite{sfetsos} and in \cite{alfonsoetal} the same classical exponents are found for phase transitions in different dimensions and driven by different mechanisms, we are led to conclude that there must be a universality at work, which would reflect the fact that we are basically analyzing a classical object: a macroscopic string.
We would like to argue that this is indeed the case. 
 
The universality of the critical exponents is related to the theory of singularity of families of functions, or ``Catastrophe theory'', see for example \cite{cata} and \cite{Chamblin:1999tk} for a
related connection in the context of charged black holes.

The basic theorem of catastrophe theory is due to Whitney and was extended by Thom and basically states that the singularities of a generic function can be of only a few types.\footnote{Let us stress that our system is ``generic'' (that is, without special symmetries or properties) by construction, since we only have it numerically. But in the case analyzed in Ref. \cite{sfetsos} the system is described analytically and is indeed generic.}
The theorem means that a generic function is ``equivalent'',\footnote{We are actually dealing with families of functions which also depend on some order parameters $a_i, i=1, ..., m$ (for us, it will be $m=2$). Two families of functions $F, \tilde F: \mathbb{R}^n \times \mathbb{R}^m \rightarrow \mathbb{R}$ are said to be equivalent if there are diffeomorphisms $y: \mathbb{R}^{n+m} \rightarrow \mathbb{R}^n,\ e: \mathbb{R}^m \rightarrow \mathbb{R}^m,\ g: \mathbb{R}^m \rightarrow \mathbb{R}$ such that $F(x,a_i)= \tilde F(y(x,a_i),e(a_i)) + g(a_i)$.} in the vicinity of the singularity, to some ``prototype'' singularity. The crucial point in this discussion is that the critical exponents appear to be invariant among the class of equivalent functions \cite{cata}. 

The prototype singularity which interests us is called the ``cusp catastrophe'' and it is defined 
by the function
\be\label{cusp}
f(x,a,b) = x^3 + a x + b\ ,
\ee
which has its critical point at $a=b=0$.\footnote{The equation of state of the Van der Waals system is recovered by the map $x=V-1,\ a=T-1,\ b=P-1$ in units of the corresponding critical values of the volume $V$, temperature $T$ and pressure $P$.}.
At the ``critical isotherm'' ($a=0$) the condition $f=x^3 + b=0$ determines the value of the critical exponent $\delta=3$ from the defining relation $b \sim x^\delta$.
Analogously, setting $b=0$, the condition $f=x^2 +a=0$ determines another critical exponent $\beta=1/2$ from the defining relation $x \sim a^\beta$.
The corresponding ``potential'' of the Landau theory is just the integral in $x$ of the function $f$ above, and is called ``unfolding'' in the catastrophe theory language.
The cusp can be seen by drawing in 3d the 2d surface $f=0$, 
and projecting it on the $(a,b)$-plane: 
in this plane, two curves meeting with a cusp at the critical point $(a=0,b=0)$ are the images of the local maxima and minima of $b(x)$ for different $a$'s.
For a detailed description of the cusp catastrophe, see for example \cite{cata}.

Remarkably, our curve $L(\tau_{tip},\mu)$ around the critical point (eq. (\ref{laround})) is precisely of the form $f(x,a,b)=0$ provided we map
\be
x \leftrightarrow \frac{\tau_{tip}}{\tau_c}-1\ , \qquad a \leftrightarrow \frac{\mu}{\mu_c}-1\ ,  \qquad b \leftrightarrow \frac{L}{L_c}-1\,.
\ee
The curve $L(\tau_{tip},\mu)$ passes from being monotonic when $\mu>\mu_c$ to developing a local minimum (and a local maximum) when $\mu<\mu_c$, where there is a first order phase transition (the transition is second order precisely when $\mu=\mu_c$). For such a family of functions Thom's theorem states that the system is equivalent to the cusp catastrophe (\ref{cusp}), and this determines its universality class.
Similar maps hold for analogous phase transitions in the quark-antiquark potential in other models with a gravity dual. Thus, the heavy quark potential in the stringy regime is equivalent to the cusp catastrophe, every time that there is a phase transition.
The cusp catastrophe is basically the Van der Waals system, so it implies that the critical exponents are the classical ones. As said, the crucial point is that the critical exponents appear to be invariant among the class of equivalent functions \cite{cata}.

Thus, we are led to conclude that \emph{the phase transition in the heavy quark potential in a theory with a supergravity dual is in the universality class of the Van der Waals system, at least in the regime of parameters where the gravity description is reliable.}
\section{A preliminary analysis of the unquenched mesonic spectrum}
\label{mesons}
\setcounter{equation}{0}
In this section we consider a probe D7-brane embedded, in the backreacted solution we have found, in a similar way as the dynamical flavor brane sources. Fluctuations on the probe brane worldvolume are mapped to mesonic modes of spin $J=0,1$. Our aim is to study how the spectrum of these fluctuations, hence the mass spectrum of the related mesons, varies with the dynamical flavor parameters $N_f,\mu$.

In general, the task of computing the spectrum, even in the quenched case, is quite difficult. This happens because, even if the 
background metric is simple, the induced metrics on the probes can become quite involved and the differential equations for the various modes can be non trivially coupled. For simplicity here we will only focus on the simplest fluctuations in the probe D7-brane action, i.e. fluctuations of the worldvolume gauge field. The equations of motion to solve will be of the form $\partial_a (e^{-\phi}\sqrt{-g} F^{ab})=0$, where $g$ stands for the induced string frame metric.

Let us start by rewriting our fully backreacted string frame metric, in a form similar
to \cite{kuper}
\bear
ds^2 &=& e^\frac{\phi}{2}\left[
h^{-\frac12}dx_{1,3}^2 + h^{\frac12}\
B^2(\tau) (d\tau^2 +(h_3 + \tilde h_3)^2) +\right. \rc
&&\left.
+ h^{\frac12}\ A^2(\tau)
\left(h_1^2 + h_2^2 + \tilde h_1^2 + \tilde h_2^2
+ \frac{2}{\cosh\tau} (h_2 \tilde h_2 - h_1 \tilde h_1)\right)
\right]\,,
\label{deformedmetric2}
\eear
where
\be
B^2(\tau) = \frac{e^{2G_3}}{9}\,\,,\qquad
\qquad
A^2(\tau) = e^{2G_2}\,,
\ee
and 
\bear
h_1 &=& -\cos\frac{\psi}{2} \sin \theta_1 d\varphi_1
+\sin\frac{\psi}{2} d\theta_1\,\,,\qquad
h_2 = -\sin\frac{\psi}{2} \sin \theta_1 d\varphi_1
-\cos\frac{\psi}{2} d\theta_1\,\,,\rc
\tilde h_1 &=& -\cos\frac{\psi}{2} \sin \theta_2 d\varphi_2
+\sin\frac{\psi}{2} d\theta_2\,\,,\qquad
\tilde h_2 = -\sin\frac{\psi}{2} \sin \theta_2 d\varphi_2
-\cos\frac{\psi}{2} d\theta_2\,\,,\rc
h_3&=&\frac{d\psi}{2} + \cos\theta_1 d\varphi_1\,\,,\qquad\qquad
\qquad\qquad
\tilde h_3=\frac{d\psi}{2} + \cos\theta_2 d\varphi_2\,\,.
\label{hdefs}
\eear
Moreover let us rewrite
\bear
\tilde h_1 &=& (h_1 - d\hat\gamma) \cos \hat\delta - (h_3 \sin\hat\gamma
+ h_2 \cos \hat\gamma) \sin \hat\delta\,\,,\rc
\tilde h_2 &=& (h_1 -d\hat\gamma)\sin\hat\delta + (h_3 \sin\hat\gamma
+ h_2 \cos \hat\gamma) \cos \hat\delta\,\,,\rc
\tilde h_3 &=& (h_3 \cos\hat \gamma -h_2 \sin \hat\gamma) + d\hat\delta\,,
\label{kuperdefs}
\eear
where we have basically replaced $\theta_2, \varphi_2$ by $\hat\gamma\in [ 0,\pi], \hat\delta \in [0,4\pi)$.

The reason why we want to make this very complicated change of variables
is that the non-chiral embedding we are interested in takes a simple
form in these coordinates. In particular, for the case of a massless embedding $z_1-z_2=0$, the D7 profile can be simply given by $\hat\gamma = 0$, $\hat\delta=$const \cite{kuper}. 

Following \cite{kuper}, let us ignore the dependence of the fluctuations
on the angles and expand the worldvolume gauge field as
\be
A(x_\mu,\tau) = e^{i k\cdot x}\left[
a(\tau) v_\mu dx^\mu + a_1(\tau) h_1 +
a_3 (\tau) h_3)\right]\,.
\ee
From the related equations of motion we find ($-k^2 = M^2$)
\bear
\partial_\tau\left(A^2 \tanh\tau
\partial_\tau a \right)+
M^2\, h\, A^2 B^2 \tanh \tau\,a = 0\,\,,\rc
\partial_\tau\left(\frac{\coth\frac{\tau}{2}}{h}
\partial_\tau a_1 \right)+\left[
-\frac12 \partial_\tau (h^{-1})-
\frac14 \frac{\tanh\frac{\tau}{2}}{h}+
M^2\,  B^2 \coth\frac{\tau}{2} \right] 
a_1 = 0\,\,,\rc
\partial_\tau\left(\frac{A^2\tanh\tau}{B^2h}
\partial_\tau a_3 \right)+\left[
- \partial_\tau (h^{-1})-
 \frac{B^2 \coth \tau}{A^2 h}+
M^2\, A^2\tanh\tau \right] 
a_3 = 0\,\,.
\eear
Notice that if we take $\phi=0$, these equations reduce
to (4.7), (4.9), (4.10) in \cite{kuper}.\footnote{Modulo a typo in equation (4.10) of \cite{kuper}.} 

\begin{figure}
 \centering
\includegraphics[width=.4\textwidth]{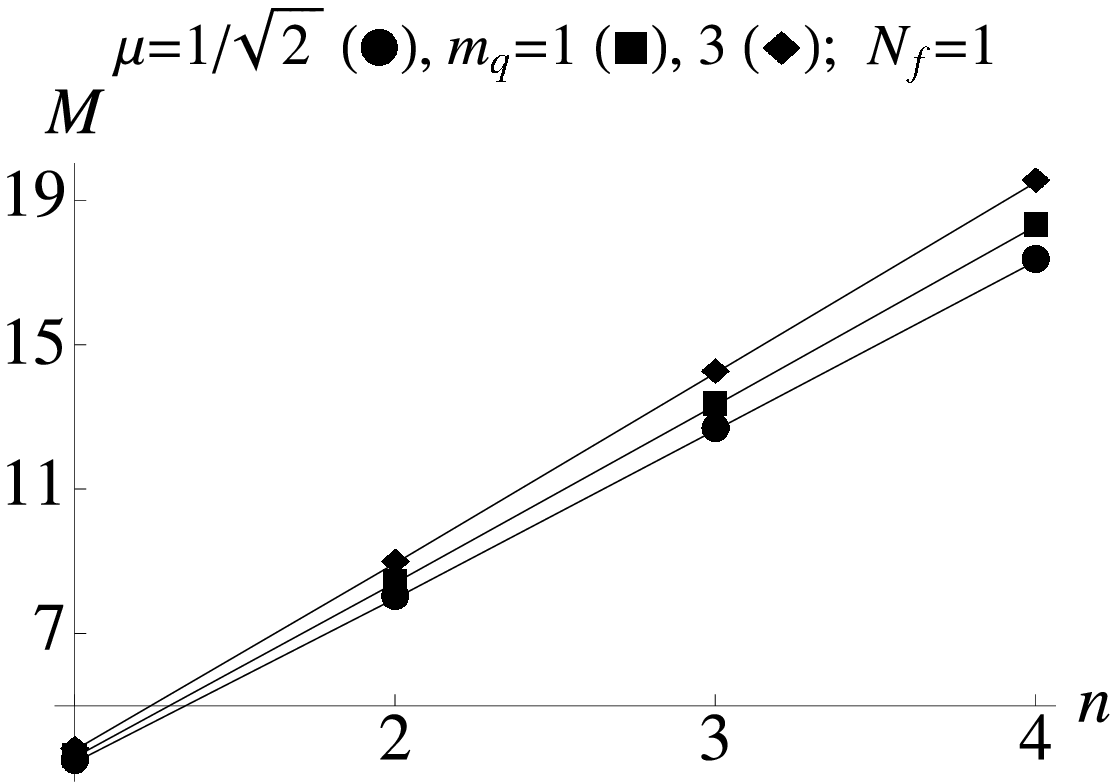}
\includegraphics[width=.4\textwidth]{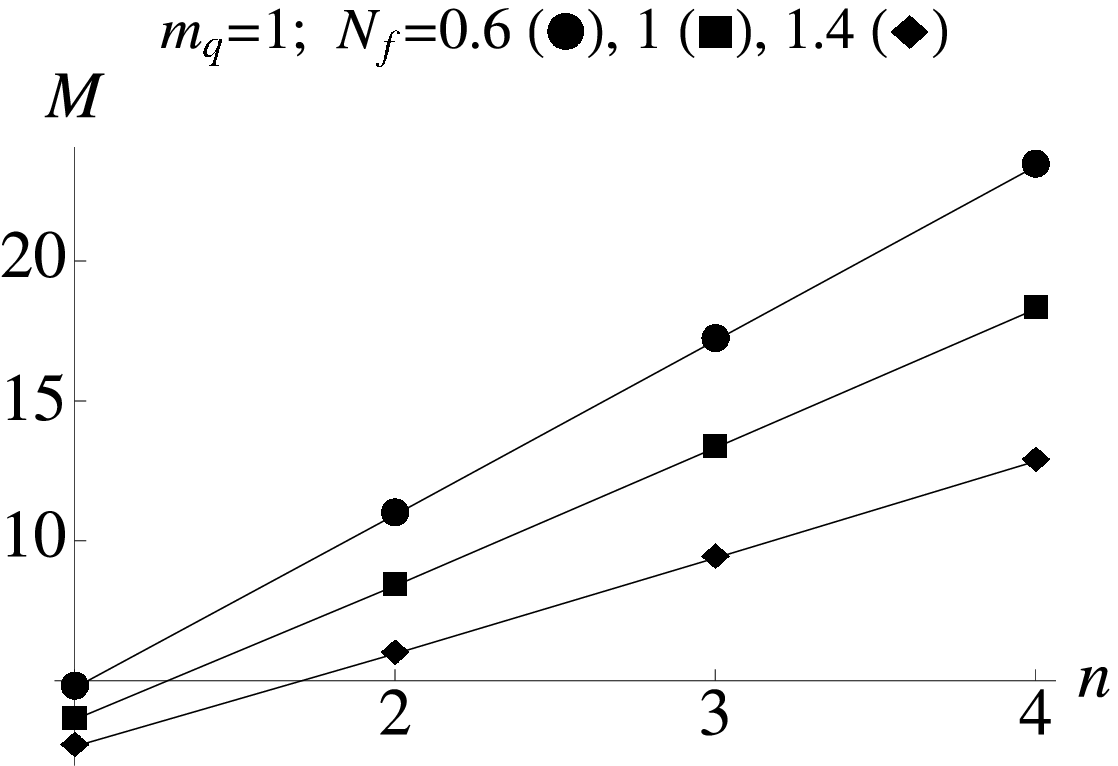}
\caption{Plots of the meson masses with $N_f$ and $m_q$ for fixed glueball (and KK) scale.}
\label{mes1}
\end{figure}

\begin{figure}
 \centering
\includegraphics[width=.4\textwidth]{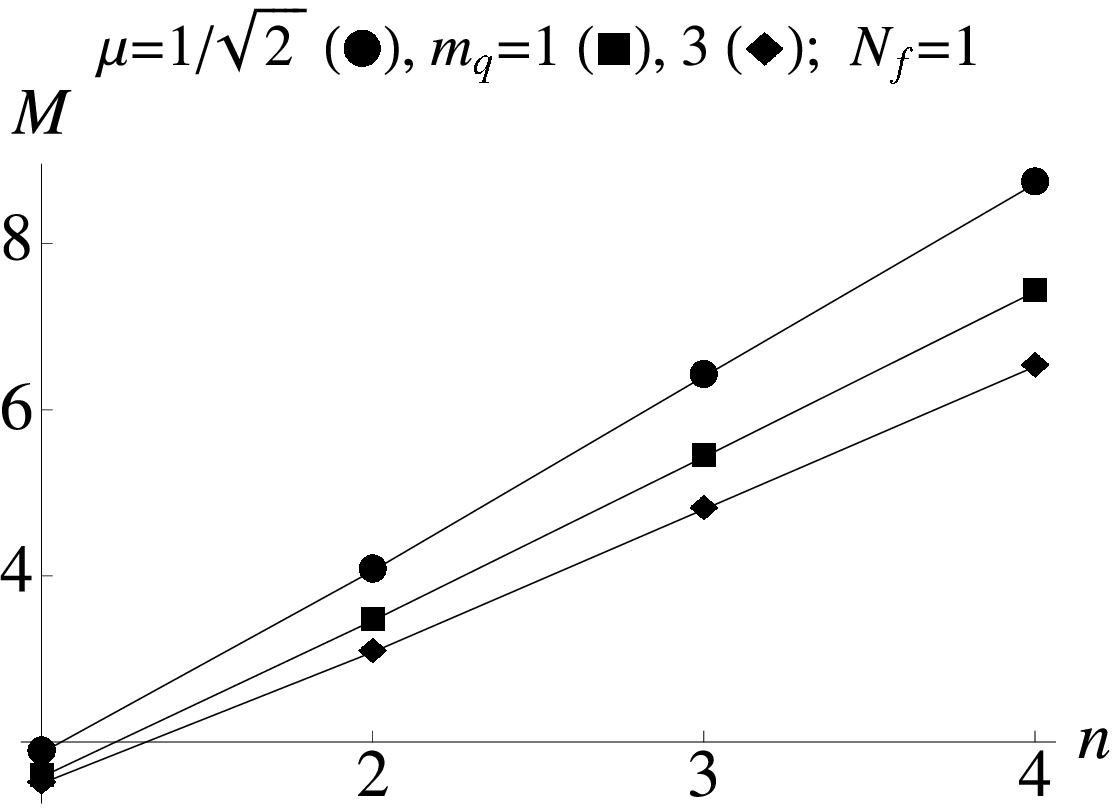}
\includegraphics[width=.4\textwidth]{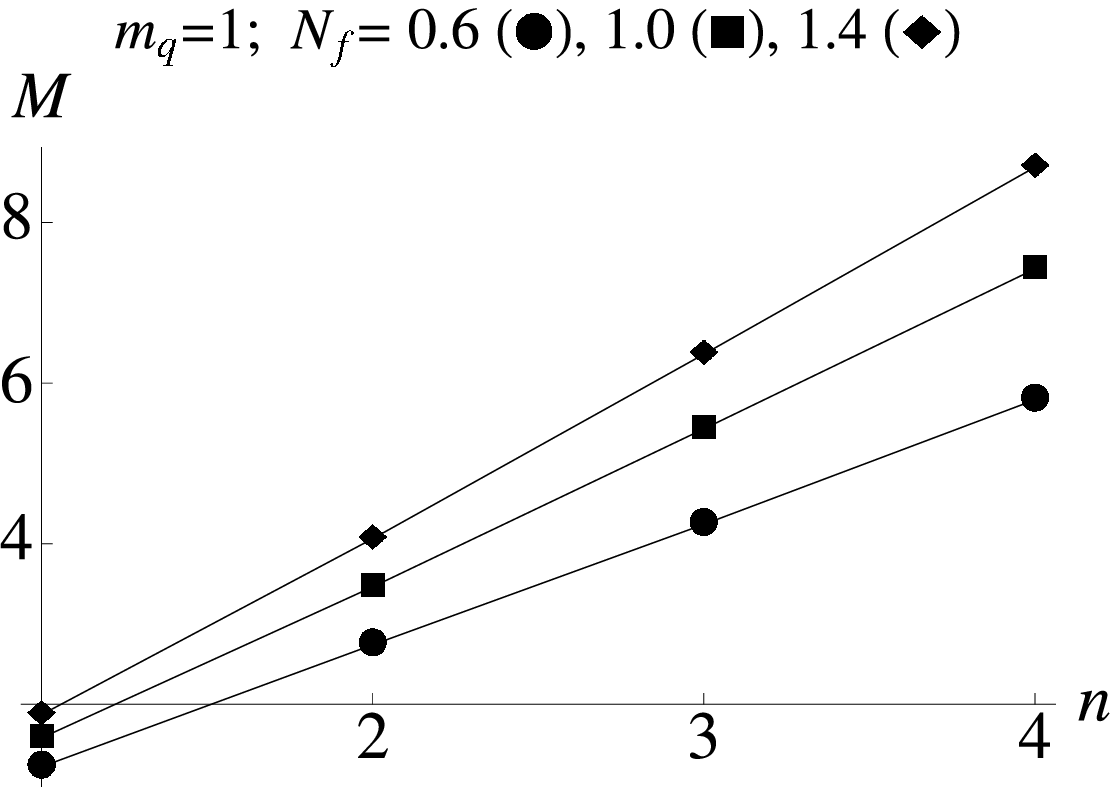}
\caption{Plots of the meson masses with $N_f$ and $m_q$ for fixed IR string tension.}
\label{mes2}
\end{figure}

We can now use the standard shooting technique to get the mesonic spectrum. Here we only focus on the $a(\tau)$ mode. We take $\tau_0$ to be fixed and impose regularity at $\tau\rightarrow0$. We cannot impose normalizability at $\tau\rightarrow\infty$ since our solutions are defined only up to $\tau_{max}$ where $h(\tau_{max})=0$. The most sensible choice seems to treat this point as the position of an infinite wall for the fluctuations and so to impose that they vanish at $\tau_{max}$. These boundary conditions render the spectrum discrete. We have examined it for the two Possibilities discussed in section \ref{wilson}, i.e. fixed glueball mass scale and fixed IR string tension. The results are shown in figure \ref{mes1} and \ref{mes2}.
The figures show that, as for the critical lengths, the behavior depends on the prescription we choose to fix some physical scale in the theory. For Possibility 1 (resp. Possibility 2) we see that the mesonic masses decrease (resp. increase) with the number of dynamical flavors and increase (resp. decrease) with their mass.

The lattice study in Ref. \cite{davies}, performed at fixed bare coupling, shows a decreasing behavior with $N_f$ of the difference between the meson masses of two distinct excitations (the $n=3$ and $n=1$ modes in our notation).
This behavior corresponds to the one of ``Possibility 2'', i.e. the constant string tension case.
Considering also the results for the screening lengths, we are led to conclude that keeping fixed the string tension is the prescription giving the most natural, possibly useful results.

\section{Summary and discussion}
\label{conclu}
The study of backreacted supergravity backgrounds offers a big opportunity to explore the effects of unquenched flavors in a holographic setup.  This analysis could shed light on new non-perturbative phenomena that take place in the planar limit of gauge theories.  In this paper we have developed this program for the case in which the dual gauge theory is the conifold theory of Klebanov and Strassler and the flavors are non-chiral and massive, with masses either larger or smaller than the dynamical IR scale $\Lambda_{IR}$. This is an interesting setup to consider, since it amounts to adding dynamical flavors to a confining gauge theory, like we do with pure Yang-Mills to get QCD.

In our approach, the dynamics of the backreacted gravity plus branes system is governed by an action in which the supergravity fields of the type IIB theory are coupled to the DBI+WZ action of the flavor D7-branes. Thus, the D7-branes act as dynamical sources of the different supergravity fields, whose  equations of motion and Bianchi identities  are modified by the presence of the flavor branes.  In the Veneziano limit  the number of flavors $N_f$ is large and we can homogeneously smear the flavor branes in their transverse space. This has the effect of substituting the $\delta$-functions source terms by continuous distributions. Amazingly, the flavor distribution function $N_f(\tau)$, that results from smearing also along the phases of the flavor mass terms, can be obtained in analytic form. The smeared D7-branes correspond to flavors having all the same mass $m$ in modulus. When $m>\Lambda_{IR}$, the distribution extends up to a certain finite value $\tau_q$ of the radial coordinate. When the mass is smaller than $\Lambda_{IR}$ the distribution extends up to $\tau=0$. It is important to keep in mind that in this case the flavor branes do not all extend up to the origin: some flavor branes end up at a finite distance from $\tau=0$ depending on the shift between the phase of the flavor mass term and that of the complex deformation parameter $\epsilon$. Thus, differently from the massless case, the origin is not a special point where all the flavor branes overlap. 

Knowing the density distribution of the flavor branes one can write an ansatz for the supergravity fields which is a direct generalization of the one for the massless case adopted in \cite{Benini:2007gx}.  The corresponding BPS equations for the different functions of the ansatz, which also solve the equations of motion,  can be integrated in closed form and the result is a generalization of both the unflavored \cite{ks} and massless flavored \cite{Benini:2007gx} backgrounds. Our new solutions have small curvature, and thus are reliable, if
$1\ll N_f\ll M$. As in the massless flavored case of \cite{Benini:2007gx} they have an UV singularity at $\tau=\tau_{max}$ where the warp factor vanishes, and a would-be Landau pole at $\tau_0>\tau_{max}$ where the dilaton diverges. However, remarkably, the massive flavored solutions are always regular at $\tau=0$ precisely like the unflavored KS solution and contrary to what happens when the unquenched quarks are massless. 

With the flavored solution at our disposal we have started a holographic analysis of the  non-perturbative dynamics of the dual planar gauge theory. In particular, we have considered the properties of bound states formed by an external heavy quark $Q$ and antiquark $\bar Q$. In the presence of dynamical flavors the $\bar Q Q$ states are metastable and can decay into heavy-light mesons. To characterize these decays we have numerically computed the  corresponding screening lengths and we have studied their behavior as functions of $N_f$ and of the mass parameters, keeping constant either the glueball mass scale or the IR string tension. We have also presented some preliminary analysis of the mesonic spectrum, obtained by analyzing regular fluctuations of a D7-brane probe (corresponding to massless flavors) in the backreacted background.

From the holographic study of the heavy quark potential, i.e. from a classical analysis of a static macroscopic open string on the background, we have uncovered the existence of quantum phase transitions, which seem to generically occur whenever there are at least two distinct physical scales in the dual gauge theory.  Our system has several scales and we have found numerical evidence that the ``connected'' part of the static potential undergoes first-order  quantum phase transitions between a Coulomb-like and a linear behavior. These transitions occur until some parameter (like the flavor mass) reaches a critical value, where  the transitions become of second order. We have evaluated the corresponding critical exponents and we have found that, in all the cases, they are given by the classical mean-field values. These results, together with those of refs. \cite{sfetsos} and \cite{alfonsoetal}, led us to conjecture the existence of a universality class for the transitions  in the static potential of every planar gauge theory having a supergravity dual. 

In this work we have just started to examine the predictions of the Klebanov-Strassler background with massive unquenched flavor. The fact that we have an IR regular solution with new parameters at our disposal (as compared with the unflavored and massless flavored solutions) opens new windows for many potentially interesting future studies. 

One problem of obvious interest, for which we have already presented here some preliminary results, is the analysis of the mesonic spectrum in the backreacted background. We would like to determine the dependence of more generic mass levels on the number of unquenched flavors and on their masses, as well as the influence of the different prescriptions to determine the parameters of the solution.

Another interesting direction of  future research is the analysis of the entanglement entropy for our  backreacted solution, following the proposal of ref.  \cite{Ryu} for theories that have a gravity dual. In this holographic formulation the quantum entanglement entropy between a region $A$ and its complement is obtained as the minimal area of a surface that approaches the boundary of $A$ at the boundary of the bulk manifold.  This proposal has been applied in ref. \cite{Klebanov:2007ws} to the study of confining (unflavored) backgrounds. The authors of \cite{Klebanov:2007ws} found that the entropy of a large class of confining models, including the Klebanov-Strassler solution, displays a quantum phase transition similar to the confinement/deconfinement transition at finite temperature. It would be very interesting to analyze how the inclusion of unquenched flavor modifies this results. In particular, by using this entanglement entropy approach we should be able to find a phase structure similar to the one uncovered here from the analysis of the Wilson loops. 

Finding a finite temperature version of our supergravity solutions would be of great interest. This would allow us to study the properties of the dual quark-gluon plasma and of the corresponding flavored black holes and to describe how the phase structure of our SQCD-like models varies with the flavor parameters.\footnote{See \cite{pallab} for recent related studies in compact models at weak coupling.}   Due to the absence of supersymmetry, the main technical problem to tackle in this case is the fact that one has to deal directly with the second-order equations of motion of the gravity plus branes system. The stability of a distribution of smeared flavor branes in this non-supersymmetric setup is an issue that should be analyzed with care. The relevance of this line of research would also be due to the fact that, since lattice gauge theory is intrinsically Euclidean, the string/gauge theory correspondence is at present the only available tool to explore real-time dynamical properties of strongly coupled QCD-like quark-gluon plasmas.

Finally, it would be important to uncover the details of the field theory which lies in the deep IR of the flavored cascading model, specially in the small (or zero) mass $0\le\mu\le1$ cases. Finding the precise maps between field theory vacua and supergravity solutions will surely require further investigation.  

We are working on some of these problems and we intend to report on them in a near future. 
\vskip 15pt
\centerline{\bf Acknowledgments}
\vskip 10pt
\noindent
We are grateful to R. Argurio, F. Benini, M. Caldarelli, F. Canoura, E. Carlon, M. D'Elia, R. Emparan, F. Ferrari, S. Kuperstein, B. Lucini, C. N\'u\~nez, J. Shock and D.  Zoakos for very useful discussions.
This work has been supported by the European Commission FP6 programme
MRTN-CT-2004 v-005104, ``Constituents, fundamental forces and symmetries in the universe''. F. B. is also supported  by the Belgian Fonds de la Recherche Fondamentale Collective (grant 2.4655.07), by the Belgian Institut Interuniversitaire des Sciences Nucl\'eaires (grant 4.4505.86) and
the Interuniversity Attraction Poles Programme (Belgian Science
Policy). A. C. is also supported by the FWO -
Vlaanderen, project G.0235.05 and by the Federal Office for
Scientific, Technical and Cultural Affairs through the Interuniversity Attraction
Poles Programme (Belgian Science Policy) P6/11-P. A. P. is also supported by a NWO VIDI grant 016.069.313 and by
INTAS contract 03-51-6346. A. R. is also supported by the MEC and  FEDER (grant FPA2005-00188), the Spanish Consolider-Ingenio 2010 Programme CPAN (CSD2007-00042) and Xunta de Galicia (Conselleria de Educacion and grant PGIDIT06PXIB206185PR).

{\it F. B. and A. L. C. would like to thank the Italian students, parents and scientists for 
their activity in support of public education and research.}

\appendix

\setcounter{equation}{0}
\renewcommand{\theequation}{\Alph{section}.\arabic{equation}}
\section{Finding $N_f(\tau)$ from the smeared embedding}
\label{appNf}
In this paper we have focused on backreacted D3-D7 solutions on the deformed conifold, the manifold defined by the equation $z_1z_2-z_3z_4=\epsilon^2$ in $\IC_4$. The D7-brane embeddings we have considered are the ``non-chiral'' ones discussed in \cite{kuper}, with representative equation $z_1-z_2=2\hat\mu$. The general family of embeddings along which we have distributed the D7-branes, is obtained through an $SO(4)$ rotation of the previous equation
\be
\bar p z_1 - p z_2 + \bar q z_3 + q z_4 = 2\hat\mu\,,
\label{genembeddingksapp}
\ee
where $p, q$ span a unit 3-sphere
\begin{equation}
p=\cos\frac{\theta}{2}e^{i(\frac{\chi+\phi}{2})}\,,\quad q=\sin\frac{\theta}{2}e^{i(\frac{\chi-\phi}{2})}\,,
\end{equation}
and $\chi \in [0,4\pi)$, $\phi \in [0,2\pi)$, $\theta \in [0,\pi]$.

To give more explicit expressions in terms of the standard deformed conifold coordinate we can use the known relations
\bear
z_1&=& \epsilon\, e^{-\frac{i}{2}(\varphi_1+\varphi_2)}\left[\cos\frac{\theta_1}{2}\cos\frac{\theta_2}{2} e^{-\frac{i}{2}\psi}e^{-\frac{\tau}{2}} + \sin\frac{\theta_2}{2}\sin\frac{\theta_1}{2} e^{\frac{i}{2}\psi}e^{\frac{\tau}{2}}\right]\,,\rc
z_2&=& \epsilon\, e^{\frac{i}{2}(\varphi_1+\varphi_2)}\left[\sin\frac{\theta_1}{2}\sin\frac{\theta_2}{2} e^{-\frac{i}{2}\psi}e^{-\frac{\tau}{2}} + \cos\frac{\theta_2}{2}\cos\frac{\theta_1}{2} e^{\frac{i}{2}\psi}e^{\frac{\tau}{2}}\right]\,,\rc
z_3&=& -\epsilon\, e^{\frac{i}{2}(\varphi_1-\varphi_2)}\left[\sin\frac{\theta_1}{2}\cos\frac{\theta_2}{2} e^{-\frac{i}{2}\psi}e^{-\frac{\tau}{2}} - \sin\frac{\theta_2}{2}\cos\frac{\theta_1}{2} e^{\frac{i}{2}\psi}e^{\frac{\tau}{2}}\right]\,,\rc
z_4&=& -\epsilon\, e^{-\frac{i}{2}(\varphi_1-\varphi_2)}\left[\sin\frac{\theta_2}{2}\cos\frac{\theta_1}{2} e^{-\frac{i}{2}\psi}e^{-\frac{\tau}{2}} - \sin\frac{\theta_1}{2}\cos\frac{\theta_2}{2} e^{\frac{i}{2}\psi}e^{\frac{\tau}{2}}\right]\,.
\eear
In the limit $\tau\rightarrow\infty$, $|\epsilon|\rightarrow 0$, with fixed $|\epsilon|^2 e^{\tau}\sim r^3$ the above expressions approach the singular conifold ones. 

Making use of these formulas we can rewrite (\ref{genembeddingksapp}) as
\be
\sinh \frac{\tau}{2} \Delta_1 - i \cosh \frac{\tau}{2}
\Delta_2 = \frac{\hat \mu}{\epsilon}\,,
\label{generalemb}
\ee
with
\bear
\Delta_1 &=& \cos\frac{\theta}{2}
\left[-\cos \frac{\theta_1}{2}\cos \frac{\theta_2}{2} \cos \xi_1 +
\sin\frac{\theta_1}{2}\sin \frac{\theta_2}{2} \cos \xi_2 \right]
+\rc
&&
+\sin\frac{\theta}{2}
\left[\sin \frac{\theta_1}{2}\cos \frac{\theta_2}{2} \cos \xi_3 +
\cos\frac{\theta_1}{2}\sin \frac{\theta_2}{2} \cos \xi_4 \right]\,\,,\rc
\Delta_2 &=& \cos\frac{\theta}{2}
\left[\cos \frac{\theta_1}{2}\cos \frac{\theta_2}{2} \sin \xi_1 +
\sin\frac{\theta_1}{2}\sin \frac{\theta_2}{2} \sin \xi_2 \right]
+\rc
&&
+\sin\frac{\theta}{2}
\left[\sin \frac{\theta_1}{2}\cos \frac{\theta_2}{2} \sin \xi_3 -
\cos\frac{\theta_1}{2}\sin \frac{\theta_2}{2} \sin \xi_4 \right]\,\,.
\label{Deltas}
\eear
We have defined
\bear
\xi_1 &=& \frac12 (\varphi_1 + \varphi_2 + \psi + \xi + \phi)\,\,,\qquad
\xi_2 = \frac12 (\varphi_1 + \varphi_2 - \psi + \xi + \phi)\,\,,\rc
\xi_3 &=& \frac12 (\varphi_1 - \varphi_2 - \psi - \xi + \phi)\,\,,\qquad
\xi_4 = \frac12 (\varphi_1 - \varphi_2 + \psi - \xi + \phi)\,\,,
\eear
satisfying the relation $\xi_1 + \xi_3 = 
\xi_2 + \xi_4$. Notice also that the $\varphi$'s only enter through
the combinations $\varphi_1 + \phi$ and $\varphi_2 + \chi$. 
Setting $\theta = \chi = \phi = 0$
(giving $z_1 - z_2 = 2\hat \mu$), the
$\Delta$'s reduce to the $\Theta$'s in (\ref{lockup}).

By explicit computation, one finds
\bear
\Delta_1^2 + \Delta_2^2=  \frac12
\Big[1+ \sin\theta_1 \sin\theta_2
\sin(\varphi_1 + \phi) \sin(\varphi_2 + \chi)+\rc
+
\cos\theta \left(\cos\theta_1 \cos\theta_2 -\sin\theta_1 \sin\theta_2
\cos(\varphi_1 + \phi) \cos(\varphi_2 + \chi)\right)
+ \rc
+\sin\theta \left(-\sin\theta_1 \cos\theta_2 \cos(\varphi_1 + \phi)
-\cos\theta_1 \sin\theta_2 \cos(\varphi_2 + \chi)\right)
\Big]\,.
\eear
Notice that the $\psi$-dependence has dropped out of this expression.
It can be checked that $\Delta_1^2
+ \Delta_2^2 \leq 1$.

From the modulus squared and phase in (\ref{generalemb}), one finds the equations
\bear
f_1&=&-2\arctan\left[\coth \left(\frac{\tau}{2}\right) \frac{\Delta_2}{\Delta_1}\right]
-2\beta + 4\pi\,n = 0 \,\,,\rc
f_2 &=& \sinh^2 \frac{\tau}{2} \Delta_1^2 +\cosh^2 \frac{\tau}{2} \Delta_2^2
-\mu^2 =0\,\,,
\label{f1f2eqs}
\eear
with 
\be
\mu \equiv \frac{|\hat \mu|}{|\epsilon|}\,,\qquad \hat\mu\equiv |\hat\mu|e^{i\,\beta}\,.
\ee
The minimal value of $\tau$ reached by one of these embeddings is the one for which the inequality $\Delta_1^2
+ \Delta_2^2 \leq 1$ is saturated. At that point, using the second equation
in (\ref{f1f2eqs}), one finds $\Delta_1^2 = \cosh^2 \frac{\tau_{min}}{2} 
- \mu^2$ and $\Delta_2^2 = -\sinh^2 \frac{\tau_{min}}{2} 
+ \mu^2$. Then, from the first expression in (\ref{f1f2eqs}), we
find the following relation among the modulus and phase of the mass term
and the minimal value $\tau_{min}$
\be
\tan^2 \beta = \coth^2 \frac{\tau_{min}}{2}\left( \frac{-\sinh^2 \frac{\tau_{min}}{2} 
+ \mu^2}{\cosh^2 \frac{\tau_{min}}{2} 
- \mu^2}\right)\,.
\ee
\subsection{Smearing the embedding}
Let us now consider the maximally symmetric smeared distribution of $N_f\gg1$ D7-branes generally embedded as above. The related density distribution $\Omega$, whose general expression is given in eq. (\ref{generalOmega}), can be rewritten, after the trivial integration over $\beta$, as
\be
\Omega = \frac{N_f}{32 \pi^3}\int \delta(f_2) \sin\theta df_1 
\wedge df_2  d\theta d\chi d\phi\,.
\label{intbeta}
\ee
Despite this being a quite non trivial integral, we know that the result is severely constrained by the symmetries to take the form given in (\ref{massive_2form}). In that expression $\Omega_{\tau\psi}$ results to be independent on $\theta_i,\varphi_i$ and so we can just pick, say, $\theta_i=\varphi_i=0$ in the corresponding expression deduced from (\ref{intbeta}). This way we get
\bear
f_1&=& -\chi - \phi - \psi  + 2\arg(\epsilon) + 2 \arctan{[\frac{-\sin{(\psi+\chi+\phi)}}{e^{-\tau}-\cos{(\psi+\chi+\phi)}}]} -2\beta +4\pi  n\,,\nonumber\\
f_2&=& 2[\cosh{\tau}-\cos{(\psi+\chi+\phi)}]|\epsilon|^2 \cos^2{\frac{\theta}{2}} - 4|\hat\mu|^2 \,,
\eear
and so
\be
(df_1\wedge df_2)_{\tau,\psi}|_{\psi=0}= -2[\cosh{\tau}+\cos{(\chi+\phi)}]|\epsilon|^2 \cos^2{\frac{\theta}{2}}\, ,
\ee
where we have put $\psi=0$ since $\Omega_{\tau\psi}$ is not expected to depend on $\psi$. Using this expression and integrating over $\theta$ we get 
\be
\Omega_{\tau\psi}=-\frac{N_f}{64\pi^3}\int \frac{4|\hat\mu|^2[\cosh\tau+\cos(\chi+\phi)]} {|\epsilon|^2[\cosh\tau-\cos(\chi+\phi)]^2}\Theta(2[\cosh\tau-\cos{(\chi+\phi)}]|\epsilon|^2 - 4|\hat\mu|^2)\, d\phi\, d\chi\,.
\label{maleficint}
\ee
This has the right limit to the expression found in the singular conifold case \cite{varna} when $\tau\rightarrow \infty, \epsilon\rightarrow 0$, $e^{\tau}|\epsilon|^2$ fixed.
Moreover, it is consistent with the embedding reaching the tip of the conifold for sufficiently small mass, also in the case where the mass is non zero: the Heaviside gives vanishing contribution for small $\tau$ only if $|\hat\mu|$ is sufficiently large (compared to $|\epsilon|$). In general, the Heaviside fixes the absolute minimal value of $\tau$ to be given by $\cosh\tau_{q}= -1 + (2|\hat\mu|^2/|\epsilon|^2)$ in agreement with the considerations in the previous sections.

Let us solve the integral above. First of all, let us redefine
\be
x=\cosh\tau, \quad \alpha=\chi+\phi, \quad \gamma=\chi-\phi\,,
\ee
such that $x_q\equiv 2\mu^2-1$. Notice also that $|\det J|=1/2$, where $J$ is the Jacobian associated to the angular change of variables above. Notice also that requiring
\be
\int_0^{4\pi}\int_0^{2\pi}d\chi\, d\phi = \frac{1}{2} \int_0^{m\pi}\int_0^{n\pi} d\alpha\, d\gamma\,,
\ee
implies $m\,n=16$. Hence, let us choose $m=n=4$ i.e. $[0,4\pi]$ as a range for both $\alpha$ and $\gamma$. Integration over $\gamma$ is trivial. From a comparison of (\ref{maleficint}) with $\Omega_{\tau\psi}$ as expected by symmetry arguments (cfr. eq. (\ref{massive_2form})) we thus arrive at the following integro-differential equation
\be
\frac{dN_f(x)}{dx}=\frac{N_f \mu^2}{2\pi}\int_0^{4\pi}\frac{(x+\cos\alpha)}{(x-\cos\alpha)^2\sqrt{x^2-1}}\Theta[x-\cos\alpha-2\mu^2]\,d\alpha\,.
\label{int2app}
\ee
In the massless case $\mu=0$ \cite{Benini:2007gx} one simply has $N_f(\tau)={\rm const}=N_f$. Notice that if $x>2\mu^2+1$, the condition imposed by the Heaviside is valid for every $\alpha$ and the calculation gets simpler (we can just erase the Heaviside in the integral). The difficult piece of calculation is in the region $2\mu^2-1<x<2\mu^2+1$. After performing the integral in $\alpha$ we get the first order equation
 (\ref{nfprimeeq}). Solving this equation we get the effective running number of flavors $N_f(\tau)$ discussed in section \ref{smedefcon}.

The above results have been cross-checked performing numerically the integration in eq. (\ref{generalOmega}), also considering other components of the density distribution form and no contradiction with the maximally symmetric expression for $\Omega$ (in particular the expected $\psi$ independence of its components) was found. 

\section{The holomorphic embeddings on the ``backreacted'' conifold}
\label{radial}
\setcounter{equation}{0}
All along the paper we have shown that the (Einstein frame) metric ansatz for the backreacted background has a standard warped $(3,1)\times 6$ form. The 6d transverse space is a ``flavor-deformation'', driven by some functions of the radial coordinate, of the deformed conifold metric. If in the standard (deformed) conifold case the $\kappa$-symmetric embedding equations for the D7-branes can be simply written as holomorphic expressions in the $z_i$, we have to ask what happens in the backreacted case. The answer to this question is simple, since it can be shown that our backreacted 6d metric is an $SU(3)$ structure metric on the deformed conifold. Hence the embedding equations for the D7-branes can be written in the same way as in the unflavored case, modulo, eventually, a difference in the radial coordinate.

To see which coordinate to choose, let us write the deformed conifold as in \cite{patse}
\be
\det W = -\frac{\epsilon^2}{2}\,,
\ee
where $W= w_i\sigma^i + w_4$ and $\sigma^i$ are the Pauli matrices. The radial coordinate $u$ of the conifold can be defined as
\be
u^2 = {\rm Tr}(W\,W^+)\,.
\label{u}
\ee
It is related to the standard deformed conifold coordinate $\tau_{DC}$, by $u^2=|\epsilon|^2\cosh\tau_{DC}$.
The family of K\"ahler metrics on the deformed conifold is given in \cite{patse} and reads
\be
ds_{6\,DC}^2=[K''(1-\frac{|\epsilon|^4}{u^4})u^4+K'u^2]\left[\frac{du^2}{u^2(1-\frac{|\epsilon|^4}{u^4})}+\frac{1}{4}(g^5)^2\right]+ ...\,,
\label{famdefcon}
\ee
where we write only the terms which are relevant for the present analysis. Now, our backreacted 6d metric reads
\be
ds_{6\,fDC}^2 = \frac{1}{9}e^{2G_3(\tau)}(d\tau^2 + g_5^2) + ...\,,
\ee
and thus requiring it belongs to the family (\ref{famdefcon}) we get
\bear
\frac{1}{9}e^{2G_3} &=& \frac{1}{4}[K''(1-\frac{|\epsilon|^4}{u^4})u^4+K'u^2]\,,\nonumber\\
\frac{1}{9}e^{2G_3}d\tau^2 &=&  [K''(1-\frac{|\epsilon|^4}{u^4})u^4+K'u^2]\frac{du^2}{u^2(1-\frac{|\epsilon|^4}{u^4})}\,.
\eear
These conditions can be fulfilled if
\be
d\tau^2 = 4 \frac{du^2}{u^2(1-\frac{|\epsilon|^4}{u^4})}\,\quad\rightarrow\quad \tau=\log\left[u^2+\sqrt{u^4-|\epsilon|^4}\right]+{\rm const}\,,
\ee
which is nothing more than
\be
u^2= |\epsilon|^2 \cosh\tau\,,
\ee
if we choose the (irrelevant) ``const'' above to be: ${\rm const}= -\log|\epsilon|^2$. From this we see that the explicit expression for the D7 embeddings on the backreacted 6d manifold is {\it exactly the same} as those on the standard deformed conifold: $\tau=\tau_{DC}$.
\section{The smeared D7-brane action}
\label{smedac}
\setcounter{equation}{0}
Let us now see which form is taken by the action for the flavor D7-branes in case we homogeneously smear them. The action is 
taken as the sum of DBI and WZ terms. Let us start with the latter, which are simpler.
\subsection{The WZ term}
Assuming that the three-form  fluxes have only components along the
internal directions,  the WZ term is
\beq
S_{WZ}\,=\,T_{D7}\,\,\sum_{N_f}\int_{{\cal M}_8}\,
\Big[\, \hat C_{8}\,+\,\hat   B_2\wedge  \hat  C_6\,+\,
{1\over 2} \hat B_2\wedge  \hat B_2 \wedge \hat C_4\,\Big]\,,
\label{WZ-localized}
\eeq
where ${\cal M}_8$ is the eight-dimensional worldvolume of the D7-branes,  $\hat B_2$ denotes the pullback to ${\cal M}_8$ of the  NSNS two-form $B_2$ and $ \hat C_{8}$,  $\hat C_{6}$ and $ \hat C_{4}$ denote the pullbacks of the corresponding RR potentials  of type IIB supergravity. 

Let us now define the two-form $\Omega$ as the  Poincare dual of ${\cal M}_8$, which for any eight-form $A_8$ satisfies
\beq
\sum_{N_f}\int_{{\cal M}_8}\,\hat A_8\,=\,\int_{{\cal M}_{10}}\,
\Omega\wedge A_8\,\,.
\eeq
In terms of $\Omega$,  we can rewrite the WZ action of the D7-branes as
\beq
S^{smeared}_{WZ}\,=\,T_{D7}\int_{{\cal M}_{10}}\,\Omega\wedge
\Big[\, C_{8}\,+\, B_2\wedge  C_6\,+\,
{1\over 2}\, B_2\wedge B_2 \wedge C_4\,\Big]\,\,.
\label{WZ-with-Omegaapp}
\eeq
It is clear from (\ref{WZ-with-Omegaapp}) that $\Omega$ determines the RR charge of the D7-branes. Notice that the wedge product of any form with  $\Omega$ naturally implements its pullback to the worldvolume ${\cal M}_8$.  Moreover, it is clear from (\ref{WZ-with-Omegaapp}) that $\Omega$ acts as a magnetic source for the RR field strengths $F_1$, $F_3$ and $F_5$. Indeed, the equations of motion of $C_8$,  $C_6$ and $C_4$ give rise to the modifications of the Bianchi identities for $F_1$, $F_3$ and $F_5$ as given in section \ref{ansatz}.
\subsection{The smeared $\kappa$-symmetry condition}
Let us rewrite the Einstein frame metric ansatz used in the paper, in terms of another radial coordinate $r$, related to $\tau$ by
$3e^{-G_3}dr=d\tau$
\bear
\label{metricr}
ds^2&=&\Big[\,h(r)\,\Big]^{-\frac{1}{2}}\,dx^2_{1,3}\,+
\,\Big[\,h(r)\,\Big]^{\frac{1}{2}}\,ds_6^2\,,\rc
ds_6^2&=& dr^2 + e^{2G_1}(\sigma_1^2+\sigma_2^2) + e^{2G_2} \bigg[
(\omega_1 + g\,\sigma_1)^2+ (\omega_2 + g\,\sigma_2)^2
\bigg]+{{e^{2G_3}}\over 9}(\omega_3 + \sigma_3)^2\,.
\eear
Let us now introduce the following tangent space basis
\bear
&&e^{x^{\mu}}\,=\,h^{-1/4}\, dx^{\mu}\,\, \qquad \mu=0,\ldots,3 \,\, , 
\qquad  e^r\,=\,h^{1/4}dr\,\, , \rc\rc
&&e^1\,=\,h^{1/4}e^{G_1}\sigma_1 \,\, , \qquad \qquad \qquad \qquad \qquad 
e^2\,=\,h^{1/4}e^{G_1}\sigma_2 \,\, , \rc\rc
&&e^{\hat{1}}\,=\,h^{1/4}e^{G_2}(\omega_1\,+\,g\sigma_1) \,\, , 
\qquad \qquad \qquad e^{\hat{2}}\,=\,h^{1/4}e^{G_2}(\omega_2\,+\,g\sigma_2) \,\, , \rc\rc
&&e^{\hat{3}}\,=\,h^{1/4}{{e^{G_3}} \over 3}(\omega_3\,+\,\sigma_3)\,\, .
\label{frame}
\eear
In this basis, the Killing spinors $\epsilon$ of the background can be written as
\beq
\epsilon\,=\,e^{{{\alpha} \over 2}\Gamma_{1\hat{1}}}h^{-1/8}\eta\,\, , 
\label{Killing}
\eeq
where $\alpha$ is an angle such that $\tan \alpha\,=\,-g e^{G_2-G_1}$ and $\eta$ is a constant spinor  satisfying  the following projection conditions
\beq
\Gamma_{x^0 x^1x^2 x^3}\,(i\sigma^2)\,\eta\,=\,\eta\,\,,\qquad\qquad
-\Gamma_{12}\,(i\sigma^2)\,\eta\,=\,\Gamma_{\hat 1\hat 2}\,(i\sigma^2)\,\eta\,=\,\eta\,\,.
\label{projections}
\eeq
In (\ref{projections}) we have employed a double spinor notation for $\eta$.  Notice that these algebraic conditions determine four independent spinors, which implies that our configurations are $1/8$ supersymmetric. 

The K\"ahler form $J$ of the backreacted deformed conifold is the two-form whose components are  the fermion bilinears $J_{\mu\nu}\,=\,h^{{1\over 4}}\,\,\bar\epsilon\,(-i\sigma_2)\,\Gamma_{\mu\nu}\,\epsilon$,  where $\mu$, $\nu$ are indices along the internal directions and $\epsilon$ is a Killing spinor normalized as $\bar\epsilon\epsilon=h^{-{1\over 4}}$. By using (\ref{Killing}) and (\ref{projections}),  the K\"ahler form of the transverse 6d manifold can be written as
\beq
h^{-{1\over 2}}J\,=\,e^{G_1+G_2}\,\Big(\,g^1\wedge g^4\,-\,g^2\wedge g^3\,\Big)\,-\,
{e^{G_3}\over 3}\,dr\wedge g^5\,\,.
\label{J-flavoredconifold}
\eeq
Making use of the equation $3(G_1'\,+\,G_2')\,=e^{G_3-G_1-G_2}$, which is a consequence of the BPS system (\ref{de}),  one can check that $h^{-{1\over 2}}J$ is closed
\beq
d\big(h^{-{1\over 2}}J\big)\,=\,0\,\,.
\eeq
The $\kappa$-symmetry condition  for the localized non-chiral embedding is \cite{kuper}
\beq
\hat J\wedge \hat B_2\,=\,0\,.
\eeq
The smeared version of this condition is
\beq
\Omega\wedge J\wedge B_2\,=\,0\,\,,
\eeq
where $\Omega$ is the D7-brane density distribution form. One can verify that, indeed, this equation is satisfied by our ansatz for $B_2$ and the K\"ahler form.
\subsection{The DBI term}
For zero worldvolume gauge fields the Einstein frame DBI term for the D7-branes can be written, in the localized case, as
\beq
S_{DBI}\,=\,-T_{D7}\,\,\sum_{N_f}\int_{{\cal M}_8}\,d^8\xi\,\,e^{\phi}\,
\sqrt{-\det\Big(\hat G_8\,+\,e^{-{\phi\over 2}}\,\hat B_2\Big)}\,,
\label{DBI-localized}
\eeq
where the $\xi$'s are coordinates that parameterize the eight-dimensional worldvolume ${\cal M}_8$ of the D7-branes and $\hat G_8$ denotes the pullback to ${\cal M}_8$ of the ten-dimensional metric $G$. 
For a background metric with standard warped form, as the one used in this paper, we can rewrite the above expression as
\beq
S_{DBI}\,=\,-T_{D7}\,\,\sum_{N_f}\int_{{\cal M}_8}\,d^4x\,d^4\zeta\,\,e^{\phi}\,
\,h^{-1} \,
\sqrt{\det\Big(\hat G_4\,+\,e^{-{\phi\over 2}}\,\hat B_2\Big)}\,,
\label{DBI-split}
\eeq
where $\zeta^{i}$ $(i=1,\cdots 4)$ are the coordinates of the four-cycle wrapped by the branes. 

The DBI action for the smeared branes takes a remarkably simple form due to the calibration condition on the wrapped cycles.
This condition allows us to rewrite (see eq. (B.10) in \cite{Benini:2007kg})
\beq
\sqrt{\det\Big(\hat G_4\,+\,e^{-{\phi\over 2}}\,\hat B_2\Big)}\,\,d^4\zeta\,=\,
{1\over 2}\,\Big(\,\hat J\wedge \hat J\,-\,e^{-\phi}\,\hat B_2\wedge \hat B_2\,\Big)\,\,.
\label{callibrated-vol}
\eeq
Using this result in (\ref{DBI-split}), we get
\beq
S_{DBI}\,=\,-{T_{D7}\over 2}\,\,\sum_{N_f}\int_{{\cal M}_8}\,d^4x\,h^{-1}\,\,
\Big(\,e^{\phi}\,\hat J\wedge \hat J\,-\,\hat B_2\wedge \hat B_2\,\Big)\,\,.
\eeq
Following the standard rule, the smeared version of this action is
\be
S_{DBI}^{smeared}\,=\,-\,{T_{D7}\over 2}\,\,
\int_{{\cal M}_{10}}\,d^{4}x\,
\Omega\wedge\,{\rm Vol (M_{1,3}})\,\wedge\,
\Big(\,e^{\phi}\, J\wedge  J\,-\, B_2\wedge  B_2\,\Big)\,.
\label{smeared-action}
\ee
\section{Matching with the massless case}
\label{matchmas}
\setcounter{equation}{0}
As we have shown in section \ref{smedefcon}, when the flavor mass vanishes, $N_f(\tau)=N_f$. In this case the integral defining $\eta(\tau)$ in (\ref{eta-def}) can be done explicitly and one can verify that ${\cal K}(\tau)$ is given by
\beq
{\cal K}(\tau)\,=\,{\Lambda(\tau)\over 4^{{1\over 3}}\,(\tau_0-\tau)^{{1\over 3}}}\,\,,
\eeq
where $\Lambda(\tau)$ is the function defined in \cite{Benini:2007gx}, namely
\beq
\Lambda(\tau)\,=\,{
\Big[\,2(\tau-\tau_0)(\tau-\sinh 2\tau)\,+\,\cosh
(2\tau)\,-\,2\tau\tau_0\, -\,1\,\Big]^{{1\over 3}}\over
\sinh\tau}\,\,.
\eeq
Moreover, if we define the new constant $\hat\epsilon$ as
\beq
\hat\epsilon^{{4\over 3}}\,=\,\Big({N_f\over 16\pi}\Big)^{{1\over 3}}\,
\epsilon^{{4\over 3}}\,\,,
\eeq
one can verify that the functions $G_1$, $G_2$ and $G_3$ reduce in this case to:
\bear
&&e^{2G_1}\,=\,{1\over  4}\,\,\hat\epsilon^{{4\over 3}}\,
{\sinh^2\tau\over \cosh\tau}\,\Lambda(\tau)\,\,,\rc\rc
&&e^{2G_2}\,=\,{1\over  4}\,\,\hat\epsilon^{{4\over 3}}\,\,\cosh\tau\,\Lambda(\tau)\,\,,\rc\rc
&&e^{2G_3}\,=\,6\,\hat\epsilon^{{4\over 3}}\,\,{\tau_0-\tau\over
\big[\,\Lambda(\tau)\,\big]^2}\,\,,
\eear
which, indeed, are the values found in \cite{Benini:2007gx}.
\section{The second order equations and consistency checks}
\label{eom}
\setcounter{equation}{0}
We verify below that the second order equations of motion for the dilaton, the graviton and the various forms are a consequence of the first-order BPS equations. Let us rewrite them using the $r$ variable introduced in appendix \ref{smedac}. For the warp factor the equation is 
\begin{equation}
 h'\, e^{2G_1+2G_2+G_3}\,=\,-
{3 \over 4}{\alpha'}^2M^2\Big[f-(f-k)F+{N_f\over 4\pi}fk\Big]\,+\,
N_0\,,
\label{Kh2-fkFapp}
\end{equation}
where $N_0$ is the integration constant we have put to zero in the paper. For the other metric functions we have 
\bear  
&&G_1'\,-\,{1 \over 6}e^{G_3-G_1-G_2}\,-\,
{3 \over 2}e^{G_2-G_1-G_3}\,+\,{3 \over
2}e^{G_1-G_2-G_3}\,=\,0\,\, , \rc\rc
&&G_2'\,-\,{1 \over 6}e^{G_3-G_1-G_2}\,+\,{3
\over 2}e^{G_2-G_1-G_3}\,-\,{3 \over
2}e^{G_1-G_2-G_3}\,=\,0\,\, , \rc\rc
&&G_3'\,+\,{1\over 3}
e^{G_3-G_1-G_2}\,-\,3\,e^{G_2-G_1- G_3}\,+\,{3 N_f\over  8\pi}\,\,e^{\phi-G_3}\,=\,0\,\, ,
 \label{de}
\eear
and, for the dilaton
\begin{equation}
\phi'\,=\,{3N_f\over 4\pi}\,e^{\phi-G_3}\,.
\label{firstorder-dilaton}
\end{equation}
The BPS equations for the functions $k$, $f$ and $F$  of the three-forms are
\bear
&& k'\,=\,3e^{\phi-G_3}\,\Bigg(\,F\,+\,{N_f\over 4\pi}\,f
\,\Bigg)\,{1+g\over 1-g}\,\,,\rc\rc
&&f'\,=\,3e^{\phi-G_3}\,\Bigg(\,
1\,-\,F\,+\,{N_f\over 4\pi}\,k\,\Bigg)\,{1-g\over 1+g}\,\,,\rc\rc
&&F'\,=\,{3 \over 2}e^{-\phi-G_3} (k-f)\,\,.
\label{kfF-with-g}
\eear
It is also interesting to recall that, in the deformed conifold case, the fibering function $g$ is related to the metric functions $G_1$ and $G_2$ as
\begin{equation}
\label{g}
g^2\,=\,1\,-\,e^{2(G_1-G_2)} \,\,.
\end{equation}
The equations written above are a consequence of the requirement of supersymmetry. The corresponding Killing spinors have been written in appendix \ref{smedac}.
\subsection{The second order equations for dilaton and forms}
The second order equations of motion for the dilaton and the forms in a setup like the one considered in the paper are (see also \cite{Benini:2007gx,Benini:2007kg})
\bear
&&\frac{1}{\sqrt{-G}} \partial_M \Big ( G^{MN}\, \sqrt{-G}\, 
\partial_N \, \phi  \Big )\,=\,e^{2\phi}\, F_{1}^2\,+\,\frac{1}{12} \Big (
e^{\phi} F^2_{3}\,-\,e^{-\phi}\,H^2_{3} \Big )\,-\,
\frac{2\kappa^2_{10}}{\sqrt{-G}} \frac{\delta}{\delta \phi} S^{smeared}_{DBI}\,,\rc
&&d\big(\,e^{2\phi}\,{}^*\,F_1\big)\,=\,-e^{\phi}\,H_3\wedge {}^*F_3\,-\,
{1\over 24}\,B_2\wedge B_2\wedge B_2\wedge B_2\wedge\Omega\,\,,\rc
&&d\big(\,e^{\phi}\,{}^*\,F_3\big)\,=\,-H_3\wedge F_5\,+\,{1\over 6}\,B_2\wedge B_2\wedge B_2\wedge\Omega\,,\rc
&&d\big(\,e^{\phi}\,{}^*\,H_3\big)\,=\,e^{\phi}\,F_1\wedge{}^*F_3\,-\,F_5\wedge F_3\,+\,
{\rm Vol (M_{1,3}})\,\wedge B_2\wedge\Omega\,\,,
\label{eomsdilforms}
\eear
where ${\rm Vol (M_{1,3}})$ is the Minkowski part of the volume element of ${\cal M}_{10}$ (for our ansatz ${\rm Vol (M_{1,3}})=h^{-1} d^4 x$). In (\ref{eomsdilforms})
 we have not included  the $F_5$ equation of motion since it coincides with the $F_5$ Bianchi identity which is solved provided the warp factor satisfies the first order equation (\ref{Kh2-fkFapp}).

It is possible to show that the solutions of the first order BPS equations solve the above equations of motion. The following hints can be useful in the proof. 

As for the dilaton e.o.m. notice that
\beq
e^{\phi} F^2_{3}\,-\,e^{-\phi}\,H^2_{3}\,=\,0\,\, ,
\label{isd}
\eeq
which is actually a consequence of the fact that, as expected from supersymmetry, the complex  three-form  $G_3\,=\,F_{(3)}\,+\,i e^{-\phi}H$ is imaginary self-dual in the internal manifold. Moreover, it is evident from $S_{DBI}^{smeared}$ that the $B_2$ field does not contribute to the equation of motion of the dilaton. 

As for the equations for the forms $F_1$ and $F_3$ notice that in our setup
\be
B_2\wedge B_2\wedge B_2\wedge B_2=0\,,\qquad B_2\wedge B_2\wedge B_2=0\,.
\ee
One can check that the equation for $F_1$ in (\ref{eomsdilforms}) is satisfied identically, as in the massless case. Moreover, but much less trivially, the equation for $F_3$ and $H_3$ are satisfied too due to the first order BPS equations. 
\subsection{Einstein equations}
The Einstein equations for our system are
\bear
R_{MN} - \frac{1}{2}G_{MN} R &=&
\frac{1}{2} \Big( \partial_M \phi \partial_N \phi -\frac{1}{2} G_{MN} 
\partial_P \phi \partial^P \phi \Big)
+ \frac{1}{2} e^{2\phi} \Big( F_M^{(1)} F_N^{(1)} 
-\frac{1}{2} G_{MN} F_1^2 \Big) + \rc\rc
&& + \frac{1}{96} F_{MPQRS}^{(5)} F_N^{(5)PQRS} \,+\,
{1\over 12}\,e^{\phi}\,\Big(\,3F^{(3)}_{MPQ}\,F^{(3)}_{N}{}^{PQ}\,
-\,{1\over 2}\, G_{MN}\,F_3^2 \Big)\,+\,\rc\rc
&&+{1\over 12}\,e^{-\phi}\,\Big(\,3H_{MPQ}^{(3)}\,H_{N}^{(3)\,PQ}
-\,{1\over 2}\, G_{MN}\,H_3^2 \Big)\,\,+\, T_{MN}\,\,,
\label{Einstein-eq}
\eear
where $T_{MN}$ is the DBI contribution to the energy-momentum tensor, namely
\beq
T_{MN}\,=\,-{2\kappa_{10}^2 \over \sqrt{-G}}\,\,
{\delta S^{smeared}_{DBI} \over \delta G^{MN}}\,\,.
\label{TDBI-definition}
\eeq
In order to check the fulfillment of (\ref{Einstein-eq}) it is essential to calculate the different components of $T_{MN}$. We will compute them by performing explicitly the derivative of the smeared DBI action (\ref{smeared-action}) with respect to the metric $G^{MN}$. Let us first consider the case in which $M$ and $N$  in (\ref{TDBI-definition}) are indices along the Minkowski directions. In this case, the dependence of 
$S^{smeared}_{DBI}$ on $G^{x^{\mu}x^{\nu}}$ comes from the  ${\rm Vol (M_{1,3}})$ volume form. The corresponding derivative is straightforward to compute and the result in flat components with respect to the basis (\ref{frame}) is
\beq
T_{\underline{x^{\mu}}\,\underline{x^{\nu}}}\,\,d^6\eta
\,=\,-{1\over 4\,h\sqrt{-G}}\,\,\eta_{\mu\nu}\,e^{\phi}\,\Omega\wedge \Big[\,J\wedge J\,-\,
e^{-\phi}\,B_2\wedge B_2\,\Big]\,\,,
\label{Tmin-smeared}
\eeq
where the $\eta$'s are the coordinates of the transverse 6d manifold. By using the explicit expressions of $\Omega$ and $J$ (eq. (\ref{J-flavoredconifold})), as well as our ansatz for $B_2$ (eq. (\ref{theansatz})), one can easily compute the wedge product of forms appearing on the right-hand side of (\ref{Tmin-smeared}). One gets
\beq
T_{\underline{x^{\mu}}\,\underline{x^{\nu}}}\,=\,-\Big[\,{N_f\over 4\pi}\,h^{-{1\over 2}}\,
e^{\phi-G_1-G_2}\,+\,
{3N_f'\over 8\pi}\,h^{-{1\over 2}}\,e^{\phi-G_3}\,+\,
{3{\alpha'}^2M^2 N_f'\over 32 \pi}\,h^{-{3\over 2}}\,
e^{-2G_1-2G_2-G_3}\,k\, f\,\Big]\,\eta_{\mu\nu}\,\,.
\label{Txx}
\eeq
Let us now obtain the components of $T_{MN}$ along the internal manifold. Clearly, the only dependence of the right-hand side of (\ref{smeared-action}) on the metric of 6d manifold comes from the K\"ahler form $J$. Then, if $\mu$, $\nu$ are coordinate indices along  the internal manifold and if $E_a^{\mu}$ are the coefficients of the corresponding inverse vierbein, one has\footnote{Let $A^{(p)}$ be an arbitrary $p$-form which, in the basis of the frame one-forms $e^{a}$, can be written as
\beq
A^{(p)}\,=\,{1\over p!}\,A^{(p)}_{a_1\cdots a_p}\,e^{a_1}\wedge\cdots e^{a_p}\,\,.
\eeq
Then, we define $\iota_{e^{a}}\,\big[A^{(p)}]$ as the following $(p-1)$-form
\beq
\iota_{e^{a}}\,\big[A^{(p)}]\,=\,{1\over (p-1)!}\,A^{(p)}_{a a_2\cdots a_p}\,
e^{a_2}\wedge\cdots e^{a_p}\,\,.
\label{iA}
\eeq
}
\beq
T_{ab}\,d^6\eta\,=\,{e^{\phi}\over 2 h\sqrt{-G}}\,
\Omega\wedge E_a^{\mu}\,E_b^{\nu}\, {\delta\over \delta G^{\mu\nu}}\,
\Big[\,J\wedge J\Big]
\,=\,-{e^{\phi}\over 2h\sqrt{-G}}\,\Omega\wedge e^{b}\,\wedge
\iota_{e^a}\,\big[\,J\,\big]\,\wedge J\,\,,
\label{T-cycle-flat}
\eeq
where $a,b$ are flat indices and $e^{a},e^{b}$ are frame one-forms along the 6d directions. In a one-form basis in which $J$ has the canonical form 
$J=e^1\wedge e^2+e^3\wedge e^4+e^5\wedge e^6$ it is straightforward to verify that the non-diagonal terms $e^b\wedge \iota_{e^a}\big[J]$, for $a\not=b$, vanish. Actually, 
 $T_{ab}$ has the following diagonal form
\beq
T_{ab}\,\,d^6\eta\,=\,-{e^{\phi}\over 4h\sqrt{-G}}\,\,\Omega \wedge(J\wedge J)_{e^a}\,\delta_{ab}\,\,,
\label{Tab-JJ}
\eeq
where $(J\wedge J)_{e^a}$  is the part of  $J\wedge J$ containing the one-form $e^a$. It is clear from (\ref{J-flavoredconifold}) that one can construct a basis in which $J$ has the canonical form  by rescaling appropriately the forms $g^1, \cdots, g^5$ and $dr$. After 
computing the wedge products of the different components of $J\wedge J$ with $\Omega$, one gets
\bear
&&\Omega\wedge (J\wedge J)_r\,=\,\Omega\wedge (J\wedge J)_{g^5}\,=\,
{h^{{1\over 2}}\sqrt{-G}\over \pi}\,e^{-G_1-G_2}\,N_f\,d^6\eta\,\,,\rc\rc
&&\Omega\wedge (J\wedge J)_{g^i}\,=\,{h^{{1\over 2}}\sqrt{-G}\over 2 \pi}\,
\Big[\,N_f\,e^{-G_1-G_2}\,+\,3N_f'\,e^{-G_3}\,\Big]\,d^6\eta\,\,,
\qquad (i=1,\cdots 4)\,\,.\qquad
\label{OmegaJJ-components}
\eear
It is now straightforward to obtain the remaining components of $T$ in  the basis
(\ref{frame}), namely
\bear
&&T_{\underline{r}\underline{r}}\,=\,T_{\hat 3\hat 3}\,=\,-{N_f\over 4\pi}\,h^{-{1\over 2}}\,e^{\phi-G_1-G_2}\,\,,\rc\rc
&&T_{ab}\,=\,T_{\hat a\hat b}\,=\,-{N_f\over 8\pi}\, h^{-{1\over 2}}\,e^{\phi-G_1-G_2}\,-\,
{3N_f'\over 8\pi}\,h^{-{1\over 2}}\,e^{\phi-G_3}\,\,,
\qquad (a,b=1,2)\,\,.
\label{Tab}
\eear
It is quite non trivial to show, using the previous ingredients, that the solutions of the BPS equations also satisfy the Einstein equations of motion. In this verification, the following remarkable identity satisfied by the three-forms of our ansatz
\bear
e^{\phi}\,F^{(3)}_{\underline{MPQ}}\,F^{(3)}_{\underline{N}}{}^{\underline{PQ}}\,+\,
e^{-\phi}\,H_{\underline{MPQ}}^{(3)}\,H^{(3)}_{\underline{N}}{}^{\underline{PQ}}\,=\,
{1\over 3}\,e^{-\phi}\,H_3^2\,\delta_{\underline{MN}}\,\,,\rc\rc
M,N\,=\,r,1,2,\hat 1,\hat 2, \hat 3\,\,,
\label{3form-id}
\eear
is quite useful. 
\section{The string breaking length}
\label{sbl}
\begin{figure}
 \centering
\includegraphics[width=.24\textwidth]{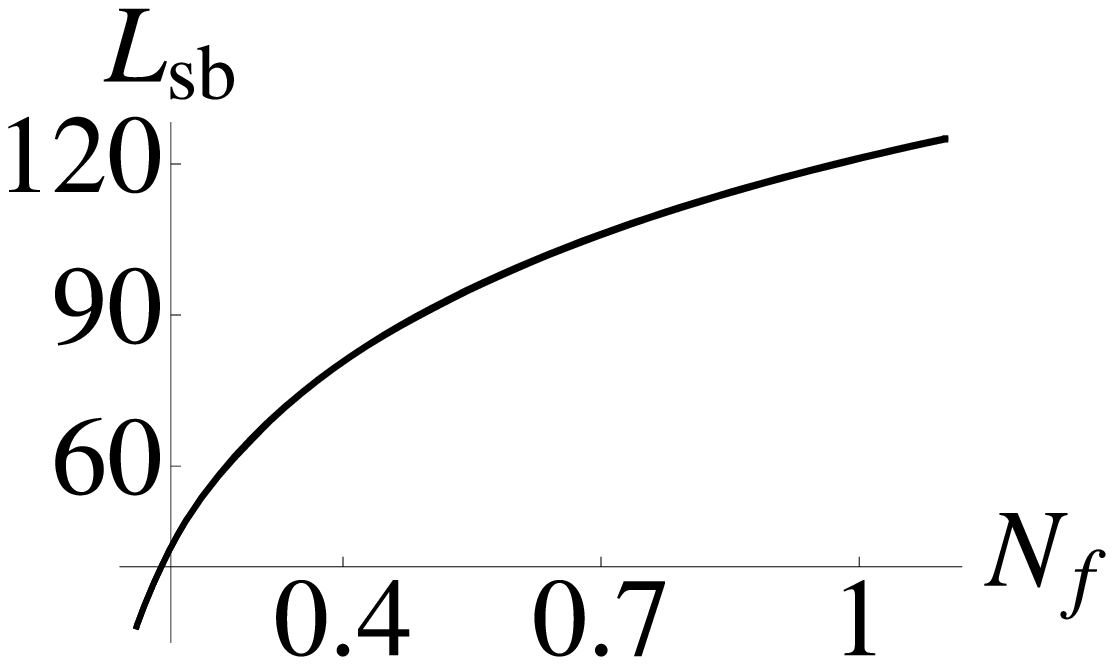}
\includegraphics[width=.24\textwidth]{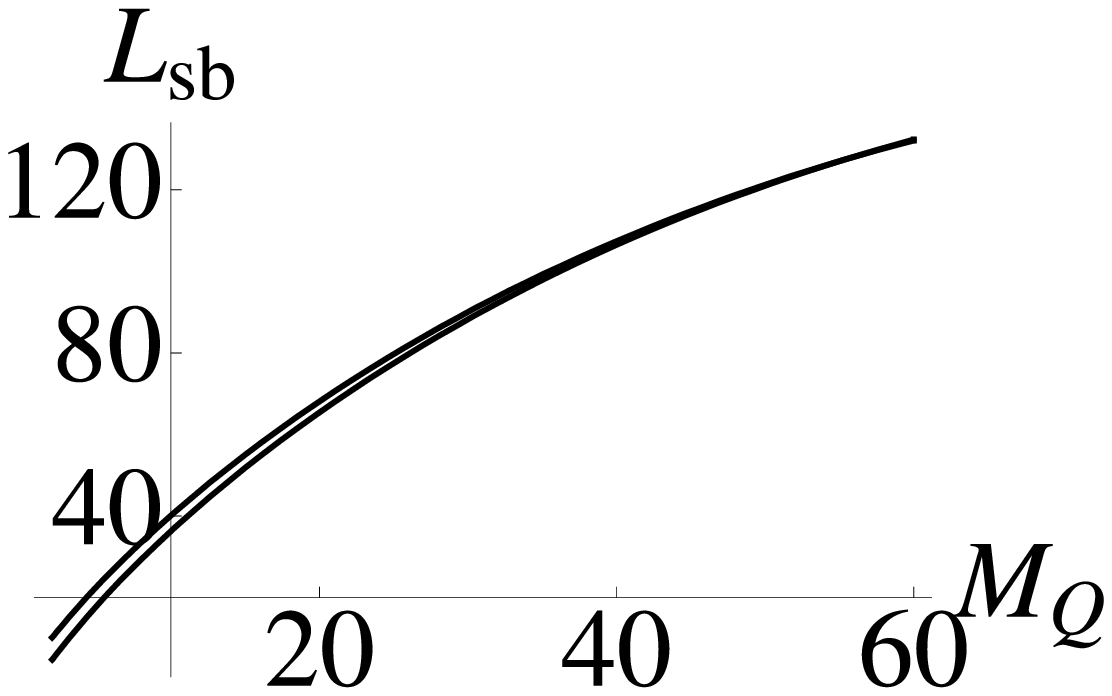}
\includegraphics[width=.24\textwidth]{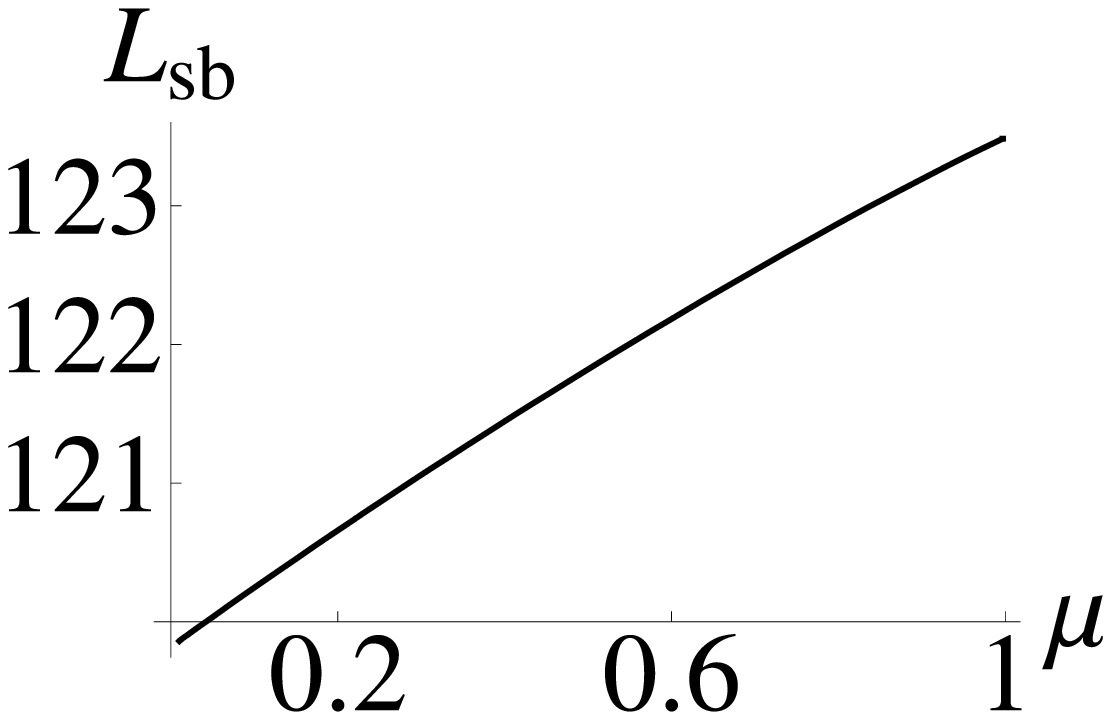}
\includegraphics[width=.24\textwidth]{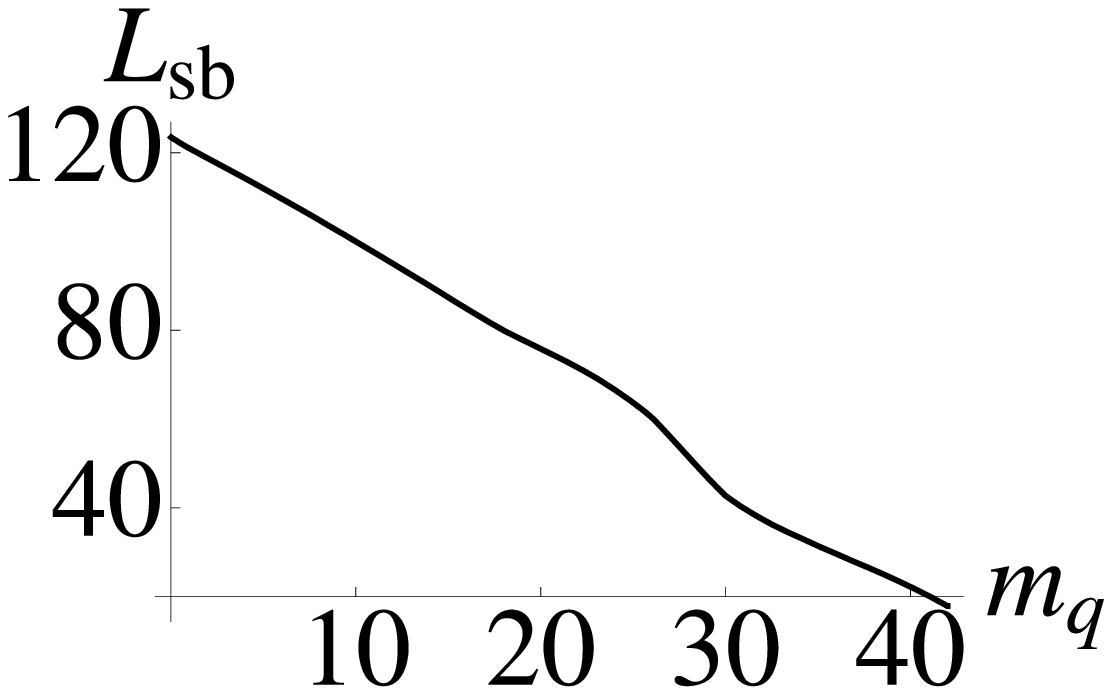}
\caption{The string breaking length for constant glueball scale $h_0$. In the first plot $M_Q=50$ and the two almost coincident lines correspond to $m_q=1$  and $\mu^2=0.125$. In the second plot $N_f=1$ and the two almost coincident lines correspond to $m_q=1$ and $\mu^2=0.125$. In the third and fourth plot $M_Q=50$ and $N_f=1$.}
\label{lsbposs1}
\end{figure}

\begin{figure}
 \centering
\includegraphics[width=.24\textwidth]{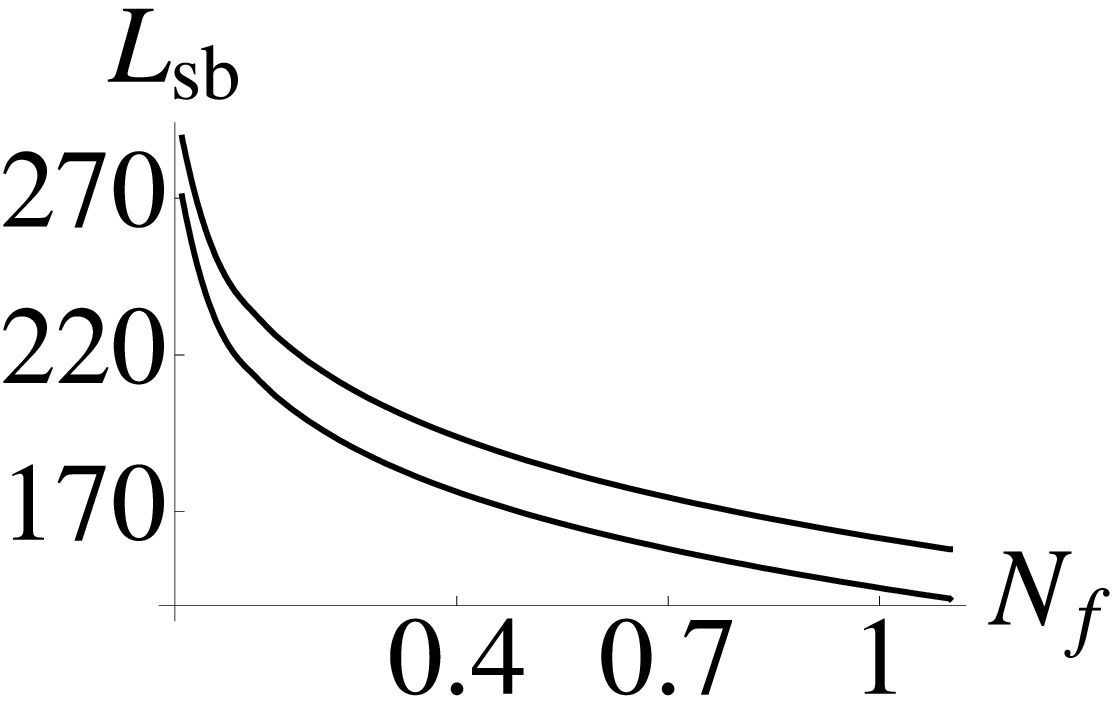}
\includegraphics[width=.24\textwidth]{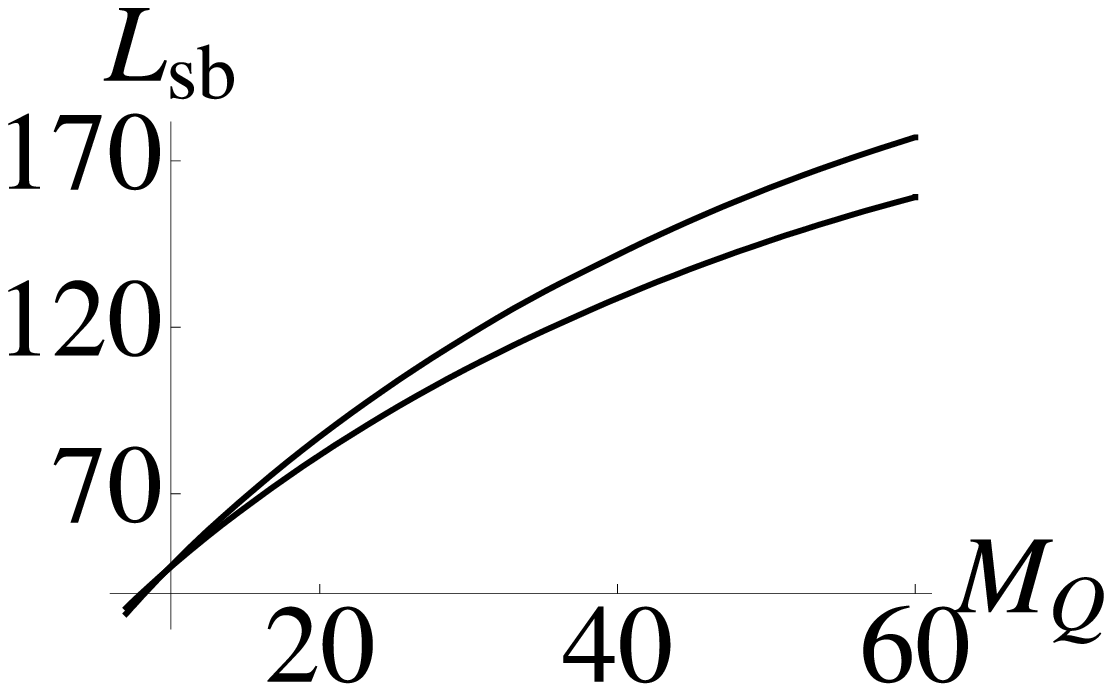}
\includegraphics[width=.24\textwidth]{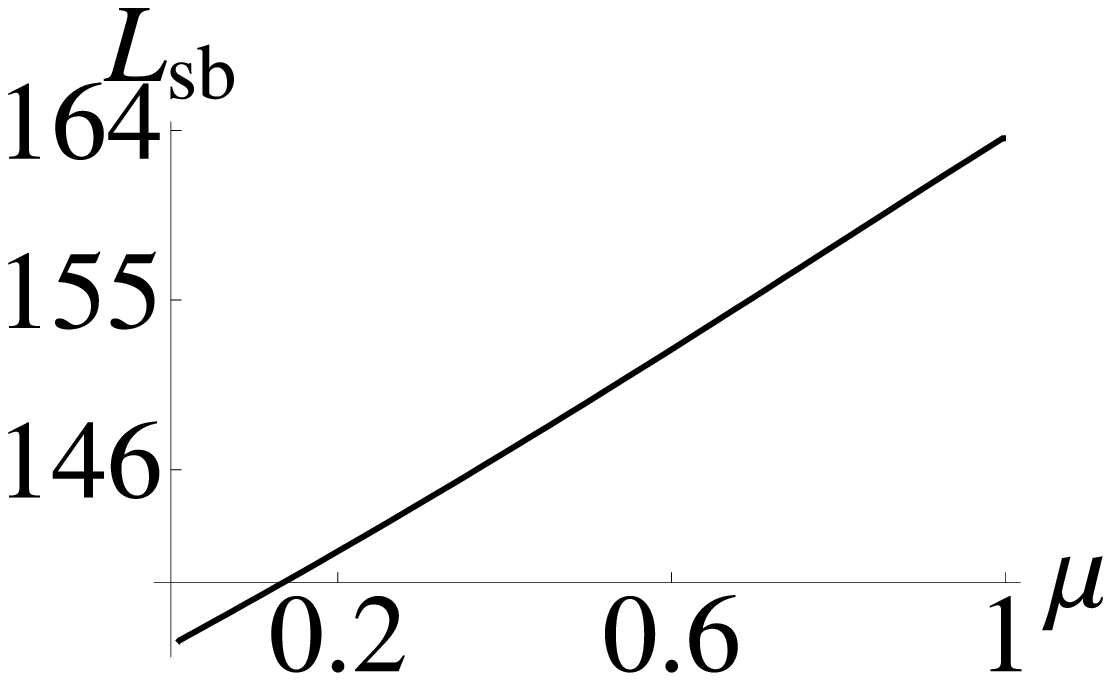}
\includegraphics[width=.24\textwidth]{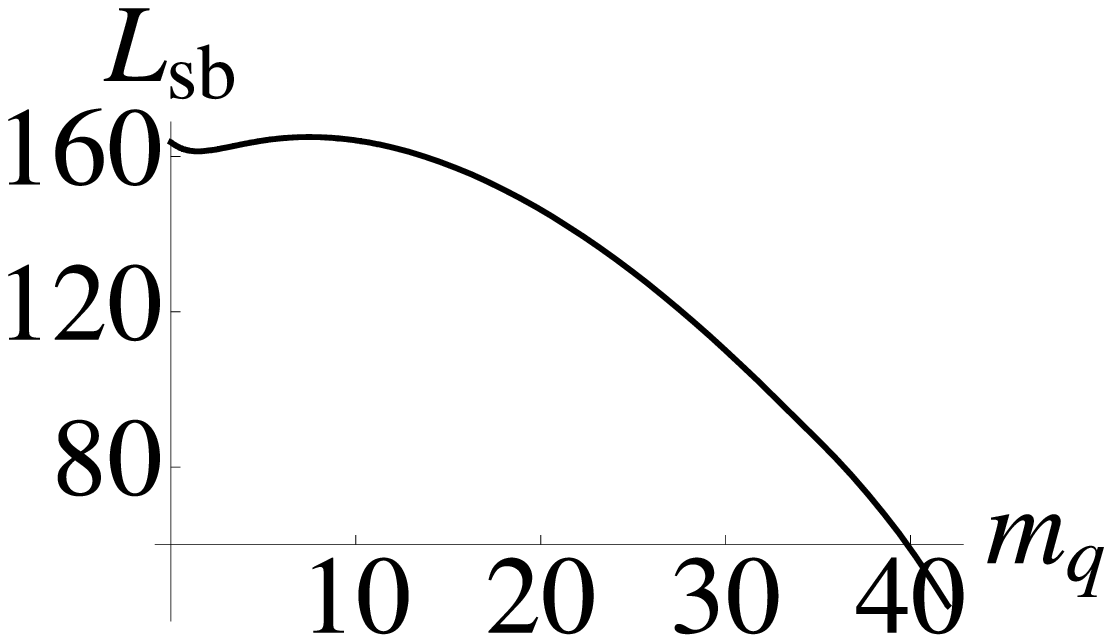}
\caption{The string breaking length for constant string tension $T$. In the first plot $M_Q=50$ and the higher (lower) line corresponds to $m_q=1$ ($\mu^2=0.125$). In the second plot $N_f=1$ and the higher (lower) line corresponds to $m_q=1$ ($\mu^2=0.125$). In the third and fourth plots $M_Q=50$ and $N_f=1$.}
\label{lsbposs2}
\end{figure}

\setcounter{equation}{0}
As it was observed in \cite{Bigazzi:2008gd,massivekw}, due to the smearing, a generic decay $\bar Q Q\rightarrow \bar Q q+\bar q Q$ is suppressed as $1/N_c$ ($1/M$ in our case). To get a decay rate which is not strongly suppressed we can consider the possibility of producing a large number of heavy-light mesons up to some reference ones. The latter are, arbitrarily, chosen as the mesons whose ``dynamical'' quarks have the same internal charges as the static ones. The string picture of the lowest energy configuration for one of these heavy-light mesons is in terms of a straight string stretching from the probe $Q$ brane to a dynamical, parallel, $q$ brane. Choosing the dynamical flavor brane as the one having the minimal possible distance $\tau_{min}=\tau_q$ from the origin, we get that the energy of the string, and so the mass of the corresponding meson, is thus given by $M_Q-m_q$. The minimal separation at which a pair of such mesons can be produced is called ``string breaking length'' $L_{sb}$, and in the present setup\footnote{For previous studies of $L_{sb}$ in a quenched D3-D7 ${\cal N}=2$ setup, see \cite{kkw}.} it is defined as
\be\label{stringbreaking}
V(L_{sb})=\cases{-2 m_q \,\,,\qquad\qquad \mu >1\,,
\cr \cr 0 \,\,,\qquad\quad\qquad\quad \mu<1\,.}
\ee
The behavior of $L_{sb}$ as a function of the flavor parameters is shown in figures \ref{lsbposs1}  (constant glueball scale) and \ref{lsbposs2}  (constant string tension), for the two ``Possibilities'' discussed in section \ref{possibilities}.

The behavior of the string breaking length $L_{sb}$ is similar to the behavior of the screening length $L_s$, apart from its dependence on $m_q$, due to the presence of the latter in the defining relation (\ref{stringbreaking}).
In fact, the string breaking length $L_{sb}$ is a monotonically increasing function of $N_f$ if the glueball scale is kept fixed (fig. \ref{lsbposs1}, first plot), while it is monotonically decreasing if it is the string tension to be kept fixed (fig. \ref{lsbposs2}, first plot).
In both cases, $L_{sb}$ is monotonically increasing with the static quark mass $M_Q$ (figs. \ref{lsbposs1}, \ref{lsbposs2}, second plots) and with the dynamical flavor mass $\mu$ in the regime $\mu<1$ (figs. \ref{lsbposs1}, \ref{lsbposs2}, third plots).
For $\mu>1$, $L_{sb}$ is monotonically decreasing with $\mu$ in the constant glueball scale case (fig. \ref{lsbposs1}, fourth plot).
In the constant string tension case, instead, $L_{sb}$ has a very peculiar behavior, displaying both a local minimum for small $m_q$ and a local maximum for intermediate values of $m_q$ (fig. \ref{lsbposs2}, fourth plot).\footnote{In the Heaviside approximation there is just the local maximum, while the local minimum is absent.}
Let us stress again that, due to the explicit presence of $m_q$ in (\ref{stringbreaking}) for $\mu>1$, the behavior of $L_{sb}$ in the latter regime needs not to be continuously connected to that in the $\mu<1$ regime.



\begin{thebibliography}{99}
\bibitem{wilson}K. G. Wilson, "Confinement of quarks", Phys. Rev. D {\bf 10}, 2445 (1974).
\bibitem{davies}C.~T.~H.~Davies {\it et al.}  [HPQCD Collaboration and UKQCD Collaboration
                  and MILC Collaboration],
  ``High-precision lattice QCD confronts experiment,''
  Phys.\ Rev.\ Lett.\  {\bf 92}, 022001 (2004)
  [arXiv:hep-lat/0304004].
\bibitem{ks}I.~R.~Klebanov and M.~J.~Strassler,
  ``Supergravity and a confining gauge theory: Duality cascades and
  chiSB-resolution of naked singularities,''
  JHEP {\bf 0008}, 052 (2000)
  [arXiv:hep-th/0007191].
 \bibitem{localized}
O.~Aharony, A.~Fayyazuddin and J.~M.~Maldacena,
  ``The large N limit of N = 2,1 field theories from three-branes in
  F-theory,''
  JHEP {\bf 9807}, 013 (1998)
  [arXiv:hep-th/9806159].
 M.~Grana and J.~Polchinski,
  ``Gauge / gravity duals with holomorphic dilaton,''
  Phys.\ Rev.\  D {\bf 65}, 126005 (2002)
  [arXiv:hep-th/0106014].
M.~Bertolini, P.~Di Vecchia, M.~Frau, A.~Lerda and R.~Marotta,
  ``N = 2 gauge theories on systems of fractional D3/D7 branes,''
  Nucl.\ Phys.\  B {\bf 621}, 157 (2002)
  [arXiv:hep-th/0107057].
B.~A.~Burrington, J.~T.~Liu, L.~A.~Pando Zayas and D.~Vaman,
  ``Holographic duals of flavored N = 1 super Yang-Mills: Beyond the probe
  approximation,''
  JHEP {\bf 0502}, 022 (2005)
  [arXiv:hep-th/0406207].
S.~A.~Cherkis and A.~Hashimoto,
  ``Supergravity solution of intersecting branes and AdS/CFT with flavor,''
  JHEP {\bf 0211}, 036 (2002)
  [arXiv:hep-th/0210105].
H.~Nastase,
  ``On Dp-Dp+4 systems, QCD dual and phenomenology,''
  arXiv:hep-th/0305069.
 B.~A.~Burrington, V.~S.~Kaplunovsky and J.~Sonnenschein,
  ``Localized Backreacted Flavor Branes in Holographic QCD,''
  JHEP {\bf 0802}, 001 (2008)
  [arXiv:0708.1234 [hep-th]].
\bibitem{vaman}J.~Erdmenger and I.~Kirsch,
  ``Mesons in gauge / gravity dual with large number of fundamental fields,''
  JHEP {\bf 0412}, 025 (2004)
  [arXiv:hep-th/0408113].
\bibitem{paris}F.~Bigazzi, R.~Casero, A.~L.~Cotrone, E.~Kiritsis and A.~Paredes,
  ``Non-critical holography and four-dimensional CFT's with fundamentals,''
  JHEP {\bf 0510}, 012 (2005)
  [arXiv:hep-th/0505140].
\bibitem{cnp} R.~Casero, C.~Nunez and A.~Paredes,
  ``Towards the string dual of N = 1 SQCD-like theories,''
  Phys.\ Rev.\  D {\bf 73}, 086005 (2006)
  [arXiv:hep-th/0602027].
\bibitem{Paredes:2006wb}
  A.~Paredes,
  ``On unquenched N = 2 holographic flavor,''
  JHEP {\bf 0612}, 032 (2006)
  [arXiv:hep-th/0610270].
 \bibitem{Benini:2006hh}
  F.~Benini, F.~Canoura, S.~Cremonesi, C.~Nunez and A.~V.~Ramallo,
  ``Unquenched flavors in the Klebanov-Witten model,''
  JHEP {\bf 0702}, 090 (2007)
  [arXiv:hep-th/0612118].
\bibitem{Bertoldi:2007sf}
  G.~Bertoldi, F.~Bigazzi, A.~L.~Cotrone and J.~D.~Edelstein,
  ``Holography and Unquenched Quark-Gluon Plasmas,''
  Phys.\ Rev.\  D {\bf 76}, 065007 (2007)
  [arXiv:hep-th/0702225].
  \bibitem{Benini:2007gx}
  F.~Benini, F.~Canoura, S.~Cremonesi, C.~Nunez and A.~V.~Ramallo,
  ``Backreacting Flavors in the Klebanov-Strassler Background,''
  JHEP {\bf 0709}, 109 (2007)
  [arXiv:0706.1238 [hep-th]].
\bibitem{Cotrone:2007qa}
  A.~L.~Cotrone, J.~M.~Pons and P.~Talavera,
  ``Notes on a SQCD-like plasma dual and holographic renormalization,''
  JHEP {\bf 0711}, 034 (2007)
  [arXiv:0706.2766 [hep-th]].
\bibitem{Casero:2007jj}
  R.~Casero, C.~Nunez and A.~Paredes,
  ``Elaborations on the String Dual to N=1 SQCD,''
  Phys.\ Rev.\  D {\bf 77}, 046003 (2008)
  [arXiv:0709.3421 [hep-th]].
  \bibitem{Benini:2007kg}F.~Benini,
  ``A chiral cascade via backreacting D7-branes with flux,'' JHEP {\bf 0810}, 051 (2008)
  [arXiv:0710.0374 [hep-th]].
\bibitem{Caceres:2007mu}
  E.~Caceres, R.~Flauger, M.~Ihl and T.~Wrase,
  ``New Supergravity Backgrounds Dual to N=1 SQCD-like Theories with $N_f=2N_c$,''
  JHEP {\bf 0803}, 020 (2008)
  [arXiv:0711.4878 [hep-th]].
\bibitem{Canoura:2008at}
  F.~Canoura, P.~Merlatti and A.~V.~Ramallo,
  ``The supergravity dual of 3d supersymmetric gauge theories with unquenched
  flavors,''
  JHEP {\bf 0805}, 011 (2008)
  [arXiv:0803.1475 [hep-th]].
  \bibitem{Bigazzi:2008gd}
  F.~Bigazzi, A.~L.~Cotrone, C.~Nunez and A.~Paredes,
  ``Heavy quark potential with dynamical flavors: a first order transition,''
  arXiv:0806.1741 [hep-th].
  \bibitem{massivekw}
F.~Bigazzi, A.~L.~Cotrone and A.~Paredes,
  ``Klebanov-Witten theory with massive dynamical flavors,''
  JHEP {\bf 0809}, 048 (2008)
  [arXiv:0807.0298 [hep-th]].
  \bibitem{carlosetal}C.~Hoyos-Badajoz, C.~Nunez and I.~Papadimitriou,
  ``Comments on the String dual to N=1 SQCD,''
Phys.\ Rev.\  D {\bf 78}, 086005 (2008)
  [arXiv:0807.3039 [hep-th]].
\bibitem{Arean:2008az}
  D.~Arean, P.~Merlatti, C.~Nunez and A.~V.~Ramallo,
  ``String duals of two-dimensional (4,4) supersymmetric gauge theories,''
  arXiv:0810.1053 [hep-th].
  \bibitem{cargese} F.~Bigazzi, A.~L.~Cotrone and A.~Paredes,
  ``Phase transitions in large N(c) heavy quark potentials,''
  arXiv:0810.4018 [hep-th].
 \bibitem{varna} F. Bigazzi, A. L. Cotrone, A. Paredes, A. Ramallo,
``Non chiral dynamical flavors and screening on the conifold'',
arXiv:0810.5220 [hep-th].
\bibitem{Gaillard:2008wt}
  J.~Gaillard and J.~Schmude,
  ``On the geometry of string duals with backreacting flavors,''
  arXiv:0811.3646 [hep-th].
 \bibitem{alfonsoetal}A.~V.~Ramallo, J.~P.~Shock and D.~Zoakos,
  ``Holographic flavor in N=4 gauge theories in 3d from wrapped branes,''
  arXiv:0812.1975 [hep-th].
  \bibitem{sfetsos}A.~Brandhuber and K.~Sfetsos,
  ``Wilson loops from multicentre and rotating branes, mass gaps and phase
  structure in gauge theories,''
  Adv.\ Theor.\ Math.\ Phys.\  {\bf 3}, 851 (1999)
  [arXiv:hep-th/9906201].
\bibitem{angelnc}D.~Arean, A.~Paredes and A.~V.~Ramallo,
  ``Adding flavor to the gravity dual of non-commutative gauge theories,''
  JHEP {\bf 0508}, 017 (2005)
  [arXiv:hep-th/0505181].
\bibitem{avramista} S.~D.~Avramis, K.~Sfetsos and K.~Siampos,
  ``Stability of strings dual to flux tubes between static quarks in N=4 SYM,''
  Nucl.\ Phys.\  B {\bf 769}, 44 (2007)
  [arXiv:hep-th/0612139];
 ``Stability of string configurations dual to quarkonium states in AdS/CFT,''
  Nucl.\ Phys.\  B {\bf 793}, 1 (2008)
  [arXiv:0706.2655 [hep-th]].
\bibitem{avramis} S.~D.~Avramis, K.~Sfetsos and D.~Zoakos,
  ``Complex marginal deformations of D3-brane geometries, their Penrose
  limits and giant gravitons,''
  Nucl.\ Phys.\  B {\bf 787}, 55 (2007)
  [arXiv:0704.2067 [hep-th]].
  \bibitem{kk}A.~Karch and E.~Katz,
  ``Adding flavor to AdS/CFT,''
  JHEP {\bf 0206}, 043 (2002)
  [arXiv:hep-th/0205236].
 \bibitem{kuper}  S.~Kuperstein,
  ``Meson spectroscopy from holomorphic probes on the warped deformed
  conifold,''
  JHEP {\bf 0503}, 014 (2005)
  [arXiv:hep-th/0411097].
   \bibitem{mn}A.~H.~Chamseddine and M.~S.~Volkov,
  ``Non-Abelian BPS monopoles in N = 4 gauged supergravity,''
  Phys.\ Rev.\ Lett.\  {\bf 79}, 3343 (1997)
  [arXiv:hep-th/9707176].  J.~M.~Maldacena and C.~Nunez,
  ``Towards the large N limit of pure N = 1 super Yang Mills,''
  Phys.\ Rev.\ Lett.\  {\bf 86}, 588 (2001)
  [arXiv:hep-th/0008001].
  \bibitem{ouyang}
  P.~Ouyang,
  ``Holomorphic D7-branes and flavored N = 1 gauge theories,''
  Nucl.\ Phys.\  B {\bf 699}, 207 (2004)
  [arXiv:hep-th/0311084].
   \bibitem{kw} I.~R.~Klebanov and E.~Witten,
  ``Superconformal field theory on threebranes at a Calabi-Yau  singularity,''
  Nucl.\ Phys.\  B {\bf 536}, 199 (1998)
  [arXiv:hep-th/9807080].
\bibitem{km}
  I.~R.~Klebanov and J.~M.~Maldacena,
  ``Superconformal gauge theories and non-critical superstrings,''
  Int.\ J.\ Mod.\ Phys.\  A {\bf 19}, 5003 (2004)
  [arXiv:hep-th/0409133].
\bibitem{kotsi}P.~Koerber and D.~Tsimpis,
  ``Supersymmetric sources, integrability and generalized-structure
  compactifications,''
  JHEP {\bf 0708}, 082 (2007)
  [arXiv:0706.1244 [hep-th]].
\bibitem{noimesons}F.~Bigazzi, A.~L.~Cotrone, A.~Paredes and A.~V.~Ramallo,
``Screening effects on meson masses from holography,''
 arXiv:0903.4747 [hep-th].
 \bibitem{maldawilson} J.~M.~Maldacena,
  ``Wilson loops in large N field theories,''
  Phys.\ Rev.\ Lett.\  {\bf 80}, 4859 (1998)
  [arXiv:hep-th/9803002].  S.~J.~Rey and J.~T.~Yee,
  ``Macroscopic strings as heavy quarks in large N gauge theory and  anti-de
  Sitter supergravity,''
  Eur.\ Phys.\ J.\  C {\bf 22}, 379 (2001)
  [arXiv:hep-th/9803001].
  \bibitem{Berg:2006xy}
  M.~Berg, M.~Haack and W.~Mueck,
  ``Glueballs vs. gluinoballs: Fluctuation spectra in non-AdS/non-CFT,''
  Nucl.\ Phys.\  B {\bf 789}, 1 (2008)
  [arXiv:hep-th/0612224].
  \bibitem{Gubser:2004qj}
  S.~S.~Gubser, C.~P.~Herzog and I.~R.~Klebanov,
  ``Symmetry breaking and axionic strings in the warped deformed conifold,''
  JHEP {\bf 0409}, 036 (2004)
  [arXiv:hep-th/0405282].
   \bibitem{Herzog:2008bp}
 C.~P.~Herzog, S.~A.~Stricker and A.~Vuorinen,
  ``Remarks on Heavy-Light Mesons from AdS/CFT,''
  JHEP {\bf 0805}, 070 (2008)
  [arXiv:0802.2956 [hep-th]].
  \bibitem{Klebanov:2000nc}
  I.~R.~Klebanov and A.~A.~Tseytlin,
  ``Gravity duals of supersymmetric SU(N) x SU(N+M) gauge theories,''
  Nucl.\ Phys.\  B {\bf 578}, 123 (2000)
  [arXiv:hep-th/0002159].
  \bibitem{lebellac}M.~Le Bellac,
  ``Quantum And Statistical Field Theory,''
{\it  Oxford, UK: Clarendon (1991)}.
\bibitem{cata}T. Poston and I. Stewart, ``Catastrophe Theory and its Applications'', Pitman publishing limited, 1978.
\bibitem{Chamblin:1999tk}
   A.~Chamblin, R.~Emparan, C.~V.~Johnson and R.~C.~Myers,
   ``Charged AdS black holes and catastrophic holography,''
   Phys.\ Rev.\  D {\bf 60}, 064018 (1999)
   [arXiv:hep-th/9902170].
\bibitem{Ryu}
  S.~Ryu and T.~Takayanagi,
  ``Holographic derivation of entanglement entropy from AdS/CFT,''
  Phys.\ Rev.\ Lett.\  {\bf 96} (2006) 181602
  [arXiv:hep-th/0603001].
  S.~Ryu and T.~Takayanagi,
  ``Aspects of holographic entanglement entropy,''
  JHEP {\bf 0608} (2006) 045
  [arXiv:hep-th/0605073].
 \bibitem{Klebanov:2007ws}
  I.~R.~Klebanov, D.~Kutasov and A.~Murugan,
``Entanglement as a Probe of Confinement,''
  Nucl.\ Phys.\  B {\bf 796} (2008) 274
  [arXiv:0709.2140 [hep-th]].
\bibitem{pallab}P.~Basu and A.~Mukherjee,
  ``Dissolved deconfinement: Phase Structure of large N gauge theories with
  fundamental matter,''
  Phys.\ Rev.\  D {\bf 78}, 045012 (2008)
  [arXiv:0803.1880 [hep-th]].
\bibitem{patse} G.~Papadopoulos and A.~A.~Tseytlin,
  ``Complex geometry of conifolds and 5-brane wrapped on 2-sphere,''
  Class.\ Quant.\ Grav.\  {\bf 18}, 1333 (2001)
  [arXiv:hep-th/0012034].
\bibitem{kkw}A.~Karch, E.~Katz and N.~Weiner,
  ``Hadron masses and screening from AdS Wilson loops,''
  Phys.\ Rev.\ Lett.\  {\bf 90}, 091601 (2003)
  [arXiv:hep-th/0211107].
\end{thebibliography}
\end{document}